%% file: Mrk501_MW2012.tex
\documentclass[longauth]{aa}  

\usepackage[switch]{lineno}

\usepackage{color}
\usepackage{float}

\usepackage{graphicx}
\usepackage{natbib,,hyperref,ifthen}

\bibpunct{(}{)}{;}{a}{}{,}
\usepackage{txfonts}
\begin{document} 


\title{The extreme HBL behaviour of Markarian~501 during 2012}

\author{
\small
M.~L.~Ahnen\inst{1} \and
S.~Ansoldi\inst{2,19} \and
L.~A.~Antonelli\inst{3} \and
C.~Arcaro\inst{4} \and
A.~Babi\'c\inst{5} \and
B.~Banerjee\inst{6} \and
P.~Bangale\inst{7} \and
U.~Barres de Almeida\inst{7,22} \and
J.~A.~Barrio\inst{8} \and
J.~Becerra Gonz\'alez\inst{9}
% ...
\and
\\
(MAGIC Collaboration)
\\
A.~Arbet-Engels\inst{1} \and
D.~Baack\inst{14} \and
M.~Balbo\inst{27} \and
A.~Biland\inst{1} \and
M.~Blank\inst{15} \and
T.~Bretz\inst{1,28} \and
K.~Bruegge\inst{14} \and
M.~Bulinski\inst{14} \and
J.~Buss\inst{14} \and
A.~Dmytriiev\inst{27} \and
D.~Dorner\inst{15} \and
S.~Einecke\inst{14} \and
D.~Elsaesser\inst{14} \and
T.~Herbst\inst{15} \and
D.~Hildebrand\inst{1} \and
L.~Kortmann\inst{14} \and
L.~Linhoff\inst{14} \and
M.~Mahlke\inst{1,28} \and
K.~Mannheim\inst{15} \and
S.~A.~Mueller\inst{1} \and
D.~Neise\inst{1} \and
A.~Neronov\inst{27} \and
M.~Noethe\inst{14} \and
J.~Oberkirch\inst{14} \and
A.~Paravac\inst{15} \and
W.~Rhode\inst{14} \and
B.~Schleicher\inst{15} \and
F.~Schulz\inst{14} \and
K.~Sedlaczek\inst{14} \and
A.~Shukla\inst{15,*} \and
V.~Sliusar\inst{27} \and
R.~Walter\inst{27} \and
\\
(FACT Collaboration)
\\
A.~Archer\inst{29} \and
W.~Benbow\inst{30} \and
R.~Bird\inst{31} \and
R.~Brose\inst{32,11} \and
J.~H.~Buckley\inst{29} \and
V.~Bugaev\inst{29} \and
J.~L.~Christiansen\inst{33} \and
W.~Cui\inst{34,35} \and
M.~K.~Daniel\inst{30} \and
A.~Falcone\inst{36} \and 
Q.~Feng\inst{37} \and 
J.~P.~Finley\inst{34} \and
G.~H.~Gillanders\inst{38} \and
O.~Gueta\inst{11} \and
D.~Hanna\inst{37} \and 
O.~Hervet\inst{39} \and
J.~Holder\inst{40} \and
G.~Hughes\inst{1,54,*} \and
M.~H\"utten\inst{11} \and
T.~B.~Humensky\inst{41} \and
C.~A.~Johnson\inst{39} \and
P.~Kaaret\inst{42} \and
P.~Kar\inst{43} \and
N.~Kelley-Hoskins\inst{11} \and
M.~Kertzman\inst{44} \and
D.~Kieda\inst{43} \and
M.~Krause\inst{11} \and
F.~Krennrich\inst{45} \and
S.~Kumar\inst{37} \and
M.~J.~Lang\inst{38} \and
T.~T.Y.~Lin\inst{37} \and 
G.~Maier\inst{11} \and
S.~McArthur\inst{34} \and
P.~Moriarty\inst{38} \and
R.~Mukherjee\inst{46} \and
S.~O'Brien\inst{47} \and
R.~A.~Ong\inst{31} \and
A.~N.~Otte\inst{48} \and
N.~Park\inst{49} \and
A.~Petrashyk\inst{41} \and
A.~Pichel\inst{50} \and
M.~Pohl\inst{32,11} \and
J.~Quinn\inst{47} \and
K.~Ragan\inst{37} \and 
P.~T.~Reynolds\inst{51} \and
G.~T.~Richards\inst{40} \and
E.~Roache\inst{30} \and
A.~C.~Rovero\inst{50} \and
C.~Rulten\inst{52} \and
I.~Sadeh\inst{11} \and
M.~Santander\inst{53} \and
G.~H.~Sembroski\inst{34} \and
K.~Shahinyan\inst{52} \and
I.~Sushch\inst{11} \and
J.~Tyler\inst{37} \and 
S.~P.~Wakely\inst{49} \and
A.~Weinstein\inst{45} \and
R.~M.~Wells\inst{45} \and
P.~Wilcox\inst{42} \and
A.~Wilhel\inst{32,11} \and
D.~A.~Williams\inst{39}
T.~J~Williamson\inst{40} \and 
B.~Zitzer\inst{37}
\\
(VERITAS Collaboration)
\\
M.~Perri\inst{55,56} \and
F.~Verrecchia\inst{55,56} \and
C.~Leto\inst{55,56} \and
M.~Villata\inst{57} \and
C.~M. Raiteri\inst{57} \and
V.~M.~Larionov\inst{58,59} \and
D.~A.~Blinov\inst{58,60,61} \and
T.~S.~Grishina\inst{58} \and 
E.~N.~Kopatskaya\inst{58} \and
E.~G.~Larionova\inst{58} \and 
A.~A.~Nikiforova\inst{58,59} \and
D.~A.~Morozova\inst{58} \and 
Yu.~V.~Troitskaya\inst{58} \and 
I.~S.~Troitsky\inst{58} \and 
O.~M.~Kurtanidze\inst{62,63,64} \and
M.~G.~Nikolashvili\inst{62} \and
S.~O.~Kurtanidze\inst{62} \and
G.~N.~Kimeridze\inst{62} \and
R.~A.~Chigladze\inst{62} \and
A.~Strigachev\inst{65} \and
A.~C.~Sadun\inst{66} \and
J.~W.~Moody\inst{67} \and
W.~P.~Chen\inst{68} \and
H.~C.~Lin\inst{68} \and 
J.~A.~Acosta-Pulido\inst{9} \and
M.~J.~Ar\'evalo\inst{9} \and
M.~I.~Carnerero\inst{57} \and
P.~A.~Gonz\'alez-Morales\inst{9} \and
A.~Manilla-Robles\inst{69} \and
H.~Jermak\inst{70} \and
I.~Steele\inst{70} \and
C.~Mundel\inst{70} \and
E.~Ben\'itez\inst{71} \and
D.~Hiriart\inst{72} \and
P.~S.~Smith\inst{73} \and
W.~Max-Moerbeck\inst{74} \and 
A.~C.~S.~Readhead\inst{75} \and
J.~L.~Richards\inst{34} \and
T.~Hovatta\inst{76} \and
A.~L\"ahteenm\"aki\inst{77,78} \and
M.~Tornikoski\inst{77} \and
J.~Tammi\inst{77} \and
M.~Georganopoulos\inst{79,80} \and 
M.~G.~Baring\inst{81}
}

\institute {ETH Zurich, CH-8093 Zurich, Switzerland
\and Universit\`a di Udine, and INFN Trieste, I-33100 Udine, Italy
\and National Institute for Astrophysics (INAF), I-00136 Rome, Italy
\and Universit\`a di Padova and INFN, I-35131 Padova, Italy
\and Croatian MAGIC Consortium: University of Rijeka, 51000 Rijeka, University of Split - FESB, 21000 Split,  University of Zagreb - FER, 10000 Zagreb, University of Osijek, 31000 Osijek and Rudjer Boskovic Institute, 10000 Zagreb, Croatia.
\and Saha Institute of Nuclear Physics, HBNI, 1/AF Bidhannagar, Salt Lake, Sector-1, Kolkata 700064, India
\and Max-Planck-Institut f\"ur Physik, D-80805 M\"unchen, Germany
\and Universidad Complutense, E-28040 Madrid, Spain
\and Inst. de Astrof\'isica de Canarias, E-38200 La Laguna, and Universidad de La Laguna, Dpto. Astrof\'isica, E-38206 La Laguna, Tenerife, Spain
\and University of \L\'od\'z, Department of Astrophysics, PL-90236 \L\'od\'z, Poland
\and Deutsches Elektronen-Synchrotron (DESY), D-15738 Zeuthen, Germany
\and Institut de F\'isica d'Altes Energies (IFAE), The Barcelona Institute of Science and Technology (BIST), E-08193 Bellaterra (Barcelona), Spain
\and Universit\`a  di Siena, and INFN Pisa, I-53100 Siena, Italy
\and Technische Universit\"at Dortmund, D-44221 Dortmund, Germany
\and Universit\"at W\"urzburg, D-97074 W\"urzburg, Germany
\and Finnish MAGIC Consortium: Tuorla Observatory and Finnish Centre of Astronomy with ESO (FINCA), University of Turku, Vaisalantie 20, FI-21500 Piikki\"o, Astronomy Division, University of Oulu, FIN-90014 University of Oulu, Finland
\newpage
\and Departament de F\'isica, and CERES-IEEC, Universitat Aut\'onoma de Barcelona, E-08193 Bellaterra, Spain
\and Universitat de Barcelona, ICC, IEEC-UB, E-08028 Barcelona, Spain
\and Japanese MAGIC Consortium: ICRR, The University of Tokyo, 277-8582 Chiba, Japan; Department of Physics, Kyoto University, 606-8502 Kyoto, Japan; Tokai University, 259-1292 Kanagawa, Japan; The University of Tokushima, 770-8502 Tokushima, Japan
\and Inst. for Nucl. Research and Nucl. Energy, Bulgarian Academy of Sciences, BG-1784 Sofia, Bulgaria
\and Universit\`a di Pisa, and INFN Pisa, I-56126 Pisa, Italy
\and now at Centro Brasileiro de Pesquisas F\'isicas (CBPF), 22290-180 URCA, Rio de Janeiro (RJ), Brasil.
\and Humboldt University of Berlin, Institut f\"ur Physik D-12489 Berlin Germany
\and also at Dipartimento di Fisica, Universit\`a di Trieste, I-34127 Trieste, Italy
\and also at Port d'Informaci\'o Cient\'ifica (PIC) E-08193 Bellaterra (Barcelona) Spain
\and also at INAF-Trieste and Dept. of Physics \& Astronomy, University of Bologna
\and University of Geneva, ISDC Data Center for Astrophysics, Chemin d\'Ecogia 16, 1290 Versoix, Switzerland
\and also at RWTH Aachen University, III. Physikalisches Institut A, Aachen, Germany
\and Department of Physics, Washington University, St. Louis, MO 63130, USA
\and Fred Lawrence Whipple Observatory, Harvard-Smithsonian Center for Astrophysics, Amado, AZ 85645, USA
\and Department of Physics and Astronomy, University of California, Los Angeles, CA 90095, USA
\and Institute of Physics and Astronomy, University of Potsdam, 14476 Potsdam-Golm, Germany
\and Physics Department, California Polytechnic State University, San Luis Obispo, CA 94307, USA
\and Department of Physics and Astronomy, Purdue University, West Lafayette, IN 47907, USA
\and Department of Physics and Center for Astrophysics, Tsinghua University, Beijing 100084, China.
\and Department of Astronomy and Astrophysics, 525 Davey Lab, Pennsylvania State University, University Park, PA 16802, USA
\and Physics Department, McGill University, Montreal, QC H3A 2T8, Canada
\and School of Physics, National University of Ireland Galway, University Road, Galway, Ireland
\and Santa Cruz Institute for Particle Physics and Department of Physics, University of California, Santa Cruz, CA 95064, USA
\and Department of Physics and Astronomy and the Bartol Research Institute, University of Delaware, Newark, DE 19716, USA
\and Physics Department, Columbia University, New York, NY 10027, USA
\and Department of Physics and Astronomy, University of Iowa, Van Allen Hall, Iowa City, IA 52242, USA
\and Department of Physics and Astronomy, University of Utah, Salt Lake City, UT 84112, USA
\and Department of Physics and Astronomy, DePauw University, Greencastle, IN 46135-0037, USA
\and Department of Physics and Astronomy, Iowa State University, Ames, IA 50011, USA
\and Department of Physics and Astronomy, Barnard College, Columbia University, NY 10027, USA
\and School of Physics, University College Dublin, Belfield, Dublin 4, Ireland
\and School of Physics and Center for Relativistic Astrophysics, Georgia Institute of Technology, 837 State Street NW, Atlanta, GA 30332-0430
\and Enrico Fermi Institute, University of Chicago, Chicago, IL 60637, USA
\newpage
\and Instituto de Astronomía y Física del Espacio (IAFE, CONICET-UBA), CC 67 - Suc. 28, (C1428ZAA) Ciudad Autónoma de Buenos Aires, Arg
\and Department of Physical Sciences, Cork Institute of Technology, Bishopstown, Cork, Ireland
\and School of Physics and Astronomy, University of Minnesota, Minneapolis, MN 55455, USA
\and Department of Physics and Astronomy, University of Alabama, Tuscaloosa, AL 35487, USA
\and now at Fred Lawrence Whipple Observatory, Harvard-Smithsonian Center for Astrophysics, Amado, AZ 85645, USA  
\and Space Science Data Center - ASI, via del Politecnico, s.n.c., I-00133, Roma, Italy
\and INAF, Osservatorio Astronomico di Roma, via di Frascati 33, I-00040 Monteporzio, Italy
\and INAF, Osservatorio Astrofisico di Torino, I-10025 Pino Torinese(TO), Italy 
\and Astronomical Institute, St. Petersburg State University, Universitetskij Pr. 28, Petrodvorets, 198504 St. Petersburg, Russia
\and Pulkovo Observatory, St.-Petersburg, Russia
\and Department of Physics and Institute for Theoretical and Computational Physics (ITCP), University of Crete, 71003, Heraklion, Greece
\and Foundation for Research and Technology - Hellas, IESL, Voutes, 7110 Heraklion, Greece
\and Abastumani Observatory, Mt. Kanobili, 0301 Abastumani, Georgia
\and Engelhardt Astronomical Observatory, Kazan Federal University, Tatarstan,  Russia
\and Xinjiang Astronomical Observatory, Chinese Academy of Sciences, Urumqi 830011, China 
\and Institute of Astronomy and National Astronomical Observatory, Bulgarian Academy of Sciences, 72 Tsarigradsko shosse Blvd., 1784 Sofia, Bulgaria
\and Department of Physics, University of Colorado Denver, Denver, Colorado, CO 80217-3364, USA
\and Department of Physics and Astronomy, Brigham Young University, Provo, Utah 84602, USA 
\and Graduate Institute of Astronomy,  National Central University, 300 Zhongda Road,  Zhongli 32001, Taiwan
\and European Southern Observatory, Karl-Schwarzschild-Str. 2, D-85748 Garching, Germany 
\and Astrophysics Research Institute, Liverpool John Moores University, IC2, 146 Brownlow Hill, Liverpool, L3 5RF.
\and Instituto de Astronom\'ia, Universidad Nacional Aut\'onoma de M\'exico, Apdo. Postal 70-264, 04510, Cd. de M\'exico, Mexico
\and Instituto de Astronom\'ia, Universidad Nacional Aut\'onoma de M\'exico, Apdo. Postal 810, 22800, Ensenada, B.C., Mexico
\and Steward Observatory, University of Arizona, Tucson, AZ 85721 USA
\and Universidad de Chile, Departamento de Astronomía, Camino El Observatorio 1515, Las Condes, Santiago, Chile
\and Cahill Centre for Astronomy and Astrophysics, California Institute of Technology, Pasadena, CA 91125, USA
\and Tuorla Observatory, Department of Physics and Astronomy, University of Turku, 20014 Turku, Finland
\and Aalto University Mets\"ahovi Radio Observatory, Mets\"ahovintie 114, 02540 Kylm\"al\"a, Finland
\and Aalto University Department of Electronics and Nanoengineering, P.O. BOX 15500, FI-00076 AALTO, Finland.
\and Department of Physics, University of Maryland Baltimore County, 1000 Hilltop Circle, Baltimore, MD 21250, USA
\and NASA Goddard Space Flight Center, Code 663, Greenbelt, MD 20771, USA
\and Department of Physics and Astronomy - MS 108, Rice University, 6100 Main Street, Houston, Texas 77251-1892, USA
\and {*} Corresponding authors: Gareth Hughes (gareth.hughes@cfa.harvard.edu), David Paneque (dpaneque@mppmu.mpg.de), Amit Shukla (amit.shukla@astro.uni-wuerzburg.de)
}


  \abstract
   {}
   {
     We aim to characterize the multiwavelength emission from Markarian~501 (Mrk~501),
     quantify the energy-dependent variability,
     study the potential multiband correlations and
     describe the temporal evolution of the broadband emission within
     leptonic theoretical scenarios. 
   }
   {A multiwavelength campaign was organized to take place between March and July of 2012.
    Excellent temporal coverage was obtained with more than 25 instruments,
    including the MAGIC, FACT and VERITAS Cherenkov telescopes, the instruments on board the \textit{Swift} and 
    \textit{Fermi} spacecraft, and the telescopes operated by the GASP-WEBT collaboration.
   }
   {
     Mrk~501 showed a very high energy (VHE) gamma-ray flux above 0.2 TeV of $\sim$0.5 
     times the Crab Nebula flux (CU) for most of the campaign.
     The highest activity occurred on 2012 June
     9, when the VHE flux was $\sim$3~CU, and the peak of the
     high-energy spectral component was found to be at $\sim$2~TeV. Both the X-ray
     and VHE gamma-ray 
     spectral slopes were measured 
     to be extremely hard, with spectral indices $\textless$ 2 during
     most of the observing campaign, regardless of the
     X-ray and VHE flux. This study reports the hardest Mrk~501 VHE spectra
     measured to date. The fractional variability was found to
     increase with energy, with the highest variability
     occurring at VHE. 
     Using the complete data set, we found correlation between the X-ray and VHE bands; 
     however, if the June 9 flare is excluded, the correlation disappears 
     (significance $\textless$ 3$\sigma$) despite the existence of
     substantial variability in the X-ray and VHE bands throughout the campaign.
   }
   { 
     The unprecedentedly hard X-ray and VHE spectra measured 
     imply that their low- and high-energy components peaked
     above 5~keV and 0.5~TeV, respectively,
     during a large fraction of
     the observing campaign, and hence that Mrk~501 behaved like an
     extreme high-frequency-peaked blazar (EHBL) throughout the 2012 observing season. This
     suggests that being an EHBL may not be a permanent characteristic
     of a blazar, but rather a state which may change over time.
     The data set acquired shows that the broadband spectral energy
     distribution (SED) of Mrk~501, and its transient evolution, is
     very complex, 
     requiring, within the framework of synchrotron self-Compton (SSC) models, 
     various emission regions for a satisfactory description. Nevertheless the 
     one-zone SSC scenario can successfully describe the segments of the SED where most energy is emitted,
     with a significant correlation 
     between the electron energy density and the VHE gamma-ray activity, suggesting that 
     most of the variability may be explained by the injection of high-energy electrons.
     The one-zone SSC scenario used reproduces the behaviour seen between
     the measured X-ray and VHE gamma-ray fluxes, and predicts that the
     correlation becomes stronger with increasing energy of the X-rays.
   }


   \maketitle


   \input{notes/introduction.tex}

   \input{notes/experiments.tex}

\input{notes/lightcurve.tex}

\input{notes/fractional_var.tex}

\input{notes/correlation.tex}

\input{notes/sed.tex}

\input{notes/discussion.tex}

\input{notes/conclusion.tex}

   \clearpage
   \input{notes/appendix_A.tex}
   \clearpage
   \input{notes/appendix_B.tex}


\clearpage
\begin{acknowledgements}
 
%
%

The MAGIC collaboration would like to thank the Instituto de Astrof\'{\i}sica de Canarias for the excellent working conditions at the Observatorio del Roque de los Muchachos in La Palma. The financial support of the German BMBF and MPG, the Italian INFN and INAF, the Swiss National Fund SNF, the ERDF under the Spanish MINECO (FPA2015-69818-P, FPA2012-36668, FPA2015-68378-P, FPA2015-69210-C6-2-R, FPA2015-69210-C6-4-R, FPA2015-69210-C6-6-R, AYA2015-71042-P, AYA2016-76012-C3-1-P, ESP2015-71662-C2-2-P, CSD2009-00064), and the Japanese JSPS and MEXT is gratefully acknowledged. This work was also supported by the Spanish Centro de Excelencia ``Severo Ochoa'' SEV-2012-0234 and SEV-2015-0548, and Unidad de Excelencia ``Mar\'{\i}a de Maeztu'' MDM-2014-0369, by the Croatian Science Foundation (HrZZ) Project 09/176 and the University of Rijeka Project 13.12.1.3.02, by the DFG Collaborative Research Centers SFB823/C4 and SFB876/C3, the Polish National Research Centre grant UMO-2016/22/M/ST9/00382 and by the Brazilian MCTIC, CNPq and FAPERJ.
\\
\\
VERITAS is supported by grants from the U.S. Department of Energy Office of Science, the U.S. National Science Foundation and the Smithsonian Institution, and by NSERC in Canada. We acknowledge the excellent work of the technical support staff at the Fred Lawrence Whipple Observatory and at the collaborating institutions in the construction and operation of the instrument.
\\
\\
The \textit{Fermi} LAT Collaboration acknowledges generous ongoing support
from a number of agencies and institutes that have supported both the
development and the operation of the LAT as well as scientific data analysis.
These include the National Aeronautics and Space Administration and the
Department of Energy in the United States, the Commissariat \`a l'Energie Atomique
and the Centre National de la Recherche Scientifique / Institut National de Physique
Nucl\'eaire et de Physique des Particules in France, the Agenzia Spaziale Italiana
and the Istituto Nazionale di Fisica Nucleare in Italy, the Ministry of Education,
Culture, Sports, Science and Technology (MEXT), High Energy Accelerator Research
Organization (KEK) and Japan Aerospace Exploration Agency (JAXA) in Japan, and
the K.~A.~Wallenberg Foundation, the Swedish Research Council and the
Swedish National Space Board in Sweden. \\

Additional support for science analysis during the operations phase is gratefully 
acknowledged from the Istituto Nazionale di Astrofisica in Italy and the Centre 
National d'\'Etudes Spatiales in France. This work performed in part under DOE 
Contract DE-AC02-76SF00515. 
\\
\\
The St.Petersburg University team acknowledges support from Russian Science Foundation grant 17-12-01029. The Abastumani team acknowledges financial support by the by Shota
Rustaveli NSF under contract FR/577/6-320/13. This research was partially supported by the Bulgarian National Science Fund of the Ministry of Education and Science under grants DN 08-1/2016 and DN 18-13/2017. The Skinakas Observatory is a collaborative project of the University of Crete, the Foundation for Research and Technology --Hellas, and the Max-Planck-Institut f\"ur Extraterrestrische Physik. We would like to thank the American Association of Variable Star Observers (AAVSO) for making some of the observations used in this study. M.I.C acknowledges financial support by PRIN-SKA-CTA-INAF 2016. This article is partly based on observations made with the IAC80 telescope operated on the island of Tenerife by the Instituto de Astrofisica de Canarias in the Spanish Observatorio del Teide. The Liverpool Telescope is operated on the island of La Palma by Liverpool John Moores University in the Spanish Observatorio del Roque de los Muchachos of the Instituto de Astrofisica de Canarias with financial support from the UK Science and Technology Facilities Council. This work is partly based upon observations carried out at the Observatorio Astron\'omico Nacional on the Sierra San Pedro M\'artir (OAN-SPM), Baja California, Mexico. 
The Steward Observatory spectropolarimetric monitoring program is supported by Fermi Guest Investigator grants NNX09AU10G and NNX12AO93G. The OVRO 40-m monitoring program is supported in part by NASA grants NNX08AW31G, NNX11A043G and NNX14AQ89G, and NSF grants AST-0808050  and AST-1109911. This publication makes use of data obtained at Mets\"{a}hovi Radio Observatory, operated by Aalto University, Finland.

\end{acknowledgements}

\bibliographystyle{aa}
\bibliography{Mrk501_MW2012.bib}

\end{document}

%% file: notes/introduction.tex
\section{Introduction}
\label{sec:introduction}

The galaxy Markarian 501 (Mrk~501; $z$=0.034) was first cataloged, 
along with Markarian 421, 
in an ultra-violet survey \citep{1972Ap......8...89M}. 
At very high energies (VHE; $E$ > 100 GeV) it was first detected 
by the pioneering Whipple imaging atmospheric-Cherenkov telescope \citep[IACT,][]{1996ApJ...456L..83Q}.

Mrk~501 is a BL Lacertae (BL Lac) object, a
member of the blazar subclass of active galactic nuclei (AGN),
the most common source class in the extragalactic VHE catalog\footnote{http://tevcat.uchicago.edu}.
Since the discovery of Mrk~501's VHE emission, it has been extensively studied across all 
wavelengths. The spectral energy distribution (SED) shows the two characteristic broad peaks,
the low-frequency peak from radio to 
X-ray and the high-frequency peak from X-ray to very high energies. 
The first peak is thought to originate from synchrotron emission. The second either from inverse-Compton scattering of electrons from the lower-energy component 
\citep{1985ApJ...298..114M,1992ApJ...397L...5M,1992A&A...256L..27D,1994ApJ...421..153S} or 
from the acceleration of hadrons which produce synchrotron emission or interact to produce pions and, in turn, gamma rays 
\citep{1993A&A...269...67M,2000NewA....5..377A,2000A&A...354..395P}.

Whilst the typical flux of Mrk~501, above 1~TeV, in a non-flaring
state, is about one-third that of the Crab Nebula (Crab units; CU)\footnote{
In this study we use the Crab Nebula VHE emission reported in
\citet{2016APh....72...76A}. The photon flux of the Crab
Nebula above 1 TeV is \mbox{2 $\times$ 10$^{-11}$ cm$^{-2}$ s$^{-1}$.}}, it has shown extraordinary flaring activity, 
the first notable examples occuring in 1997 \citep{1997ApJ...487L.143C,1999A&A...350...17D,1999ApJ...518..693Q}.
Another such flaring episode in the same year \citep{1999A&A...349...11A} showed the flux above 2~TeV ranged from a fraction of 1~CU
to 10~CU, with an average of 3~CU, and the doubling timescale was found to be as short as 15 
hours. In the same period the \textit{BeppoSAX} X-ray satellite reported a hundredfold increase in 
the energy of the synchrotron peak in coincidence with a hardening of the spectrum.

Mrk~501 is an excellent object with which to study blazar phenomena because it
is bright and nearby, which permits significant detections in
relatively short observing times in essentially all energy bands. Therefore,
the absorption of gamma rays in the extragalactic background light (EBL, \citet{2013APh....43..112D,2007A&A...475L...9A,2015MNRAS.451..611B}),
although not negligible, plays a relatively small role below a few
TeV. The flux attenuation factor, $\exp(-\tau)$, at a photon energy of 5~TeV is smaller than 0.5 (for z=0.034) for most EBL models 
\citep{2008A&A...487..837F,2011MNRAS.410.2556D,2012MNRAS.422.3189G}.
In 2008, an extensive multi-instrument program was organised in order to
perform an objective (unbiased by flaring states) and detailed study of the temporal evolution, over many years, of
the broadband emission of Mrk~501 \citep[see e.g.][]{2011ApJ...727..129A,2015A&A...573A..50A,2015ApJ...812...65F,2017A&A...603A..31A}.

Here, we report 
on one of those campaigns, that took place in 2012 and serendipitously observed the largest 
flare since 1997.  This paper is organized as follows: In Section \ref{sec:experiments} the experiments that took part 
in the campaign are described along with their data analysis. Section \ref{sec:lightcurves} 
describes the multiwavelength light curves from these instruments and is followed by Sections 
\ref{sec:variability} and \ref{sec:corr}, in which the multiband
variability and related correlations are characterized.
Section \ref{sec:sed} characterizes the broadband SED within a
standard leptonic scenario, and in Section \ref{sec:discussion}, we discuss the
implications of the osbservational results reported in this
paper. Finally, in Section \ref{sec:conclussion}, we make some concluding remarks.

%% file: notes/experiments.tex
\section{Participating Instruments}
\label{sec:experiments}

   \input{notes/magic.tex}

\input{notes/veritas.tex}

\input{notes/fact.tex}

\input{notes/fermi.tex}

\input{notes/swift.tex}

\input{notes/optical.tex}

\input{notes/radio.tex}

\begin{figure*}
   \centering
   \includegraphics[height=48.5pc,width=46pc]{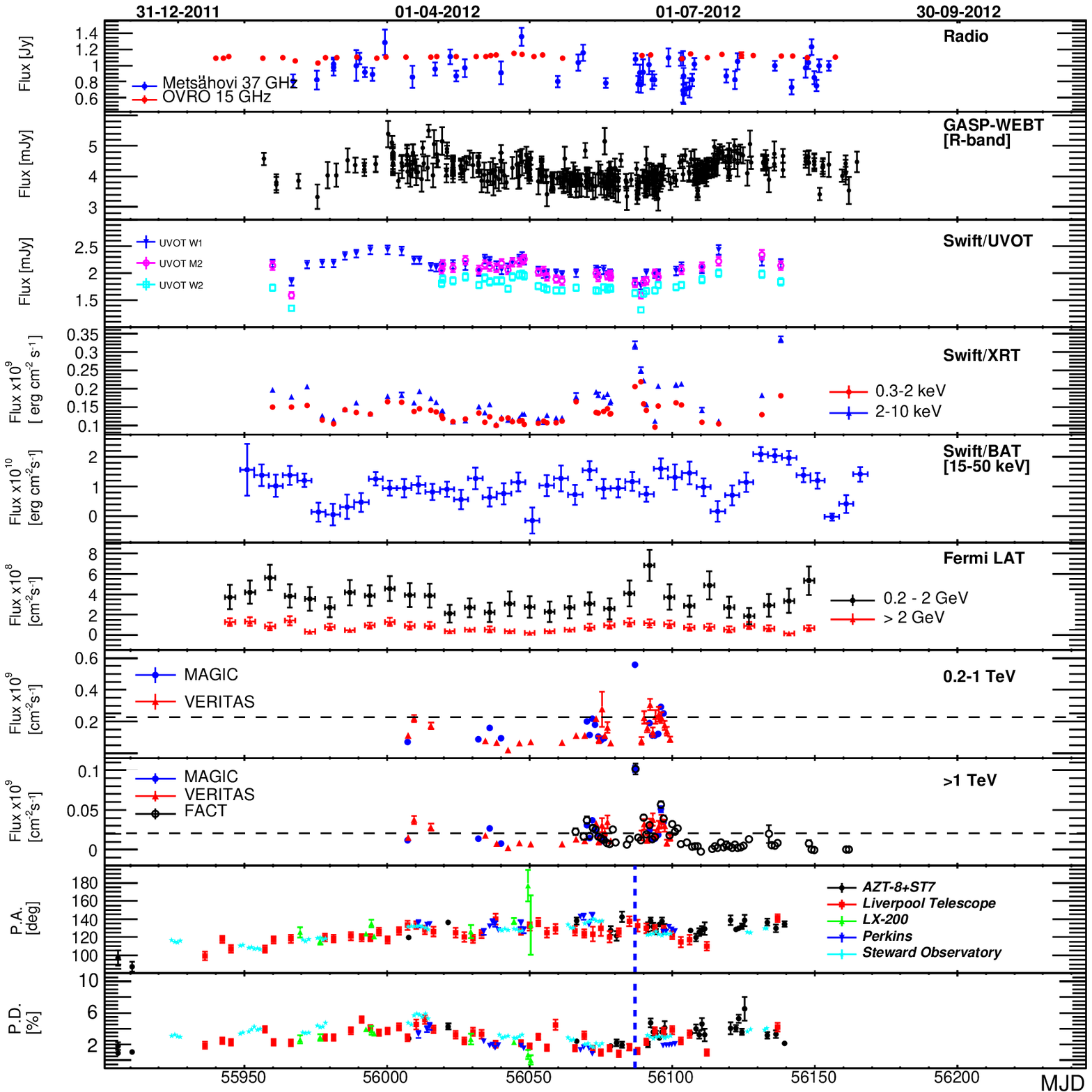}
   \caption{Multiwavelength light curve for Mrk~501 during
     the 2012 campaign. 
The bottom two panels report the electric vector polarization angle (P.A.) and polarization degree (P.D.).
The correspondence between the instruments and the measured quantities is given in the legends.
The horizontal dashed line in the VHE light curves represents 1 CU as reported in
\citet{2016APh....72...76A}, and the blue vertical dotted lines in the panels with the
polarization light curves depict the day of the large VHE flare (MJD~56087).
}
      \label{fig:lightcurvecombinedTeV}
   \end{figure*}

%% file: notes/magic.tex
\subsection{MAGIC}

The Major Atmospheric Gamma-ray Imaging Cherenkov Telescopes (MAGIC) comprise two telescopes located at the Observatorio del Roque de Los 
Muchachos, La Palma, Canary Islands, Spain (2.2~km a.s.l., 28$^{\circ}$ 45$^{\prime}$ N 17$^{\circ}$ 54$^{\prime}$ W). Both 
telescopes are 17~m in diameter and have a parabolic dish. The system is able to detect air showers initiated by gamma rays in the 
energy range from $\sim$50~GeV to $\sim$50~TeV. 

During 2011 and 2012 the readout systems of both telescopes were upgraded, 
and the camera of MAGIC-I (operational since 2003) was replaced, increasing the density of pixels.
This 
resulted in a telescope performance enabling a detection of a $\sim$0.7\% Crab Nebula-like source 
within 50 hours, or a 5\% Crab-like flux in 1 hour of observation. The systematic uncertainties in the 
spectral measurements for a Crab-like point source were estimated to be 11\% in the normalization factor (at $\sim$200 GeV) 
and 0.15 in the power-law slope. The systematic uncertainty in the absolute energy determination is estimated to be 15\%. 
Further details about the performance of the MAGIC telescopes after the hardware upgrade in 2011--2012 can be found in 
\cite{2016APh....72...76A}. The data were analyzed using MARS, the
standard analysis package of MAGIC \citep{Zanin2013-MARS,2016APh....72...76A}.

The data from March and April 2012 were taken in stereo mode, 
whilst the data taken in May and June 2012 were taken with MAGIC-II
operating as a single telescope due to a technical issue which
precluded the operation of MAGIC-I.

%% file: notes/veritas.tex
\subsection{VERITAS}

The VERITAS experiment (Very Energetic Radiation Imaging Telescope Array System) is an array of 
IACTs located at the Fred Lawrence Whipple Observatory in 
southern Arizona (1.3~km a.s.l., N 31$^\circ$40$^{\prime}$, W 110$^\circ$57$^{\prime}$). It 
consists of four Davies-Cotton-type telescopes.
Full array operations 
began in September 2007. Each telescope has a focal length and dish diameter of 12~m.
The total effective mirror area is 106~m$^{2}$ and the camera of each telescope is made up of 499 photomultiplier tubes (PMTs). A single pixel 
has a field of view of 0.15$^{\circ}$. The system operates in the 
energy range from $\sim$100~GeV to $\sim$50~TeV. 

VERITAS has also undergone several upgrades. 
In 2009 one of the telescopes was moved in order to make the array more symmetric.
During the summer of 2012 the VERITAS cameras were upgraded by replacing all of the 
photo-multiplier tubes \citep{2013arXiv1308.4849D}.
For more details on the VERITAS instrument see
\citet{2008AIPC.1085..657H}.

The performance of VERITAS is characterized by a sensitivity of
$\sim$1\% of the Crab Nebula flux to detect (at 5$\sigma$) a
point-like source  in 25 hours of observation, which is equivalent to detecting
(at 5$\sigma$) a $\sim$5\% Crab flux in 1 hour. The uncertainty on the 
VERITAS energy calibration is approximately 20\%. The systematic uncertainty on 
reconstructed spectral indices is estimated at $\pm$0.2, independent of the source 
spectral index, according to studies of \citet{Madhavan2013}. Further details 
about the performance of VERITAS can be found on the VERITAS 
website\footnote{\url{http://veritas.sao.arizona.edu/specifications}}.

%% file: notes/fact.tex
\subsection{FACT}


The First G-APD Cherenkov Telescope (FACT) is
the first Cherenkov telescope to use silicon photomultipliers (SiPM/G-APD) as photodetectors.
As such, the camera consists of 1440 
G-APD sensors, each
with a field of view of 0.11$^{\circ}$ providing a total field of view of 4.5$^{\circ}$.
FACT is located next to the MAGIC telescopes at the Observatorio del Roque de Los Muchachos.
The telescope makes use of the old HEGRA CT3 
\citep{1998NewAR..42..547M} mount, and has a focal length of 5~m and an 
effective dish diameter of $\sim$3~m. The telescope operates in the 
energy range from $\sim$0.8~TeV to $\sim$50~TeV. 
For more details about the design and experimental setup see \cite{2013JInst...8P6008A}.



Since 2012, FACT has been continuously monitoring known TeV blazars, 
including Mrk~501 and Mrk~421. 
FACT provides a dense sampling rate by focusing on a subset of sources
and the ability of the instrument to operate safely during nights of bright ambient light. 
The data are analyzed and processed immediately, and results are available publicly 
online \footnote{\texttt{http://fact-project.org/monitoring/}} within minutes of the observation 
\citep{2013arXiv1311.0478D,2014arXiv1407.1988B,2015arXiv150202582D}.

%% file: notes/fermi.tex
\subsection{\textit{Fermi} LAT}

The Large Area Telescope (LAT, \citet{2009ApJ...697.1071A}) is a pair-conversion 
telescope ({\it Fermi Gamma-ray Space Telescope}) operating in the
energy range from $\sim$30~MeV to $>$TeV. \textit{Fermi} scans the 
sky continuously, completing one scan every 3~hours. The \textit{Fermi}-LAT data presented 
in this paper cover the period from 2011 December 29 (MJD 55924)
to 2012 August 13  (MJD 56152). The data were analyzed using the standard \textit{Fermi} analysis software tools 
(version \textit{v10r1p1}), using the \textit{P8R2\_SOURCE\_V6} response function. 
Events with energy above 0.2 GeV and coming from a 10$^{\circ}$ region of interest (ROI) around Mrk 501 were 
selected, with a 100$^{\circ}$ zenith-angle cut to avoid contamination from the Earth's limb. Two 
background templates were used to model the diffuse Galactic and
isotropic extragalactic background, \textit{gll\_iem\_v06.fits} and
\textit{iso\_P8R2\_SOURCE\_V6\_v06.txt}, 
respectively\footnote{http://fermi.gsfc.nasa.gov/ssc/data/access/lat/BackgroundModels.html}.
All point sources in the third {\it Fermi}-LAT source catalog
\citep[3FGL,][]{2015ApJS..218...23A} located in the 10$^{\circ}$ ROI
and an additional surrounding 5$^{\circ}$-wide annulus were included
in the model. In the unbinned likelihood  fit, the  spectral parameters were set to
the values from the 3FGL,  while  the normalization parameters of the
nine sources within the ROI identified as variable  were
left free. The normalisation of the diffuse components, as well as the
the model parameters related to Mrk~501 were also left free.

Because of the moderate sensitivity of \textit{Fermi}-LAT to detect
Mrk~501 (especially when the source is not flaring), we performed the
unbinned likelihood analysis on one-week time intervals for determining the light
curves in the two energy bands 0.2--2~GeV and $>$2~GeV reported in
Section~\ref{sec:lightcurves}. In both cases we fixed the PL index to 1.75, as was 
done in \cite{Mrk501MW2009_Variability}. On the other hand, in order to increase the
simultaneity with the VHE data, we used 3-day time intervals (centered at the night of the VHE
observations) for most of the unbinned likelihood spectral analyses reported in
Section~\ref{sec:sed}. For those spectral analyses, we performed first
the PL fit in the range from 0.2~GeV to 300~GeV (see spectral
results in Table~\ref{tab:fermiIndex}). Then, we performed the
unbinned likelihood analysis
in three energy bins (split equally in log space from 0.2~GeV to
300~GeV) where the PL index was fixed to the value retrieved from
the spectral fit to the full energy range. Flux upper  limits at 95\% confidence level were calculated whenever
the test statistic (TS) value\footnote{The TS value quantifies the probability of having a point gamma-ray source at
the location specified. It is roughly the square of the significance
value \citep{1996ApJ...461..396M}.}  for the source 
was below 4 \footnote{A TS value of 4 corresponds to a $\sim$2$\sigma$ flux measurement, which is a commonly used threshold for flux measurements of known sources.}.

%% file: notes/swift.tex
\subsection{\textit{Swift}}

The study reported in this paper makes use of the three instruments on
board the \textit{Neil Gehrels Swift Gamma-ray Burst Observatory}
\citep{2004ApJ...611.1005G}; namely the 
Burst Alert Telescope \citep[BAT,][]{2005ApJ...633L..77M}, the X-ray
Telescope \citep[XRT,][]{2005SSRv..120..165B} and the Ultraviolet/Optical 
Telescope \citep[UVOT,][]{2005SSRv..120...95R}.

The 15-50 keV hard X-ray fluxes from BAT were retrieved from the transient monitor results
provided by the \textit{Swift}/BAT team \citep{0067-0049-209-1-14}\footnote{See
\url{http://swift.gsfc.nasa.gov/results/transients/}}, where we made a
weighted average of all the observations performed within temporal
bins of five days. The BAT count rates are converted to energy flux
using that 0.00022 counts cm$^{-2}$ s$^{-1}$ corresponds to
1.26$\times$10$^{-11}$erg~cm$^{-2}$ s$^{-1}$ \citep{Krimm:2013lwa}. 
This conversion is strictly correct only
for sources with the Crab Nebula spectral index in the BAT energy
domain ($\Gamma$=2.1), but the systematic error for sources with
different indices is small and often negligible in comparison with the
statistical uncertainties, as reported in \citet{0067-0049-209-1-14}. 

The XRT and UVOT data come from
dedicated observations organized and performed within the framework of
the planned extensive multi-instrument campaign. In this study we consider the 52 Mrk~501 observations  performed
between 2012 February 2 (MJD 55959) and 2012 July 30 (MJD 56138). All observations were carried out in
the Windowed Timing (WT) readout mode, with an average exposure of 0.9~ks. The data were  
processed using the XRTDAS software package (v.2.9.3), which was developed by the ASI Science Data 
Center and released by HEASARC in the HEASoft package (v.6.15.1).  The data are  
calibrated and cleaned with standard filtering criteria using the \textit{xrtpipeline} task and 
calibration files available from the \textit{Swift}/XRT CALDB (version 20140120). For the spectral 
analysis, events are selected within a 20-pixel ($\sim$46~arcsecond) radius, which contains 90\% of the 
point-spread function (PSF). 
The background was estimated from a nearby circular region with a radius of 40~pixels. Corrections for 
the PSF and CCD defects are applied from response files generated using the \textit{xrtmkarf} task 
and the cumulative exposure map. Before the spectra are fitted the 0.3-10 keV data are binned to 
ensure that there are at least 20~counts in each energy bin. The spectra are then corrected for 
absorption with a neutral-hydrogen column density fixed to the Galactic 21-cm value in the 
direction of Mrk 501, namely 1.55 $\times$ 10$^{20}$ cm$^{-2}$ \citep{2005A&A...440..775K}.

\textit{Swift}/UVOT made between 31 and 52 measurements, depending on the filter used.
The data telemetry volume was reduced using 
the \textit{image mode}, where the photon timing information is discarded and the image is directly 
accumulated on-board. 
In this paper we considered UVOT image data taken within the same observations
acquired by XRT. Here we use the UV lenticular filters, W1, M2 and W2,  
which are the ones that are not affected by the strong flux of the host galaxy.
We evaluated the photometry of the source according to the
recipe in \citet{2008MNRAS.383..627P}, extracting source counts with an aperture
of 5~arcsecond radius and an annular background aperture with inner and outer radii
of 20~arcsecond and 30~arcsecond. The count rates were converted to fluxes using the
updated calibrations \citep{2011AIPC.1358..373B}. Flux values were then corrected for mean Galactic extinction using an $E (B - V )$
value of 0.017 \citep{2011ApJ...737..103S} for the UVOT filter effective
wavelength and the mean Galactic interstellar extinction curve in \citet{1999PASP..111...63F}.

%% file: notes/optical.tex
\subsection{Optical Instruments}
\label{optical:description}

Optical data in the R band were provided by various telescopes around the world, including the ones from 
the GASP-WEBT program \citep{2008A&A...481L..79V,2009A&A...504L...9V}. In this paper we report observations 
performed in the R band from the following observatories: 
Crimean Astrophysical Observatory, St. Petersburg, 
Sierra de San Pedro M\`{a}rtir,
Roque de los Muchachos (KVA), 
Teide (IAC80), 
Lulin (SLT), 
Rozhen (60cm), 
Abastumani (70 cm), 
Skinakas, 
the robotic telescope network AAVSOnet, 
ROVOR and 
iTelescopes. 
The calibration was performed using the stars 
reported by \citet{1998A&AS..130..305V}, the Galactic extinction was corrected using the coefficients given 
in \citet{2011ApJ...737..103S}, and the flux from the host galaxy in the R band was estimated using 
\citet{2007A&A...475..199N} for the apertures of 5 arcsecond and 7.5 arcsecond used by the various 
instruments. The reported fluxes include instrument-specific offsets of a few mJy, owing to the different 
filter spectral responses and analysis procedures of the various optical data sets (e.g. for signal and 
background extraction) in combination with the strong host-galaxy contribution, which is about 2/3 of the 
total flux measured in the R band.  The offsets applied are the following ones: Abastumani = 3.0 mJy; San 
Pedro Martir = -1.8 mJy; Teide = 2.1 mJy; Rozhen = 3.9 mJy; Skinakas = 0.8 mJy; AAVSOnet = -3.8 mJy; 
iTelescopes = -2.5 mJy; ROVOR = -2.7 mJy. These offsets were determined using several of the GASP-WEBT instruments as 
reference, and scaling the other instruments (using simultaneous observations) to match them.  
Additionally, a point-wise fluctuation of 0.2 mJy ($\sim$0.01 mag) was added in quadrature to the 
statistical errors in order to account for 
potential day-to-day differences for observations with the same instrument.

We also report on polarization measurements from five facilities: Lowell Observatory (Perkins telescope), 
St. Petersburg (LX-200), Crimean (AZT-8+ST7), Steward Observatory (2.3m Bok and 1.54m Kuiper telescopes) and Roque de los 
Muchachos (Liverpool telescope).  All polarization measurements 
were obtained from R band imaging polarimetry, except for the measurements from Steward Observatory, which 
are derived from spectropolarimetry between 4000~$\r{A}$ and 7550~$\r{A}$ with a resolution of 
$\sim$15~$\r{A}$. The 
reported values are constructed from the median \textit{Q}/\textit{I} and \textit{U}/\textit{I} 
in the 5000--7000~$\r{A}$ band. The 
effective wavelength of this bandpass is similar to the Kron-Cousins R band. The wavelength dependence 
in the polarization of Mrk~501 seen in the spectro-polarimetry is small and does not significantly affect 
the variability analysis of the various instruments presented here, as can be deduced from the good 
agreement between all the instruments shown in the bottom panels of Figure~\ref{fig:lightcurvecombinedTeV}. 
The details related to the observations and analysis of the polarization data are reported by
\cite{2008A&A...492..389L,2009arXiv0912.3621S,2010ApJ...715..362J,2016MNRAS.462.4267J}.

%% file: notes/radio.tex
\subsection{Radio Observations}

We report here radio observations from telescopes at the Mets\"{a}hovi Radio Observatory 
and the Owens Valley Radio Observatory (OVRO).
The 14~m Mets\"{a}hovi Radio telescope operates at 37 GHz and the OVRO at 15 GHz.
Details of the observation strategies can be found in \citet{1998A&AS..132..305T} and 
\citet{2011ApJS..194...29R}. For both instruments Mrk~501 is a point source, and therefore the 
measurements represent an integration of the full source extension,
which is much larger than the region that is expected to produce the
blazar emission at optical/X-ray and gamma-ray energies that we wish
to study.
However, as reported by \citet{2011ApJ...741...30A}, there is a correlation between radio and GeV 
emission of blazars. 
In the case of Mrk~501,
\citet{Mrk501MW2009_Variability} showed that the radio core emission
increased during a period of high gamma-ray activity,
therefore 
part of the radio emission seems to be related to the gamma-ray component, and should be considered when studying the blazar emission.

%% file: notes/lightcurve.tex
\section{Multiwavelength Light Curve}
\label{sec:lightcurves}

Figure \ref{fig:lightcurvecombinedTeV} shows a complete set of light curves for all 
participating instruments, from radio to VHE.
The first panel from the top shows the radio data from the Mets\"{a}hovi and Owens Valley radio 
observatories.
Each data point represents the average over one night of observations.
Optical data in the R-band, after host galaxy subtraction as prescribed in section 2.6, 
are shown in the second panel. The light curve also
shows very little variability, with just a slow change in flux of
about $\sim$10-20\% on timescales of many tens of days. 
When compared to the 13 years of optical observations from the
Tuorla group\footnote{See \url{http://users.utu.fi/kani/1m/Mkn_501_jy.html}}, one can note
that in 2012 the flux was at a historic minimum.
The ultraviolet data from the \textit{Swift}/UVOT
are presented in the third panel and follows the same pattern as the
R band fluxes. Overall, the low-frequency observations (radio to ultraviolet) show little variation during this period.

In the X-ray band the \textit{Swift}/XRT and BAT light curves show a
large amount of variation, occurring on timescales of days (i.e. much
faster than those in the optical band). 
The \textit{Swift}/XRT points represent nightly fluxes derived from 
$\sim$1~ks observations (where the error bars are smaller than the markers), 
while  the \textit{Swift}/BAT points are
the weighted average of all measurements performed within
5-day intervals. 
On the day of June 9 2012 (MJD 56087) 
a flare is observed where the \textit{Swift}/XRT flux reached 3.2 $\times$ 10$^{-10}$
erg cm$^{-2}$ s$^{-1}$ 
in the 2--10~keV band. Interestingly, the largest flux point in the 0.3--2~keV band 
occurs two days later, indicating 
that the X-ray activity can have a different variability pattern
below and above 2~keV.

The \textit{Fermi}-LAT light curves, which are binned in 7~day time intervals, show some
mild variability. The ability to detect small amplitude variability at these energies is 
strongly limited by the relatively large statistical uncertainty in the flux measurements.

The seventh and eighth panels of Figure~\ref{fig:lightcurvecombinedTeV} show the VHE light curves 
from MAGIC, VERITAS and FACT.
Here we split the VHE information from MAGIC and VERITAS into two bands, 
from 200~GeV to 1~TeV and above 1~TeV.
Each point represents a nightly average, with the 18 MAGIC
observations, obtained from  an average observation of
1.25~hours, and the 28 VERITAS data points, obtained
from an average observation of 0.5~hours.
The VHE emission is highly variable, with 
the average in both bands being approximately 0.7~CU above 1~TeV.
The largest VHE flux is observed on 2012 June 9, where the light
curves show a very clear flare (which is also visible in the X-ray
light curve) with a 0.2--1~TeV flux of 5.6 $\times$ 10$^{-10}$
cm$^{-2}$ s$^{-1}$ (2.8 CU),  and the $>$1~TeV flux reaching 1.0 $\times$ 
10$^{-10}$ cm$^{-2}$ s$^{-1}$ (4.9 CU). 
Unfortunately, VERITAS was not scheduled to observe Mrk~501 on June 9.

FACT observed Mrk~501 for an average of 3.3~hours per night over 73 nights during the 
campaign. As with the other TeV instruments, the data shown are binned nightly.
The FACT fluxes reported in Figure \ref{fig:lightcurvecombinedTeV}  were obtained with a 
first-order polynomial that relates the MAGIC flux (ph cm$^{-2}$ s$^{-1}$ above 1 TeV) and the FACT 
excess rates (events/hour), as explained in Appendix~A.

Measurements of the degree of optical linear polarization and its position
angle are displayed in the bottom panels of Figure~\ref{fig:lightcurvecombinedTeV}.  
As with the optical photometry, the polarization shows only mild variations on time scales of
weeks to months during the campaign.  Variations of the degree of
polarization are muted by the strong contribution of unpolarized starlight
from the host galaxy falling within the observation apertures.
At these optical flux levels and with the apertures used for the ground-based
polarimetry, the optical flux from the host galaxy is about 2/3 of the
flux measured, and hence the intrinsic polarization of the blazar is about a factor of
three higher than observed. 
Different instruments used somewhat different apertures and optical
bands, which implies that the
contribution of the host galaxy to the optical flux and polarization
degree will be somewhat different for the different instruments. 
Since the host galaxy is not subtracted, this leads to small offsets
(at the level of $\sim$1\%) in the measurements of the degree of
polarization. The position angle of the polarization (which is not
affected by the host galaxy)  remains at 120-140$^{\circ}$ for more than a month before and after the VHE
flare.  For comparison, the position angle of the 15~GHz VLBI jet is at
$\sim$150$^{\circ}$ \citep{2009AJ....137.3718L}.  Overall, there is no
apparent optical signature, either in flux or linear polarization, that
can be associated with the gamma-ray activity observed in Mrk~501 during
2012.

%% file: notes/fractional_var.tex
\section{Fractional Variability}
\label{sec:variability}

 \begin{figure*}
   \centering
   \includegraphics[width=20pc]{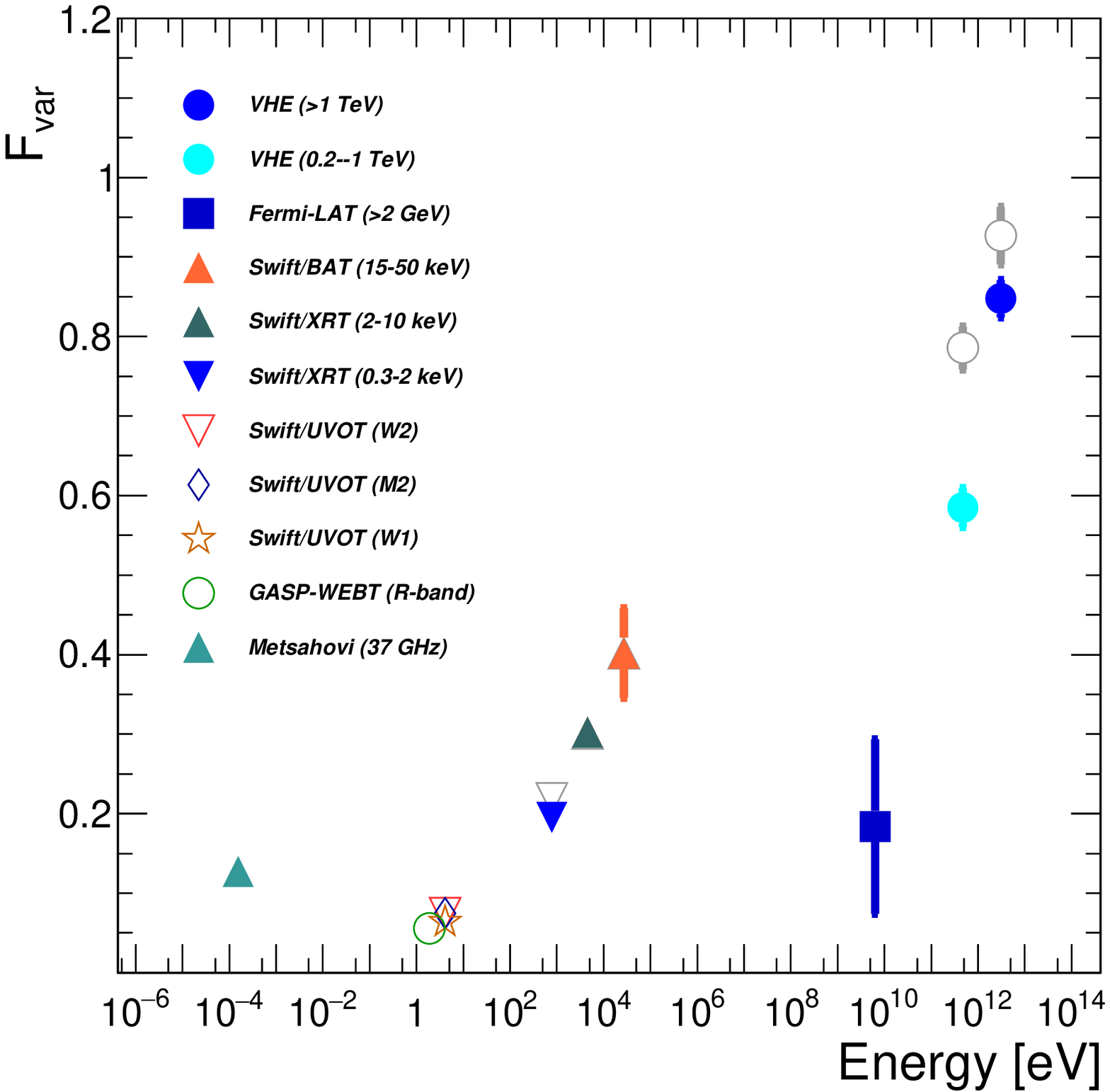}
   \includegraphics[width=20pc]{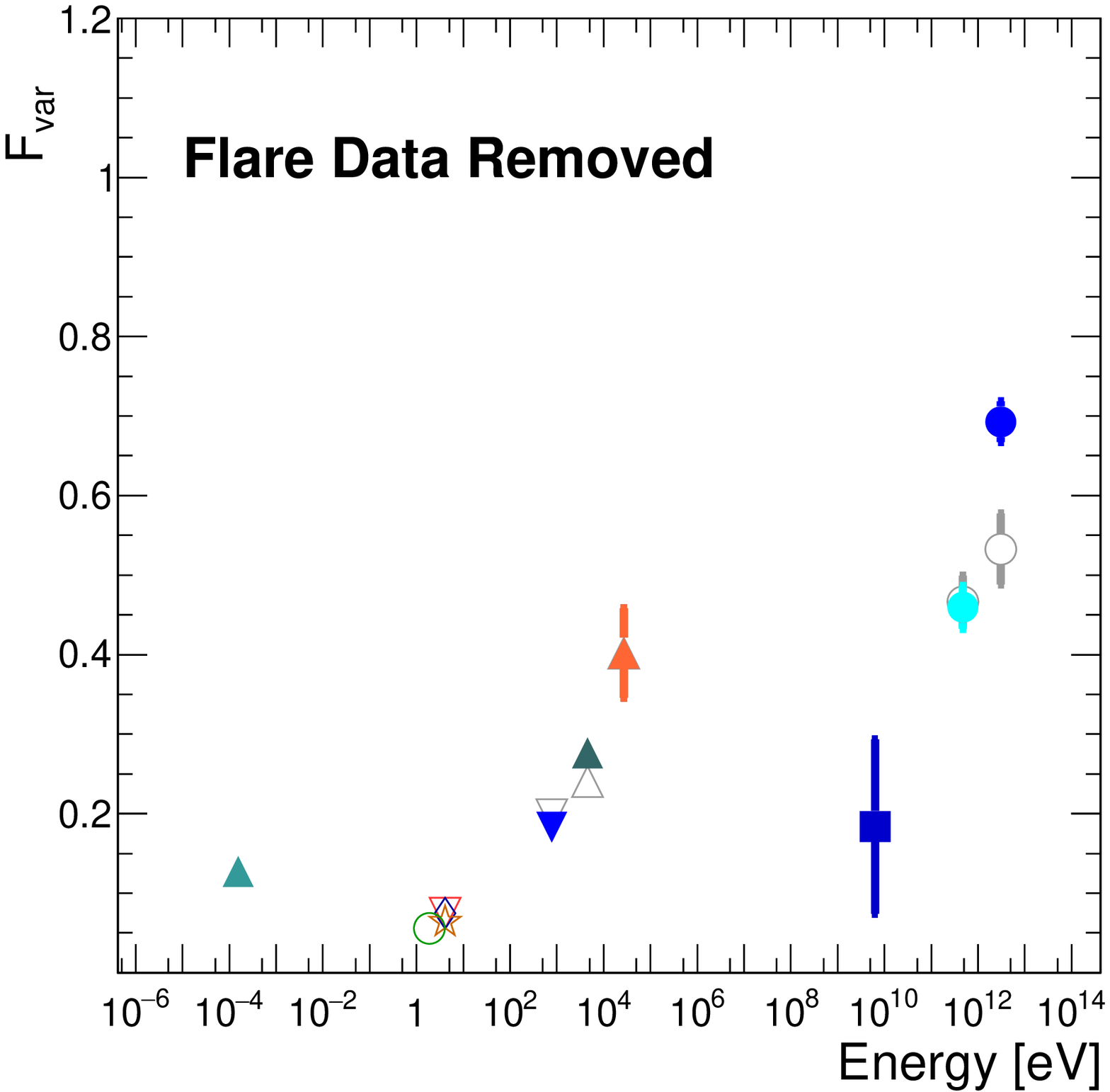}
   \caption{Fractional variability $F_\text{{var}}$ for each instrument as a function of energy.
	    Left panel includes all data, while the right panel includes
	    all data except for the day of the VHE flare (MJD 56087).
            $F_\text{{var}}$ values computed with X-ray and VHE data
            taken within the same night are shown with gray open markers.
            }
              \label{fig:fvar_all}
   \end{figure*}

In order to characterize the variability at each wavelength we follow the prescription of 
\cite{2003MNRAS.345.1271V} where the fractional variability ($F_\text{{var}}$) is defined as

  \begin{equation}
    F_{var} = \sqrt{\frac{S^{2} - \langle \sigma_{err}^{2} \rangle }{ {\langle x \rangle}^2}}
  \end{equation}
Here \textit{S} is the standard deviation of the flux measurement, $\langle\sigma_{err}^{2}\rangle$ 
the mean squared error and ${\langle x \rangle}^{2}$ the square of the
average  photon flux.
The error on $F_\text{{var}}$ is estimated following the prescription of
\cite{2008MNRAS.389.1427P}, as described by \cite{2015A&A...573A..50A}

  \begin{equation}
    \Delta F_\text{{var}} = \sqrt{ F_\text{{var}}^{2} + err( \sigma_\text{{NXS}}^{2} ) } - F_\text{{var}}
  \end{equation}
and $err( \sigma_\text{{NXS}}^{2} )$ is taken from equation 11 in \cite{2003MNRAS.345.1271V} 

  \begin{equation}
    err( \sigma_\text{{NXS}}^{2} ) = \sqrt{ \
\left( \sqrt{\frac{2}{N}} \frac{\langle \sigma^{2}_\text{{err}} \rangle}{\langle x \rangle ^{2}} \right)^{2} \
+ \
\left( \sqrt{\frac{\langle \sigma^{2}_\text{{err}} \rangle}{N}} \frac{2F_\text{{var}}}{\langle x \rangle} \right)^{2} \
}
  \end{equation}
where \textit{N} is the number of flux measurements. This method, commonly used to quantify the variability, has the 
caveat that the resulting $F_\text{{var}}$ and its related 
uncertainty depend on the instrument sensitivity and the observing strategy performed. For instance, densely 
sampled light curves with small uncertainties in the flux measurements may allow us to see flux variations that are 
hidden otherwise, and hence may yield a larger $F_\text{{var}}$ and/or smaller uncertainties in the calculated values 
of $F_\text{{var}}$. This introduces differences in the ability to detect variability in the different 
energy bands. Issues regarding the application of this method, in the context of multiwavelength campaigns,
are discussed by \citet{2014A&A...572A.121A,2015A&A...573A..50A,2015A&A...576A.126A}. In the multi-instrument dataset 
presented in this case, the sensitivity of the instruments {\it Swift}/BAT and {\it Fermi}-LAT precludes the detection of 
Mrk~501 on hour timescales, and hence integration over several days is required (and still yields flux measurements 
with relatively large uncertainties). This means that the {\it Swift}/BAT and {\it Fermi}-LAT $F_\text{{var}}$ values are not 
directly comparable to those of the other 
instruments, for which $F_\text{{var}}$ values computed with nightly observations (and typically smaller error bars) are reported.

Figure \ref{fig:fvar_all} shows the $F_\text{{var}}$ as a function of energy.
The left panel uses all the data presented in Figure \ref{fig:lightcurvecombinedTeV}, with the exception of nights where
there were simultaneous FACT and MAGIC data. In these cases the FACT data are removed. 
The figure displays $F_\text{{var}}$ values for those bands with positive excess variance 
($S^2$ larger than \mbox{$<\sigma_{err}>^2$}); a negative excess
variance is interpreted as 
absence of variability either because there was no variability or because the instruments 
were not sensitive enough to detect it. We obtained negative excess variances
for the 15 GHz radio fluxes measured with OVRO and the 0.2-2~GeV fluxes
measured with {\it Fermi}-LAT. 
The right panel shows the same data except for the flare day (MJD 56087), which has been removed from the multi-instrument dataset, and hence 
shows a more typical behavior of the source
during the 2012 multi-instrument campaign. 

Figure \ref{fig:fvar_all} also reports the values of $F_\text{{var}}$ obtained by using the
X-ray/VHE observations taken simultaneously\footnote{The $F_\text{{var}}$ in the radio and optical bands does not change much when selecting sub-samples of the full dataset because the 
variability in these energy bands is small and the flux variations have longer timescales, in comparison with those 
from the X-ray and VHE bands.}. Additionally, the right panel in Figure \ref{fig:fvar_all} also shows that, 
when the large VHE flare from MJD 56087 is removed, the
$F_\text{{var}}$ changes substantially in the VHE gamma-ray band
(e.g. from 0.93$\pm$0.04 down to 0.53$\pm$0.05  above 1 TeV) but the 
variability changes mildly in the X-ray band (e.g. from
0.301$\pm$0.003 to 0.241$\pm$0.003 at 2-10~keV). In both panels there is a
general increase of the fractional variability with increasing
energy of the emission. These results will be further discussed in
Section~\ref{discussion:variability}.

%% file: notes/correlation.tex
\section{Correlation between the X-ray and VHE gamma-ray emission}
\label{sec:corr}

  \begin{figure*}
   \centering
      \includegraphics[width=20pc]{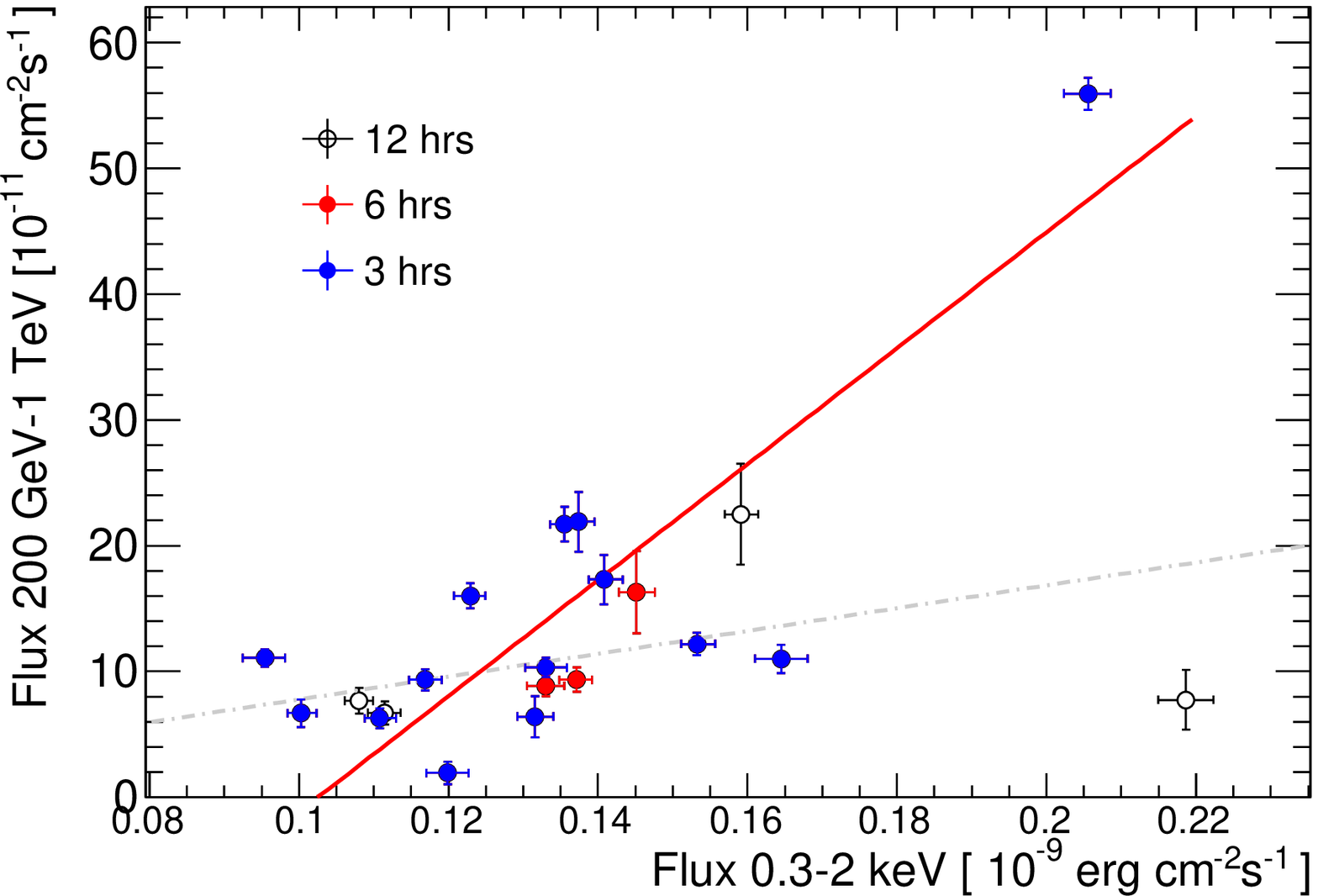}
      \includegraphics[width=20pc]{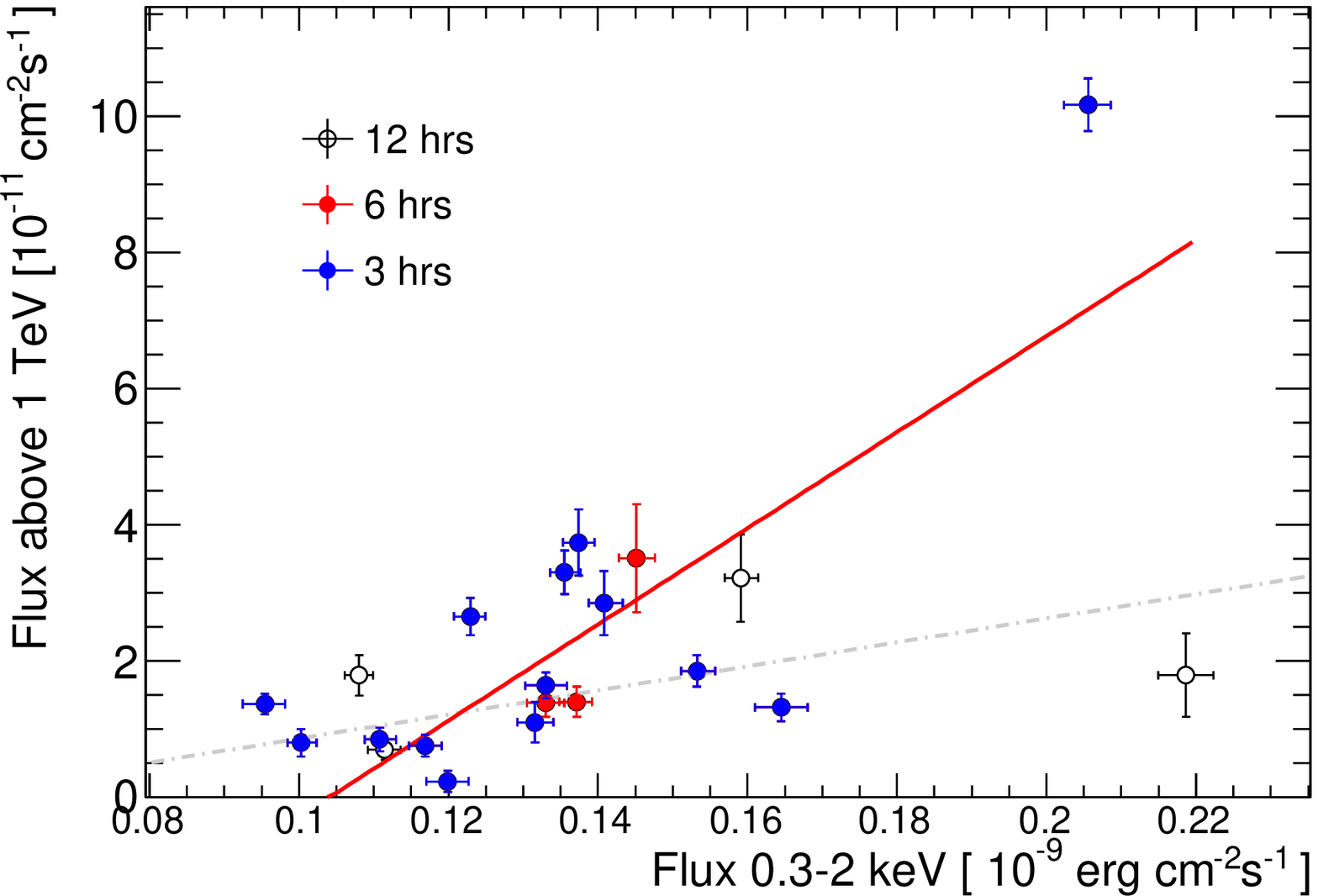}
      \includegraphics[width=20pc]{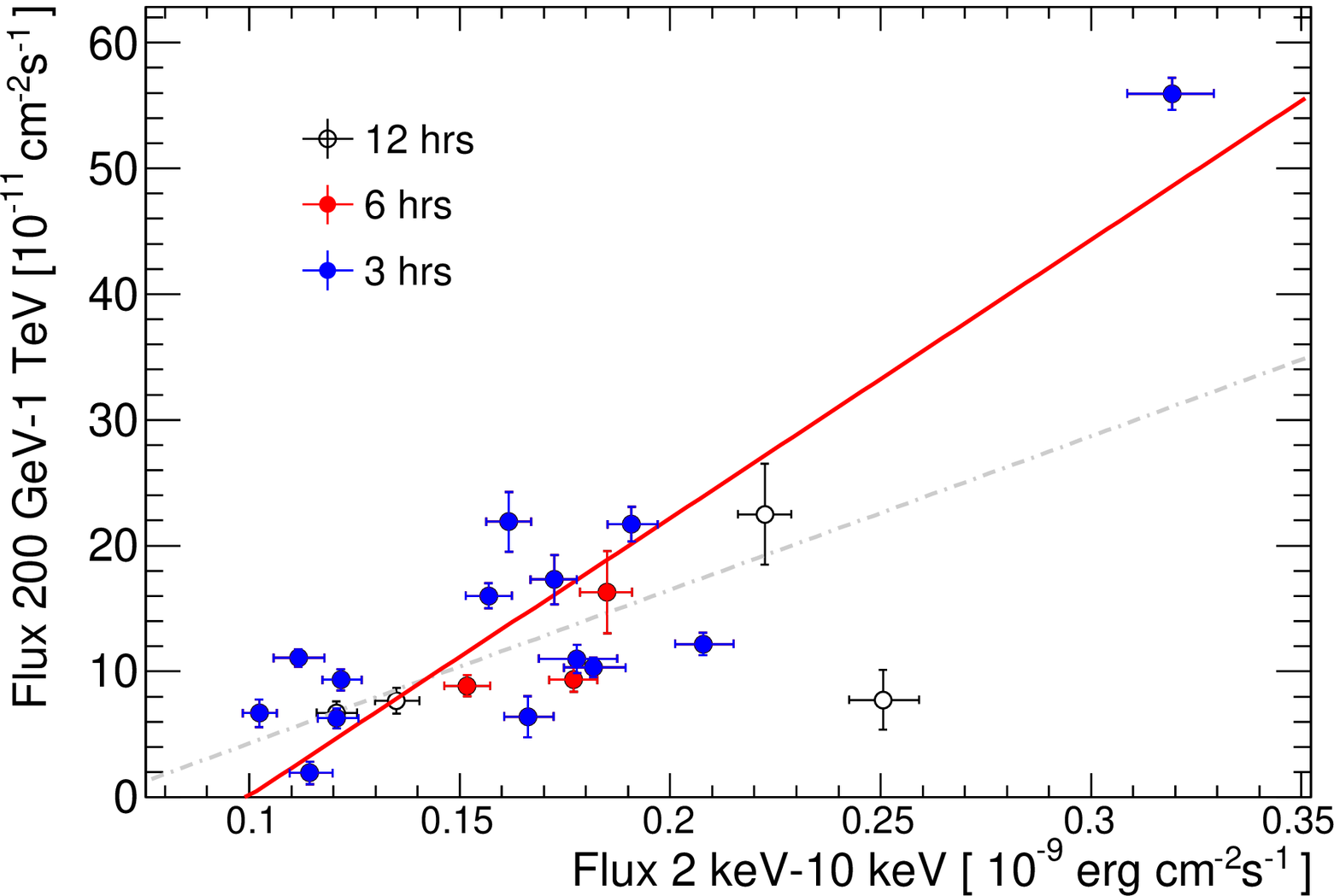}
      \includegraphics[width=20pc]{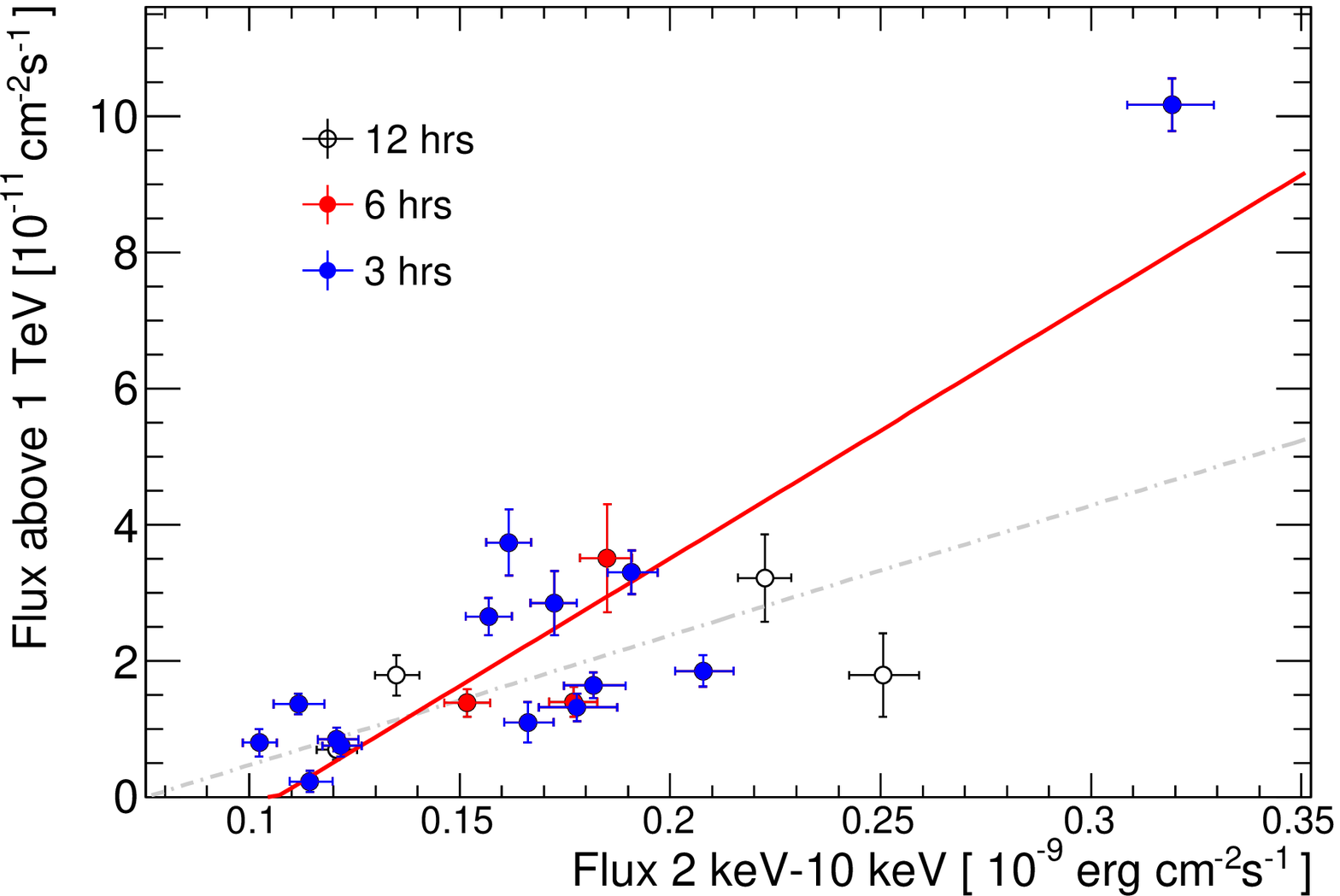}
   \caption{VHE flux as a function of the \textit{Swift}/XRT flux for the energy ranges shown.
            Open circles represent data that were taken within 12 hours of each other, red circles within
            6 hours and blue circles within 3 hours.
	    In each case, linear fits to the closed
            circle points (6 hours or less) are depicted with a red
            line (when considering the June 9 flare) and with a
            dotted-dashed grey line (when excluding the June 9
            flare). 
           }
     \label{fig:corrtev}
   \end{figure*}

\begin{table*}
\caption{Correlation Results: VHE vs X-ray flux. See Section \ref{sec:corr} and 
Figure \ref{fig:corrtev}. 
The normalised slope is the gradient of the fit in Figure \ref{fig:corrtev}, divided by the 
ratio of the average of each distribution, in order to create a dimensionless scaling factor.
Pearson correlation function 1~$\sigma$ errors and the
significance of the correlation are calculated following \citet{2002nrca.book.....P}.
Discrete correlation function (DCF) and errors are calculated as prescribed in \cite{1988ApJ...333..646E}.
}
\label{tab:Corr1}      
\centering                          
\begin{tabular}{l | l | l }        
\hline                 
& VHE (0.2 -- 1~TeV) & VHE ($>$ 1~TeV) \\
\hline & \begin{tabular}{l l l} Normalized & Pearson correlation& \\
  Slope of fit & coefficient ($\sigma$) & DCF \\
  \end{tabular}
& \begin{tabular}{l l l} Normalized & Pearson correlation& \\
  Slope of fit & coefficient ($\sigma$) & DCF \\
  \end{tabular} \\
\hline \textit{Swift}/XRT (0.3 -- 2~keV) & \begin{tabular}{l l l}
4.34$\pm$0.19 & \hspace{.05cm} 0.76$^{+0.10}_{-0.15}$ (3.7) \hspace{0.1cm} & 0.72$\pm$0.59 \\
  \end{tabular} & \begin{tabular}{l l l}
4.14$\pm$0.27 & \hspace{.05cm} 0.78$^{+0.10}_{-0.15}$ (3.9) \hspace{0.1cm} & 0.74$\pm$0.59 \\
  \end{tabular} \\
Excluding Flare & \begin{tabular}{l l l}
1.01$\pm$0.21 & \hspace{.05cm} 0.38$^{+0.24}_{-0.30}$ (1.4) \hspace{0.1cm} & 0.37$\pm$0.14 \\
  \end{tabular} & \begin{tabular}{l l l}
1.28$\pm$0.25 & \hspace{.05cm} 0.39$^{+0.23}_{-0.29}$ (1.6) \hspace{0.1cm} & 0.42$\pm$0.17 \\
  \end{tabular} \\
\textit{Swift}/XRT (2 -- 10~keV) & \begin{tabular}{l l l}
2.57$\pm$0.13 & \hspace{.05cm} 0.87$^{+0.06}_{-0.10}$ (4.7) \hspace{0.1cm} & 0.81$\pm$0.64 \\
  \end{tabular} & \begin{tabular}{l l l}
2.72$\pm$0.16 & \hspace{.05cm} 0.88$^{+0.06}_{-0.10}$ (4.9) \hspace{0.1cm} & 0.83$\pm$0.64 \\
  \end{tabular} \\
Excluding Flare & \begin{tabular}{l l l}
1.64$\pm$0.20 & \hspace{.05cm} 0.56$^{+0.18}_{-0.25}$ (2.2) \hspace{0.1cm} & 0.54$\pm$0.21 \\
  \end{tabular} & \begin{tabular}{l l l}
1.66$\pm$0.21 & \hspace{.05cm} 0.59$^{+0.16}_{-0.24}$ (2.5) \hspace{0.1cm} & 0.60$\pm$0.20 \\
  \end{tabular} \\
\hline                                   
\end{tabular}
\end{table*}

This section focuses on the cross-correlation between the X-ray and
VHE emission, which are the energy bands with the largest variability
in the emission of Mrk~501 (as shown in Figure~\ref{fig:fvar_all}).
Figure \ref{fig:corrtev} shows the integral flux for the two VHE ranges,
0.2--1~TeV and $>$1~TeV,
plotted against that for the two \textit{Swift}/XRT flux bands, 0.3--2~keV and 2--10~keV.
The symbols are color-coded depending on the time difference
between the observations: 3, 6 or 12 hours. The correlation studies
are performed with data taken within 6 hours  (the red and blue
symbols), which is approximately the largest temporal coverage provided by a Cherenkov
telescope for one source during one night.

\begin{figure*}
   \centering
      \includegraphics[width=20pc]{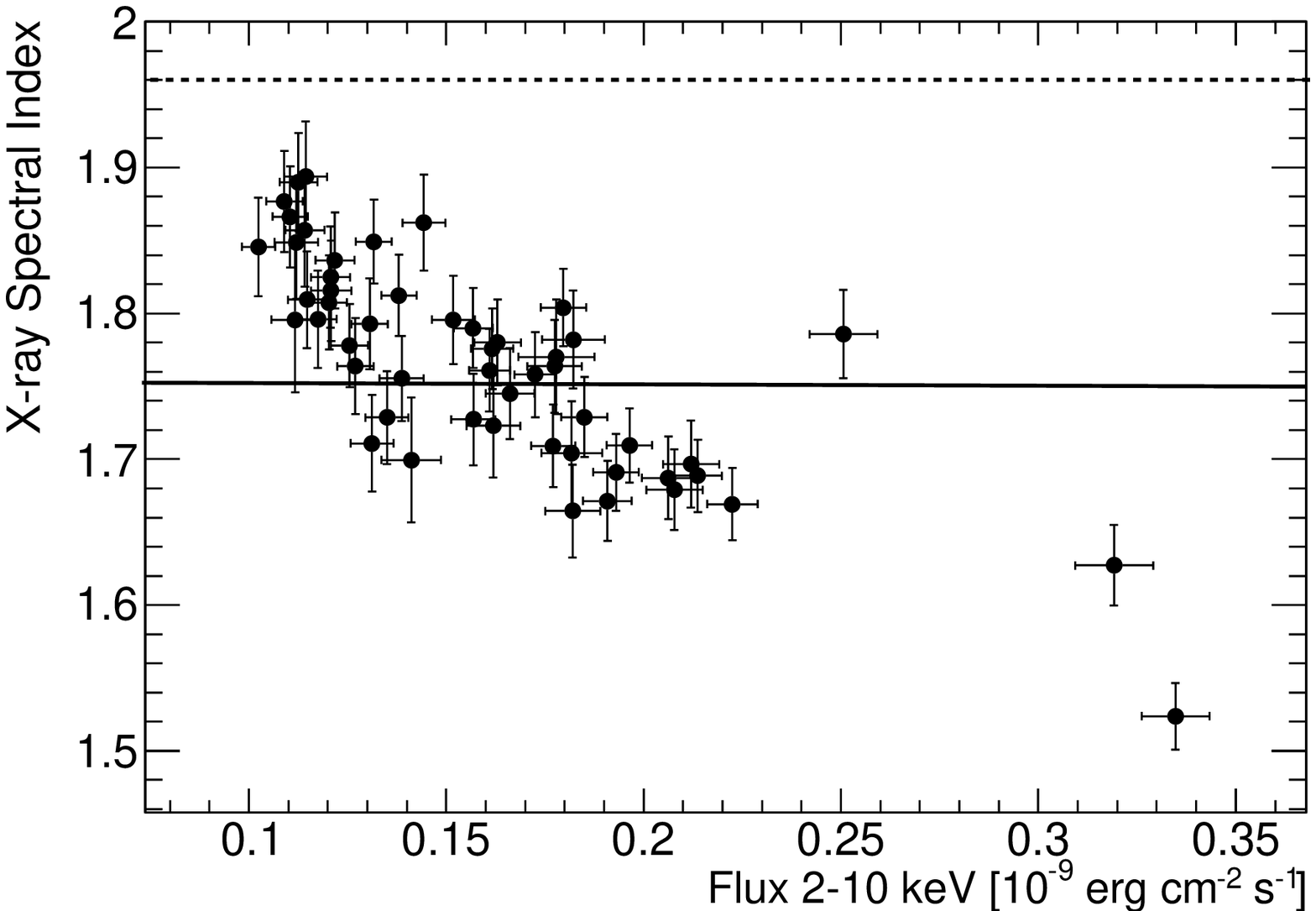}
      \includegraphics[width=20pc]{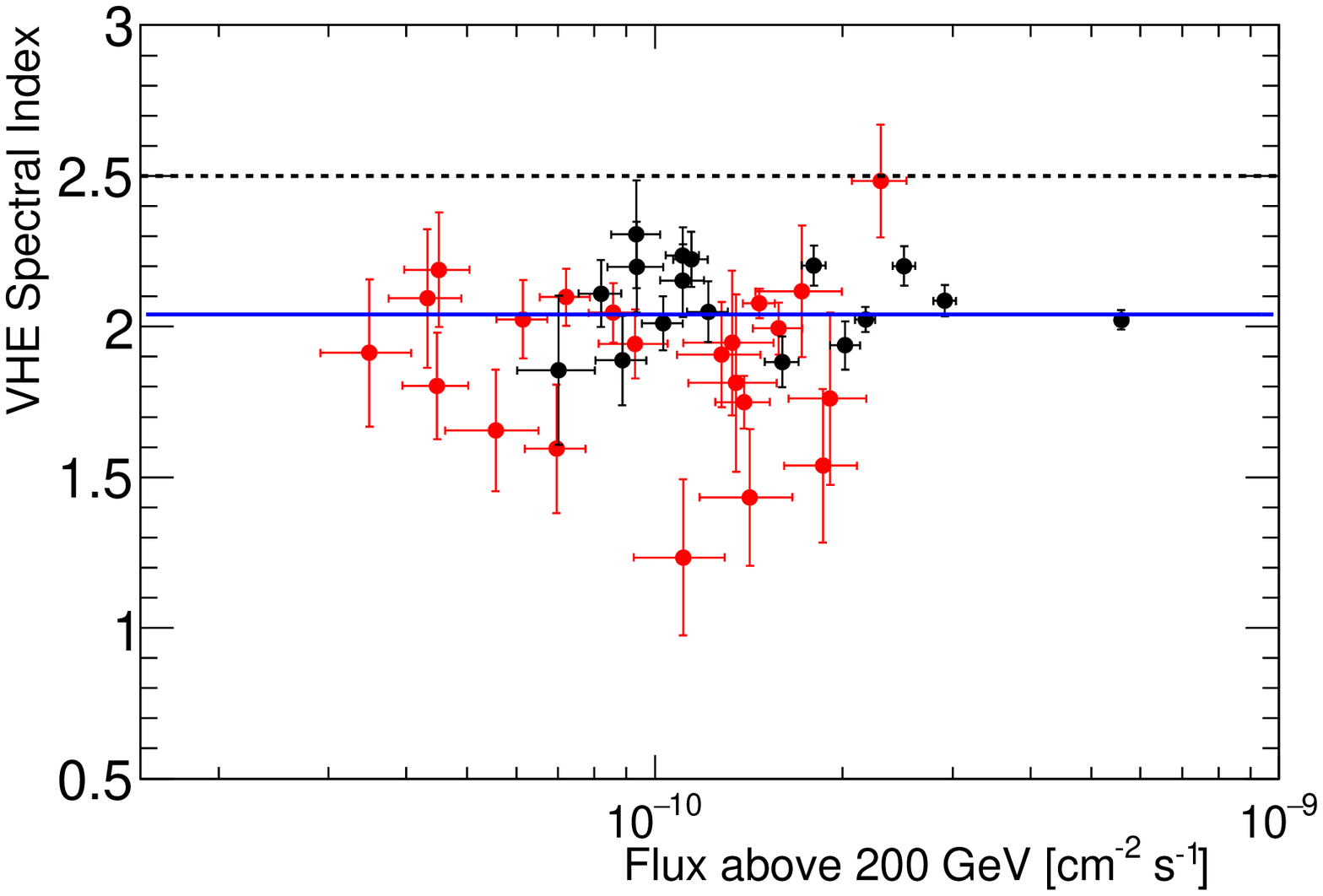}
   \caption{Left Panel: \textit{Swift}/XRT  X-ray power-law spectral index  vs flux in the 2-10 keV band.
   Right Panel: Measured VHE power-law spectral index vs VHE
   Flux above 0.2 TeV. Red points represent VERITAS data and
            black points MAGIC data. The data are EBL corrected using \citet{2008A&A...487..837F}.
            The  \textit{Swift}/XRT spectrum of Mrk\,501 is often curved, and can
be described at keV energies with a spectral index that is typically between
1.8 and 2.1, while the VHE spectral index measured with MAGIC and
VERITAS during typical non-flaring activity is about 2.5
\citep{2011ApJ...727..129A,2015A&A...573A..50A}. The typical spectral
indices at X-ray and VHE are marked with a dashed line. For
comparison purposes, the panels depict with solid lines the result of a fit with a
            constant to the X-ray and VHE spectral indices. 
            }
      \label{fig:index_xrt_pl}
   \end{figure*}

Three methods were used to test for correlation in each of the four panels shown
in Figure \ref{fig:corrtev}, and the results are shown in Table~\ref{tab:Corr1}.  A Pearson's correlation test 
was applied to the data and a maximum correlation of 4.9$\sigma$
is found 
between the higher-energy component of the X-ray band (2--10~keV) and the higher-energy 
component of the VHE band ( $\textgreater$ 1~TeV).
However, this falls to 2.5~$\sigma$ when the day of the flare is removed.
We also quantified the correlations using the discrete correlation
function \citep[DCF,][]{1988ApJ...333..646E} which has the advantage
over the Pearson correlation that 
the errors in the individual flux measurements (which contribute to the dispersion in the flux values) 
are naturally taken into account.
Using the data shown in Figure \ref{fig:corrtev} the correlation for
the two higher-energy bands of the X-ray and VHE light curves yields
0.83$\pm$0.64 when using all data, and 0.60$\pm$0.20
after removing the June 9 flare.  The three-times-larger error in the DCF
when the big VHE flare is included is due to the fact that the error in the DCF is
given by the dispersion in the individual (for a given pair of
X-ray/VHE data points) unbinned discrete correlation function,
and this single (flaring) data point
deviates substantially from the behavior of the others.
The DCF value for the data without the flaring activity corresponds to a
marginal correlation at the level of 3$\sigma$, which 
is consistent with the Pearson
correlation analysis. 
The DCF method is often used to look for a time delay between the
emission at different wavelengths. 
Such a search was carried out for the two X-ray and VHE gamma-ray bands, and no significant delay was found.
Neither a linear (shown in
Figure~\ref{fig:corrtev}) nor a quadratic fit function describes the data
well; the linear fit of the highest-energy component in each band, 
gives a $\chi^{2}$/DoF of 148.7/15.
In summary, this correlation study yields only a marginal correlation,
which is greatest when comparing the high-energy X-ray and higher-energy VHE gamma-ray components.

In Figure \ref{fig:index_xrt_pl}  we present the correlation between the
spectral index (derived from a power-law fit) and the integral flux for both 
the \textit{Swift}/XRT and VHE data. The spectral fit results
with power-law functions are reported in Tables~\ref{tab:magicIndex},
\ref{tab:veritasIndex}, \ref{tab:xrtIndex} (in Appendix~B),
respectively for MAGIC, VERITAS and \textit{Swift}/XRT.
At X-rays, the source shows the harder-when-brighter 
behavior reported several times for Mrk~501
\citep[e.g.][]{1998ApJ...492L..17P,2007ApJ...669..862A}, but such behaviour is not
observed in the VHE domain during the observing campaign in 2012.
The Mrk~501 spectra measured in the X-ray and VHE ranges were harder than previously observed, during both
high and low activity. The very hard X-ray and VHE gamma-ray spectra
observed during the full campaign will be further discussed in Section~\ref{Mrk501vsEHBL}.

%% file: notes/sed.tex
\section{Temporal evolution of the broadband spectral energy distribution}
\label{sec:sed}

In order to model the data, several time-resolved spectral energy distributions were formed.
Spectral measurements were selected in cases where a \textit{Swift}/XRT spectrum and a MAGIC/VERITAS 
spectrum were obtained within 6 hours of each other (i.e. from observations performed during the same night).
This allowed 17 distinct SEDs to be constructed, spanning three months.
The mean absolute time difference between the X-ray and VHE data are
1.2~hours, with the maximum time difference being 4.0~hours. 
Because of the substantially lower variability at radio and
optical (see Figure~\ref{fig:fvar_all}) in comparison to that at
X-rays and VHE gamma-rays, strict simultaneity in these
bands is not relevant. Nevertheless, the \textit{Swift}/UVOT data are naturally simultaneous to that of
\textit{Swift}/XRT, and the high sampling performed by optical
instruments provides a flux measurement well within half day of
the X-ray and VHE observations.

\subsection{Theoretical model and fitting methodology}

The broadband SED of Mrk~501 has previously been modeled well using one-zone synchrotron self-Compton (SSC)
scenarios during high and low activity 
\citep{2001ApJ...554..725T,2011ApJ...727..129A,2015A&A...573A..50A,2015ApJ...812...65F}. 
The emission is assumed to come from a spherical region, containing a population of relativistic electrons, traveling along 
the jet. The region has a radius $R$, is permeated by a magnetic field 
of strength $B$ and is moving relativistically with a Doppler factor $\delta$.
The electron energy distribution (EED) is assumed 
to have an energy density $U_e$, and be parameterized by a broken power law with index
$p_{1}$ from $\gamma_{1}$ to $\gamma_{b}$ and $p_{2}$ from
$\gamma_{b}$ to $\gamma_{2}$, where $\gamma_{i}$ is Lorentz factor of the electrons.

A $\chi^{2}$-minimization fit was performed to find the best-fit SED model to the observed spectra. 
An SSC code developed by \cite{2004ApJ...601..151K} was incorporated into the XSPEC spectral fitting 
software \citep{1996ASPC..101...17A} as an external model to perform the minimization 
using the Levenburg-Marquadt algorithm\footnote{{\scriptsize \url{https://heasarc.gsfc.nasa.gov/xanadu/xspec/manual/XSappendixLocal.html}}}.

In order to decrease the degeneracy among the model parameters, and after inspecting the 17 broadband SEDs, 
we decided to fix the values of the 
parameters $\gamma_\text{{min}}$, $\gamma_\text{{max}}$, \textit{R} and $\delta$, 
and to set the location of $\gamma_\text{{brk}}$ to be the cooling 
break, along with a canonical index change of 1 at $\gamma_\text{{brk}}$ (i.e. $p_2-p_1=1$).  
The parameters $\gamma_{min}$,$\gamma_{max}$ are very difficult to constrain with the available broadband SED, 
as described in \citet{Mrk501MW2009_Variability}, and it was decided to fix them to 3$\times$10$^2$ and 8$\times$10$^6$ (log$\gamma$=2.5 and 6.9), 
which are reasonable values used in the literature \citep[see][]{2011ApJ...727..129A,2015A&A...573A..50A}.
Additionally, the values of $\delta$ and \textit{R} were fixed to reasonable
values that could successfully describe the data and ensure a minimum 
variability timescale of 1~day, as no intra-night variability was observed, making this
the fastest variability observed during the three-month period considered in this paper.
A $\delta$ $\sim$10  (which results in \textit{R}$\sim$2.65$\times$10$^{16}$cm by variability arguments) 
is a suitable value used to model the emission of
high-peaked BL Lacs such as Mrk 501 \citep[e.g.][]{Mrk501MW2009_Variability}, though it is larger than the modest bulk 
Lorentz factors suggested by Very Long Baseline Array measurements \citep{2010ApJ...723.1150P,2004ApJ...600..115P,2002ApJ...579L..67E}.

First, we fit the synchrotron peak to adjust the characteristics of the EED and \textit{B} field. The synchrotron 
peak is more accurately determined than
the inverse-Compton, and has a more direct relation to the EED. Then, we fit the inverse-Compton peak, using all parameters from the fit to 
the synchrotron peak, and leaving the electron energy density \textit{U}$_\text{{e}}$ as the only free parameter. After that, we fit the broadband SED using the parameter values from 
the previous step as starting values. Lastly, we perform a broadband SED fit, using the parameter values from the previous step as starting 
values, and loosen slightly the condition that the cooling break occurs at $\gamma_\text{{brk}}$, and that the indices in the EED change by 
exactly 1.0. In this last step, we allow \textit{B} and $\gamma_\text{{brk}}$ to vary within $\pm$2\%, and \textit{p}$_\text{{1}}$ and 
\textit{p}$_\text{{2}}$ to 
vary within $\pm$1\% of the values obtained from the previous step. This last step in the fitting procedure provides a non-negligible 
improvement in the data-model agreement, with minimal (a few \%) departures from the canonical values of $\gamma_\text{{brk}}$ and spectral--index 
change within the one-zone SSC scenario.

\subsection{Model results}

\begin{figure*}
   \centering
     \includegraphics[height=11.5pc,width=21pc]{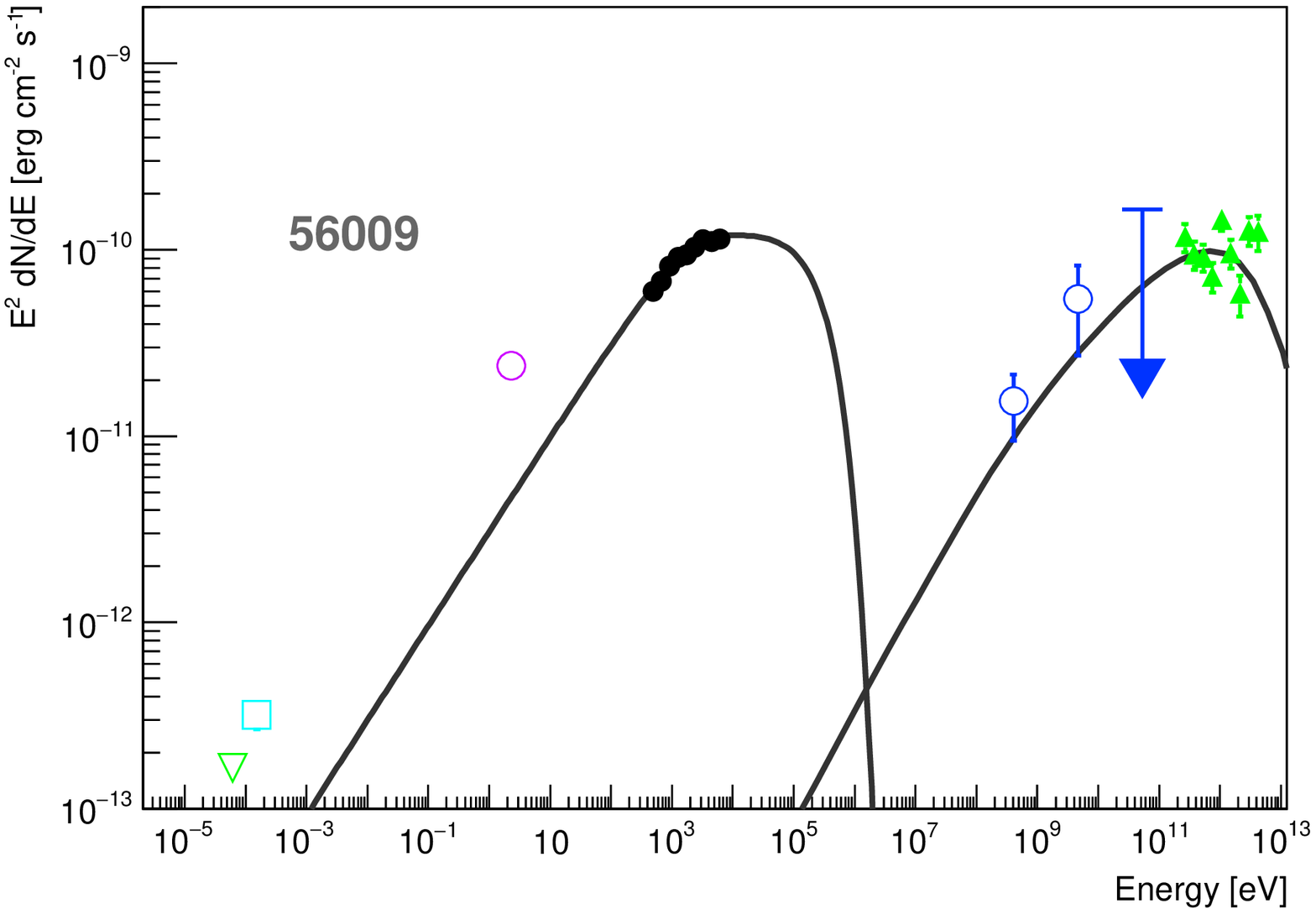}
     \includegraphics[height=11.5pc,width=21pc]{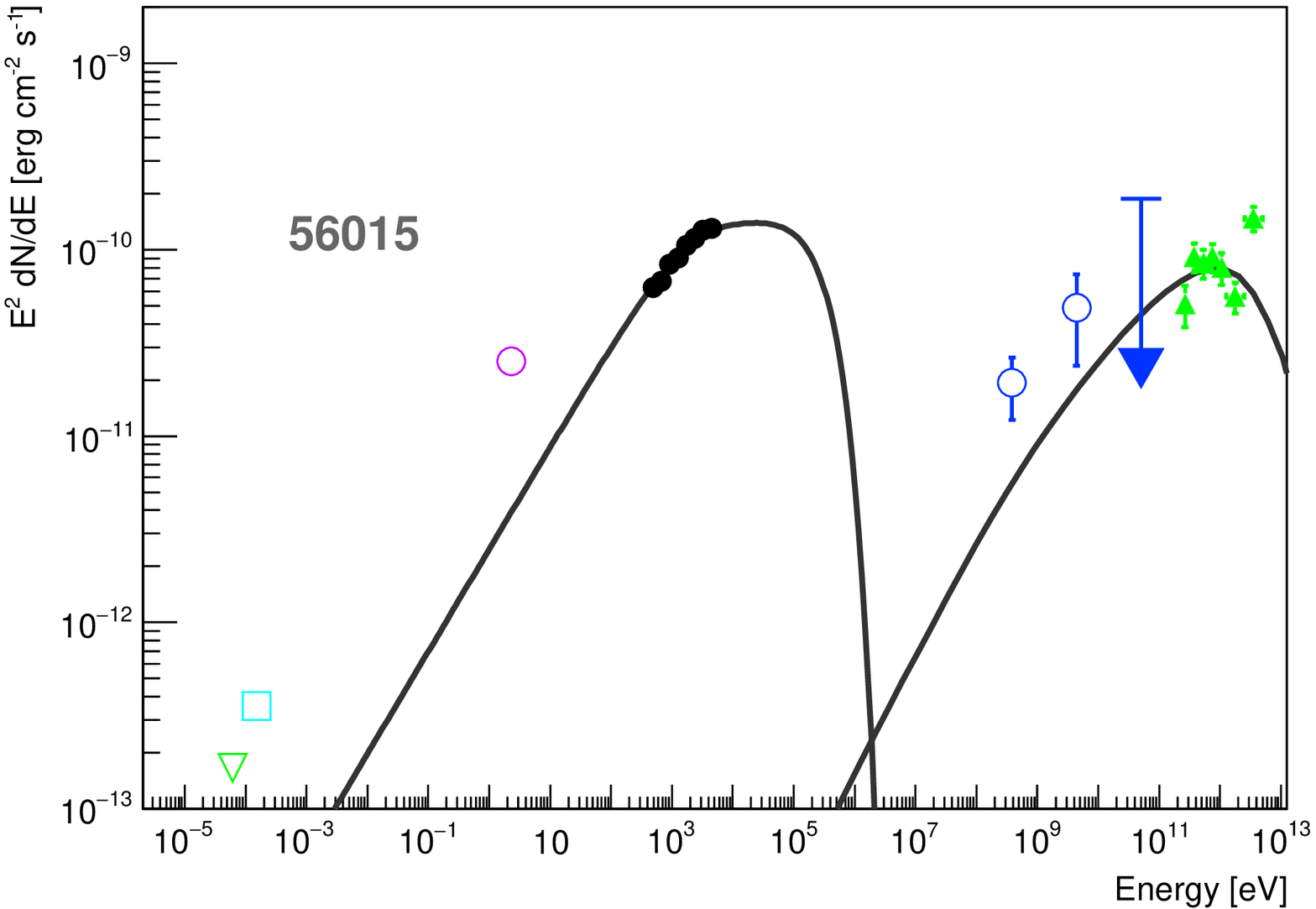}
     \includegraphics[height=11.5pc,width=21pc]{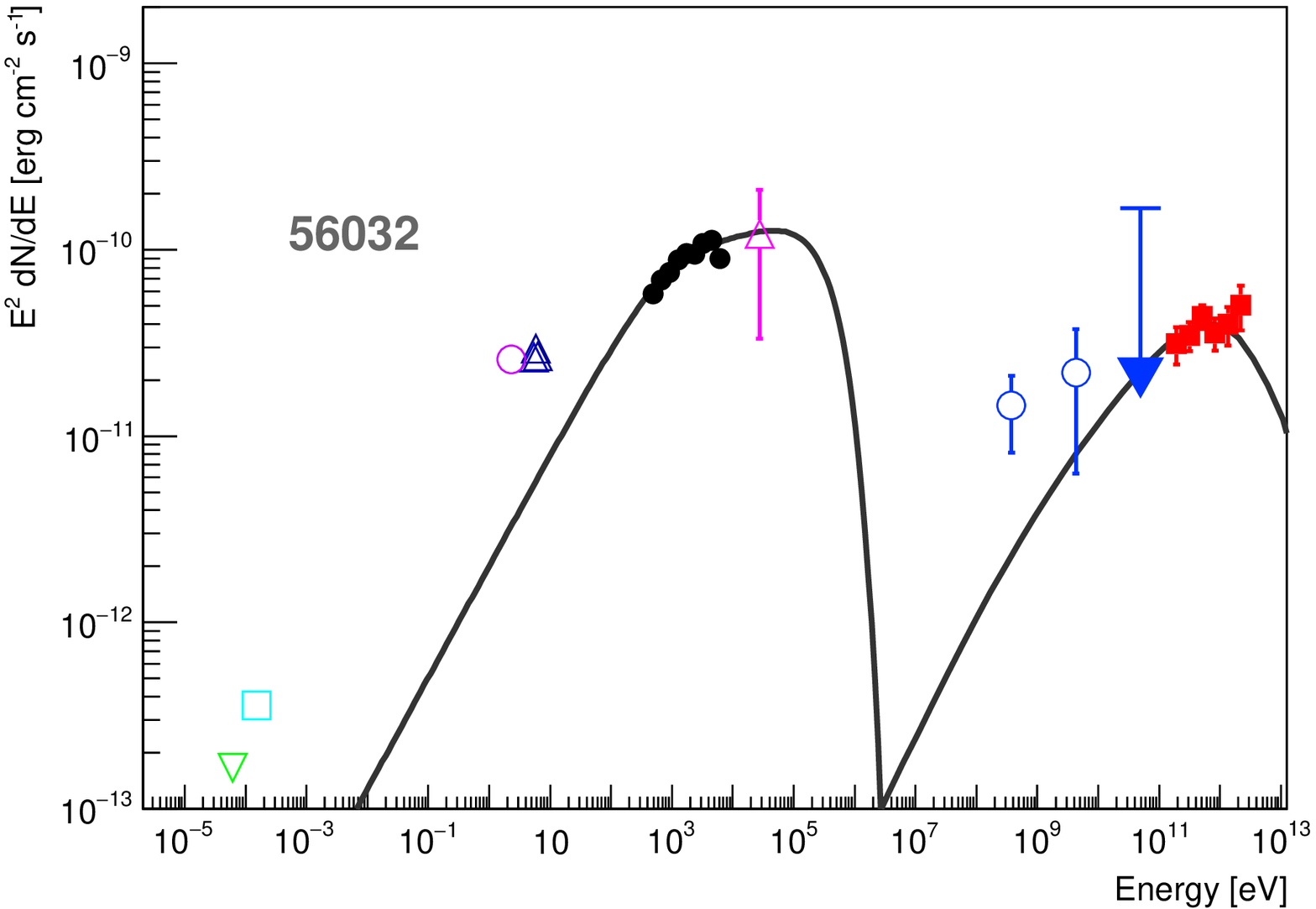}
     \includegraphics[height=11.5pc,width=21pc]{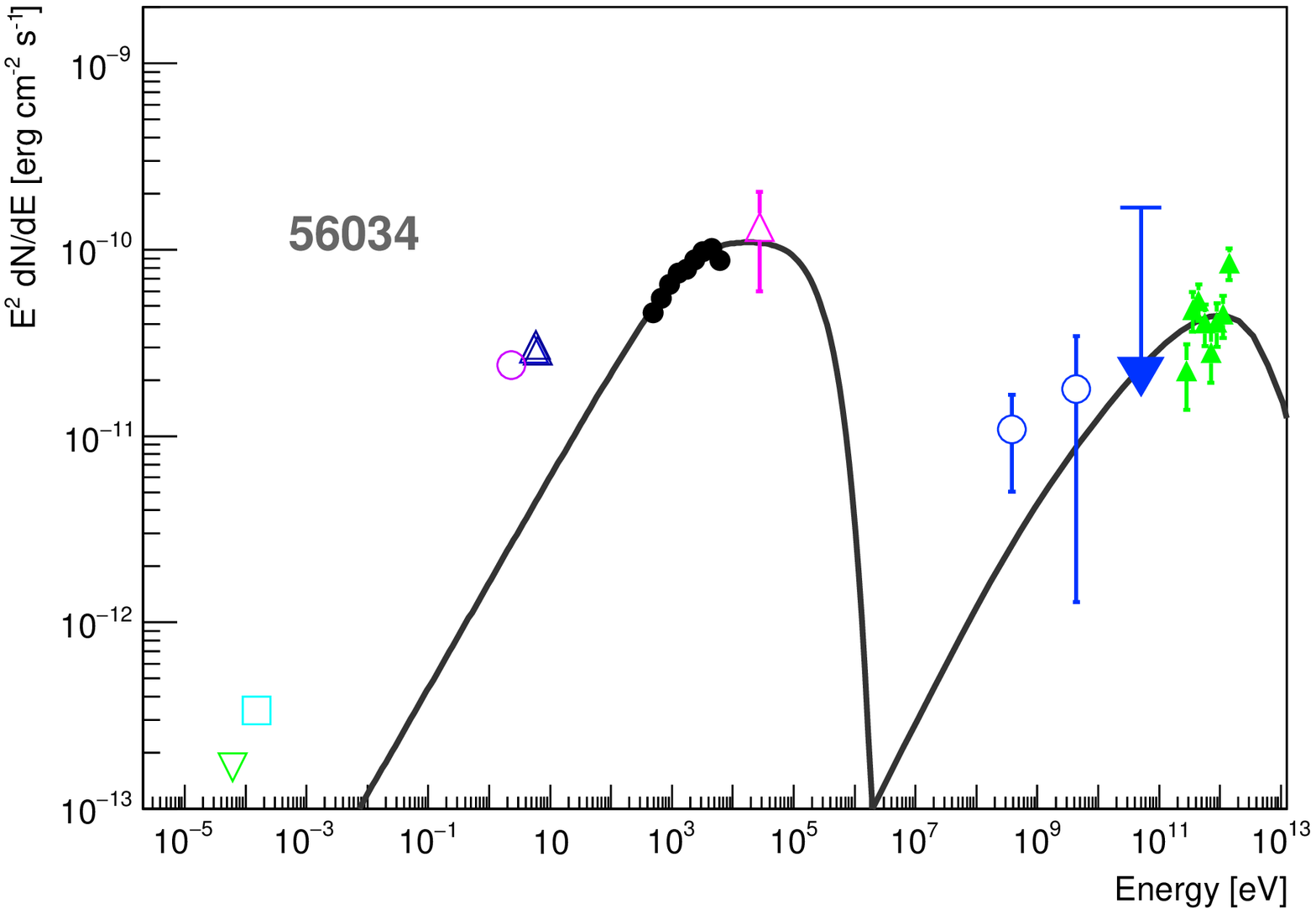}
     \includegraphics[height=11.5pc,width=21pc]{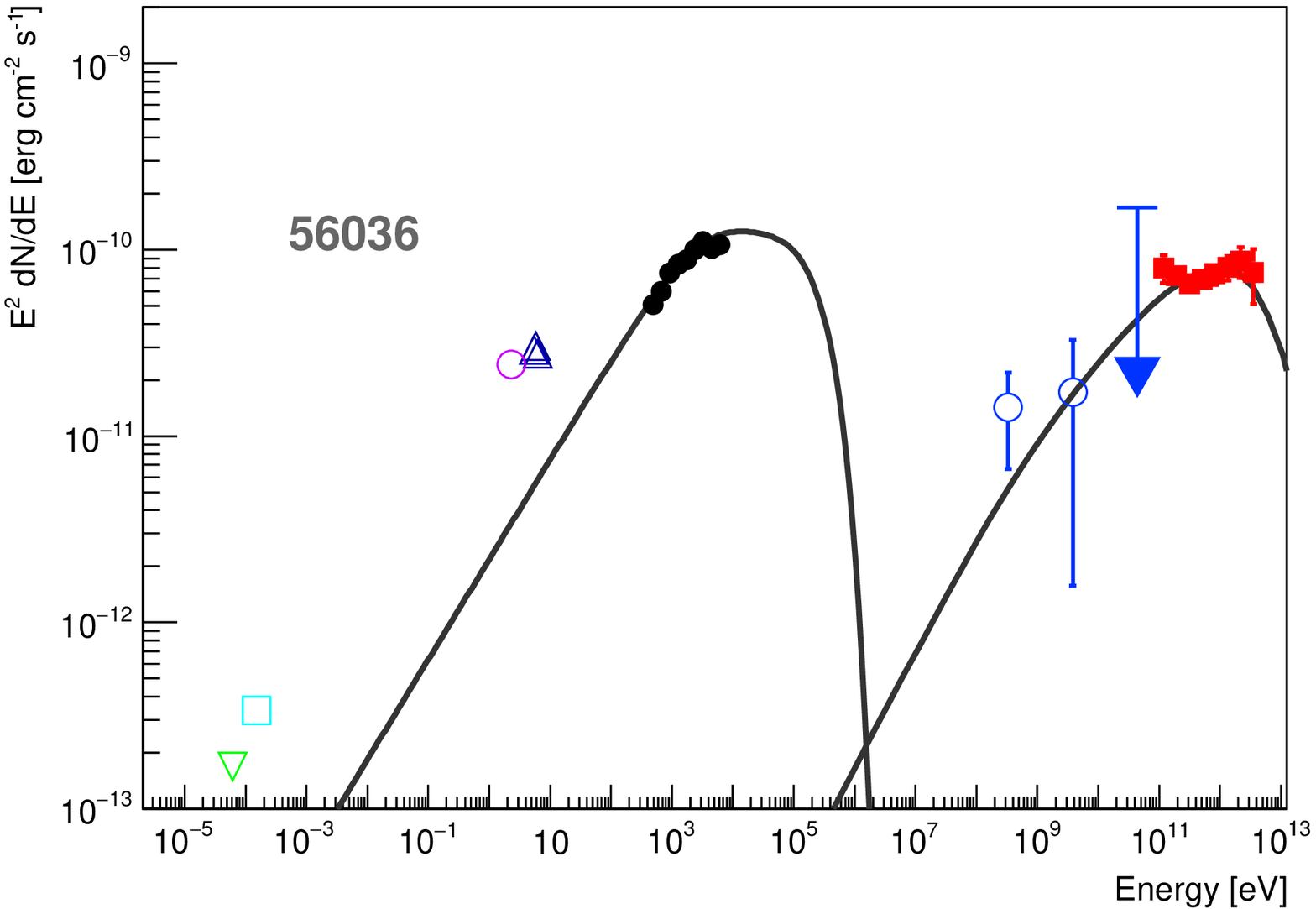}
     \includegraphics[height=11.5pc,width=21pc]{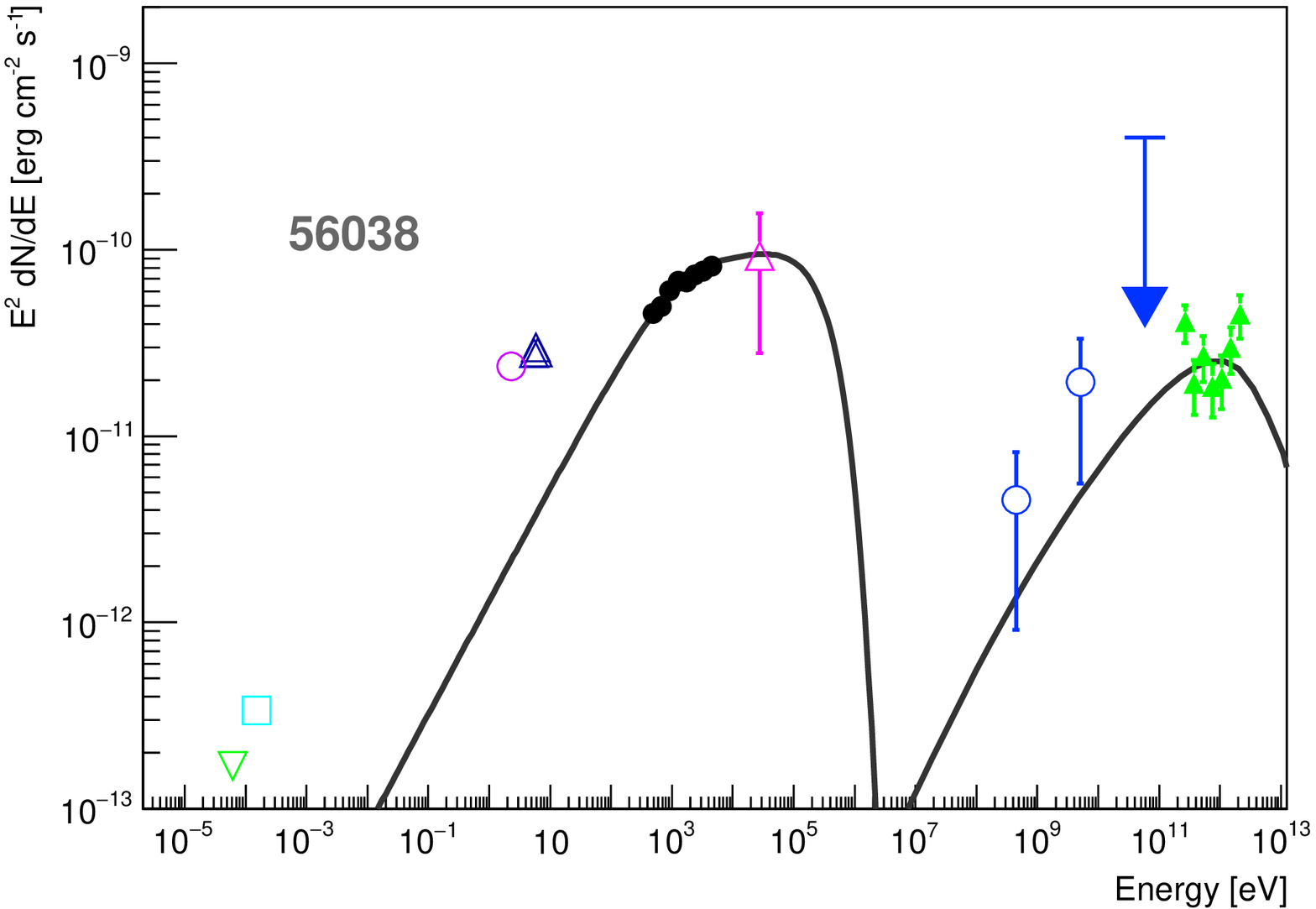}
     \includegraphics[height=11.5pc,width=21pc]{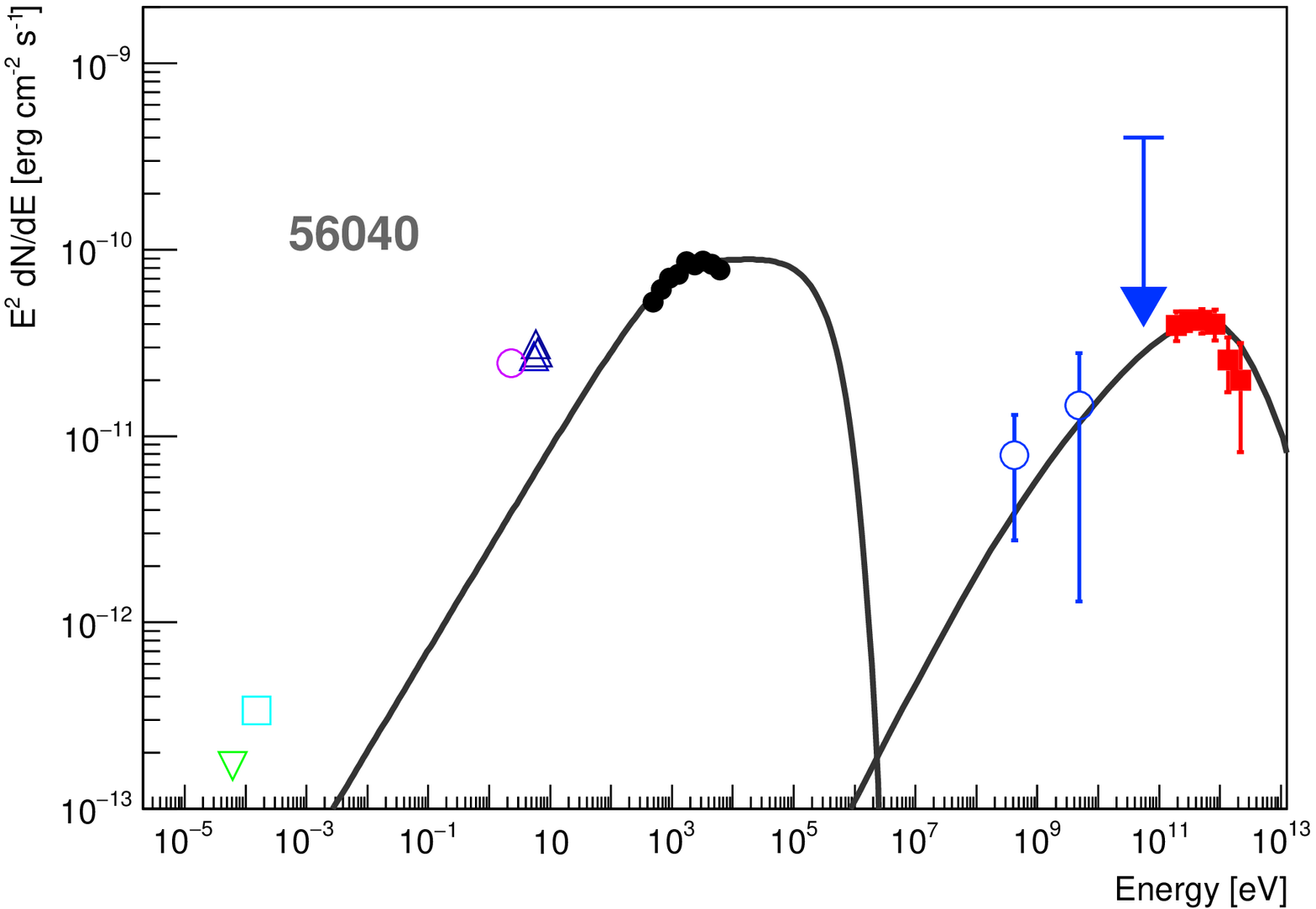}
     \includegraphics[height=11.5pc,width=21pc]{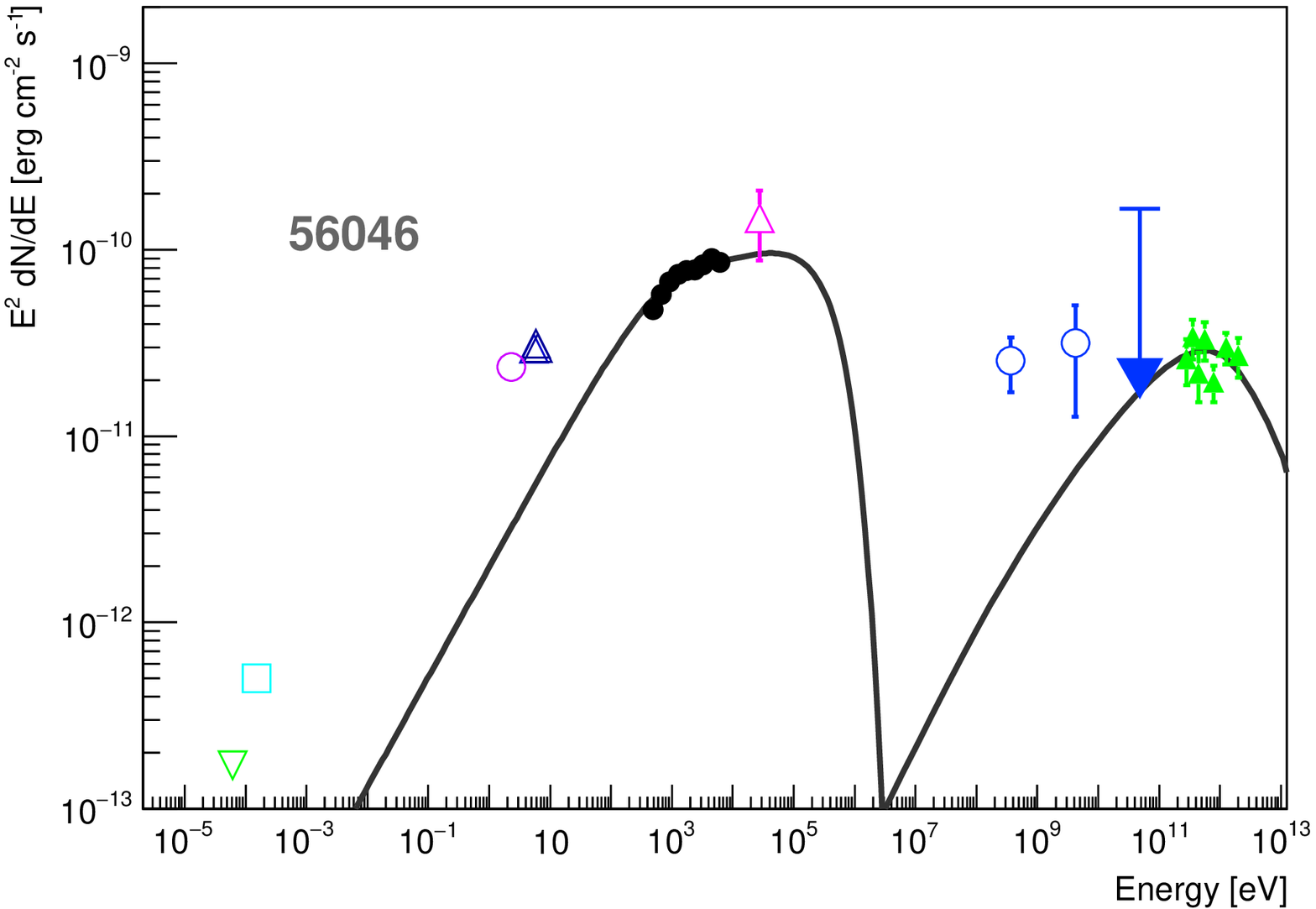}
     \caption{ Spectral energy distributions (SED) for 8 observations
       between MJD 56009 and MJD 57046.
               The markers match the following experiments; the 
               green open triangle OVRO (radio 15 GHz), blue open square Mets$\"{a}$hovi (radio 37 GHz),
               red open circle (R--band optical, corrected for host galaxy),
               blue open triangles \textit{Swift}/UVOT (UV),
               black filled circles \textit{Swift}/XRT (X--ray),
               pink open triangles \textit{Swift}/BAT (X--ray),
               blue open circles \textit{Fermi}--LAT (gamma rays) and 
               red/green filled squares/triangles MAGIC/VERITAS (VHE gamma rays).
	       VHE data are EBL--corrected using \citet{2008A&A...487..837F}.
               The BAT energy flux relates to a one-day 
               average, while the \textit{Fermi}--LAT energy flux relates to
               three-day average centered at the VHE observation. 
	       Filled markers are those fit by the theoretical model, while open markers are not.
	       The black line represents the best fit with a one--zone SSC model,
               with the results of the fit reported in Table \ref{tab:sed1}.
             }
     \label{fig:sedall}
   \end{figure*}

\begin{figure*}
   \centering
     \includegraphics[height=11.5pc,width=21pc]{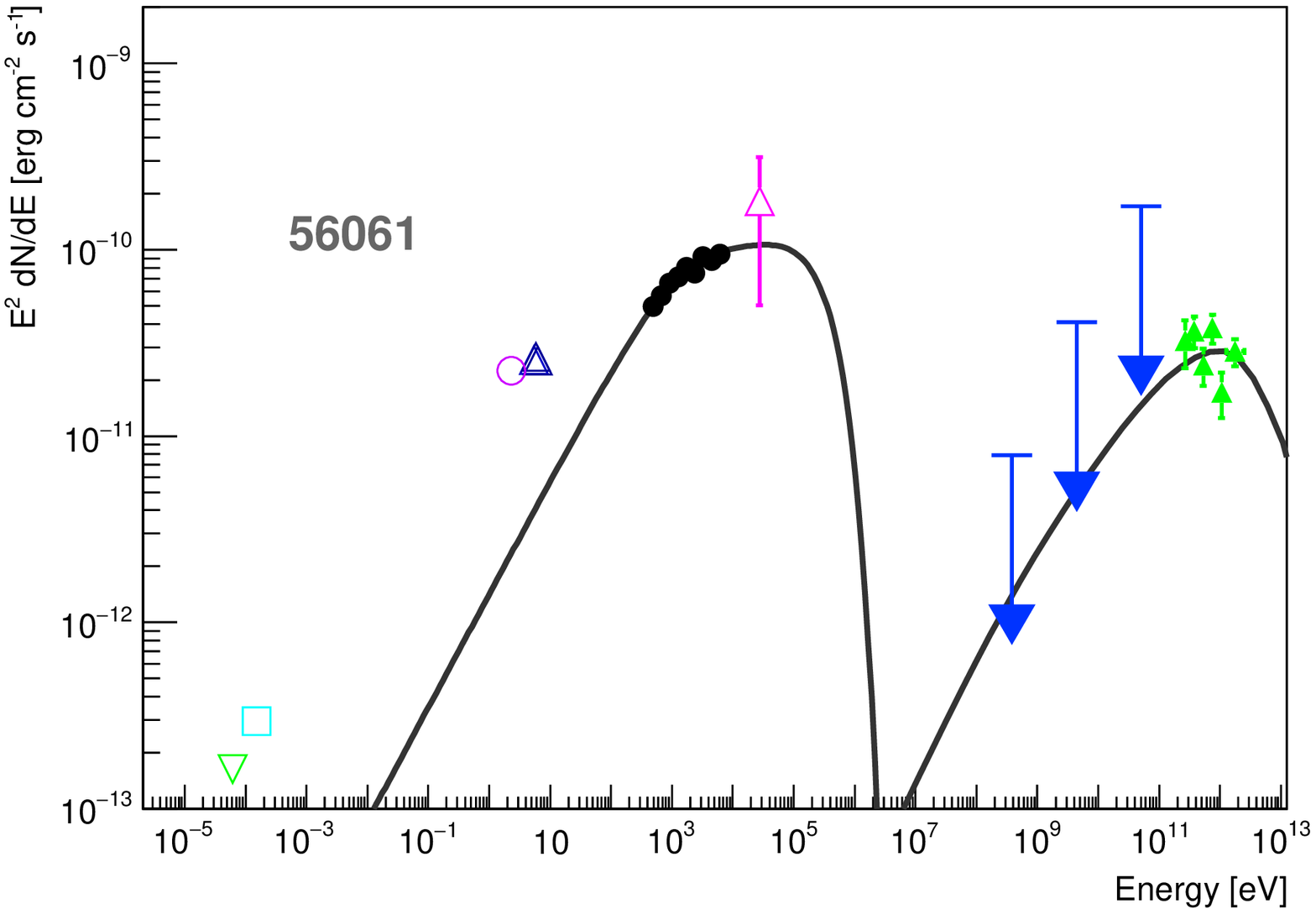}
     \includegraphics[height=11.5pc,width=21pc]{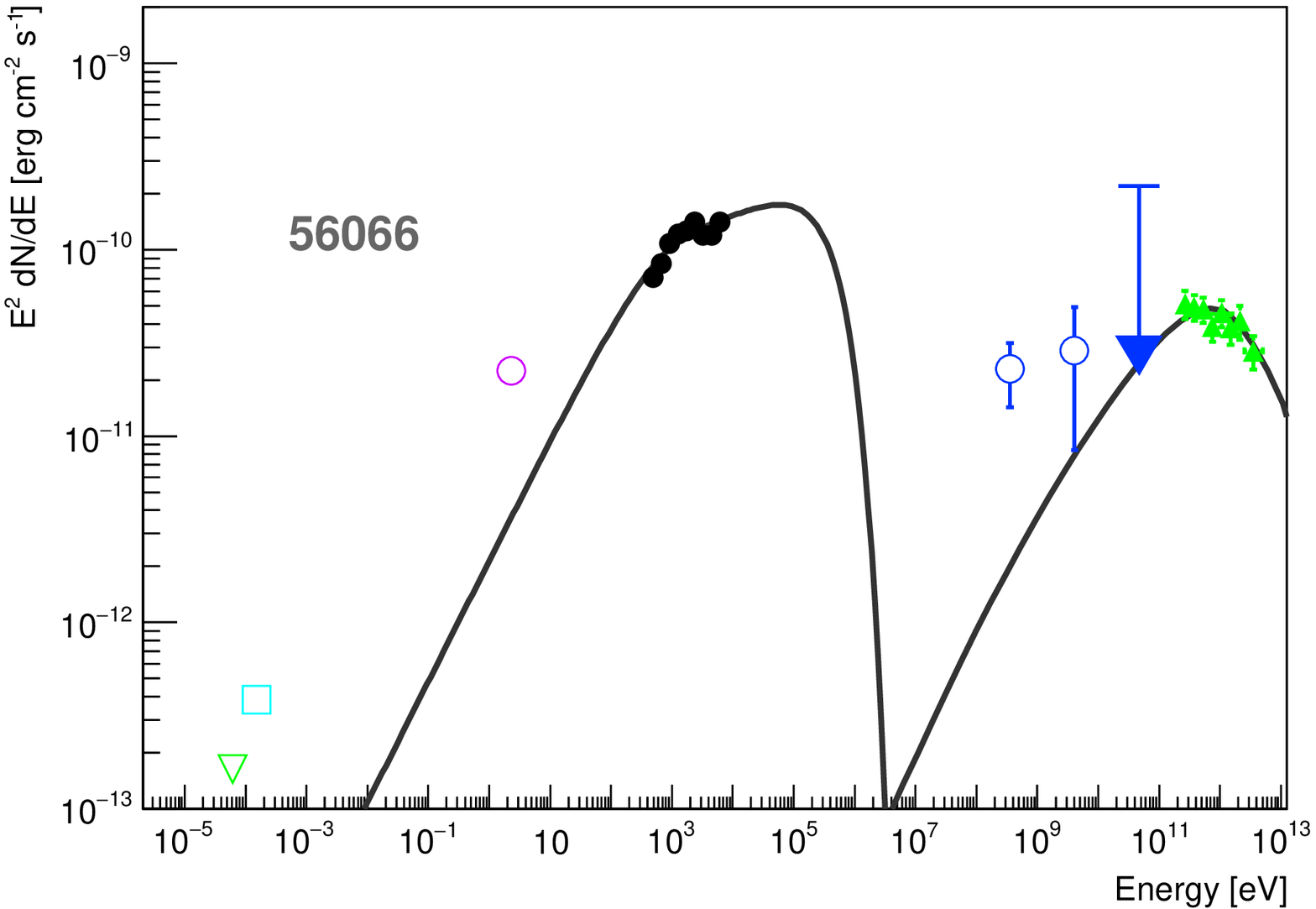}
     \includegraphics[height=11.5pc,width=21pc]{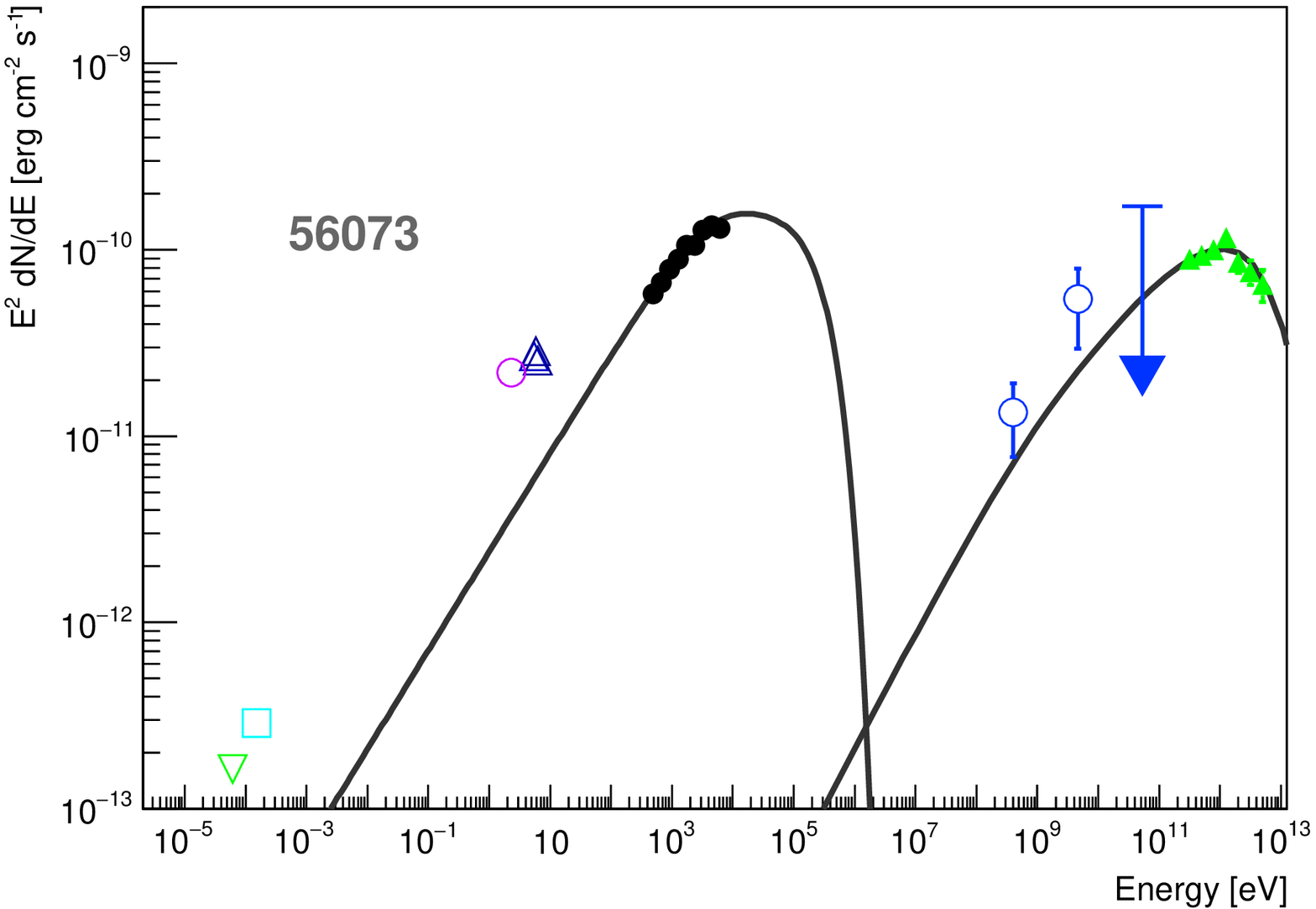}
     \includegraphics[height=11.5pc,width=21pc]{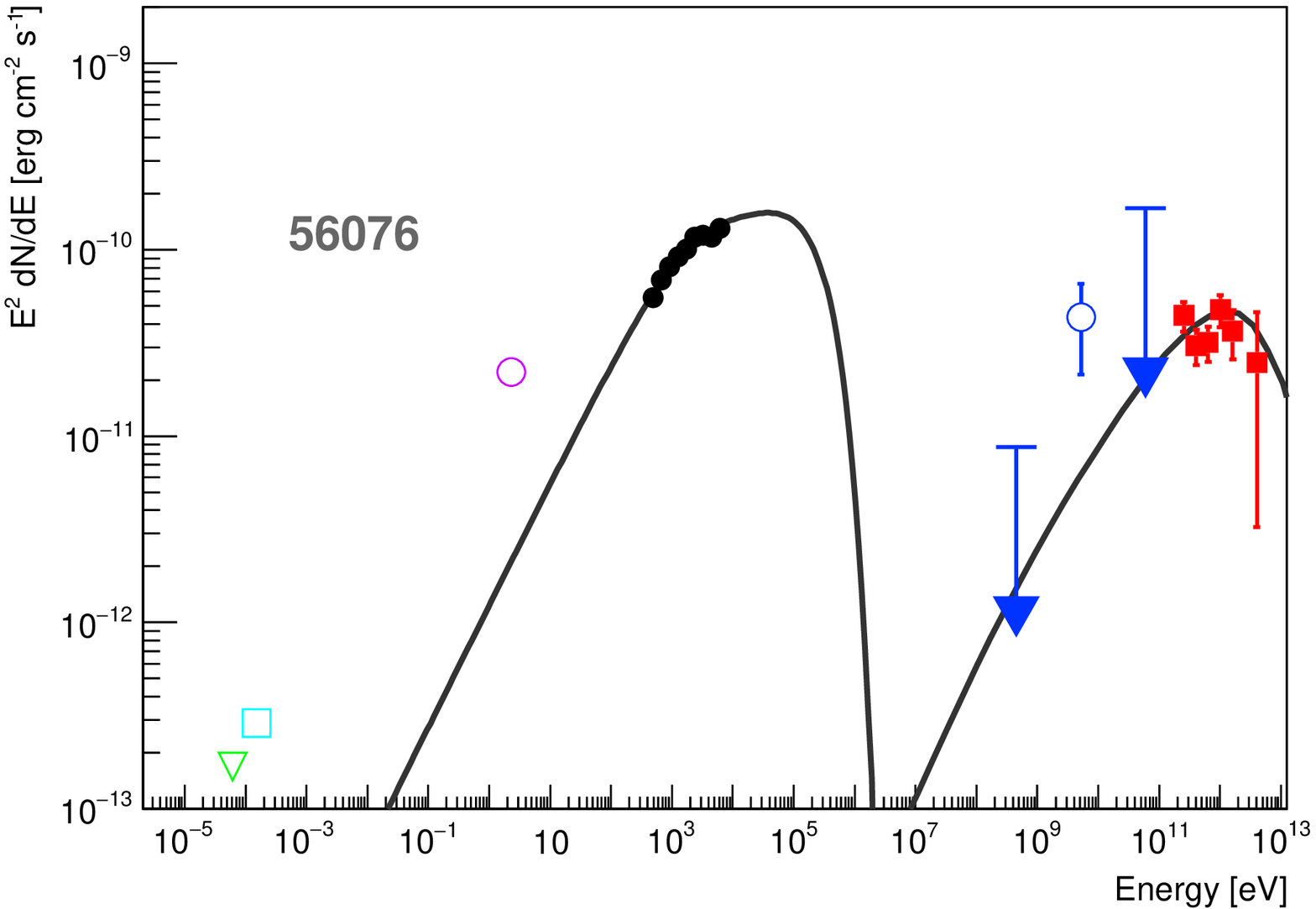}
     \includegraphics[height=11.5pc,width=21pc]{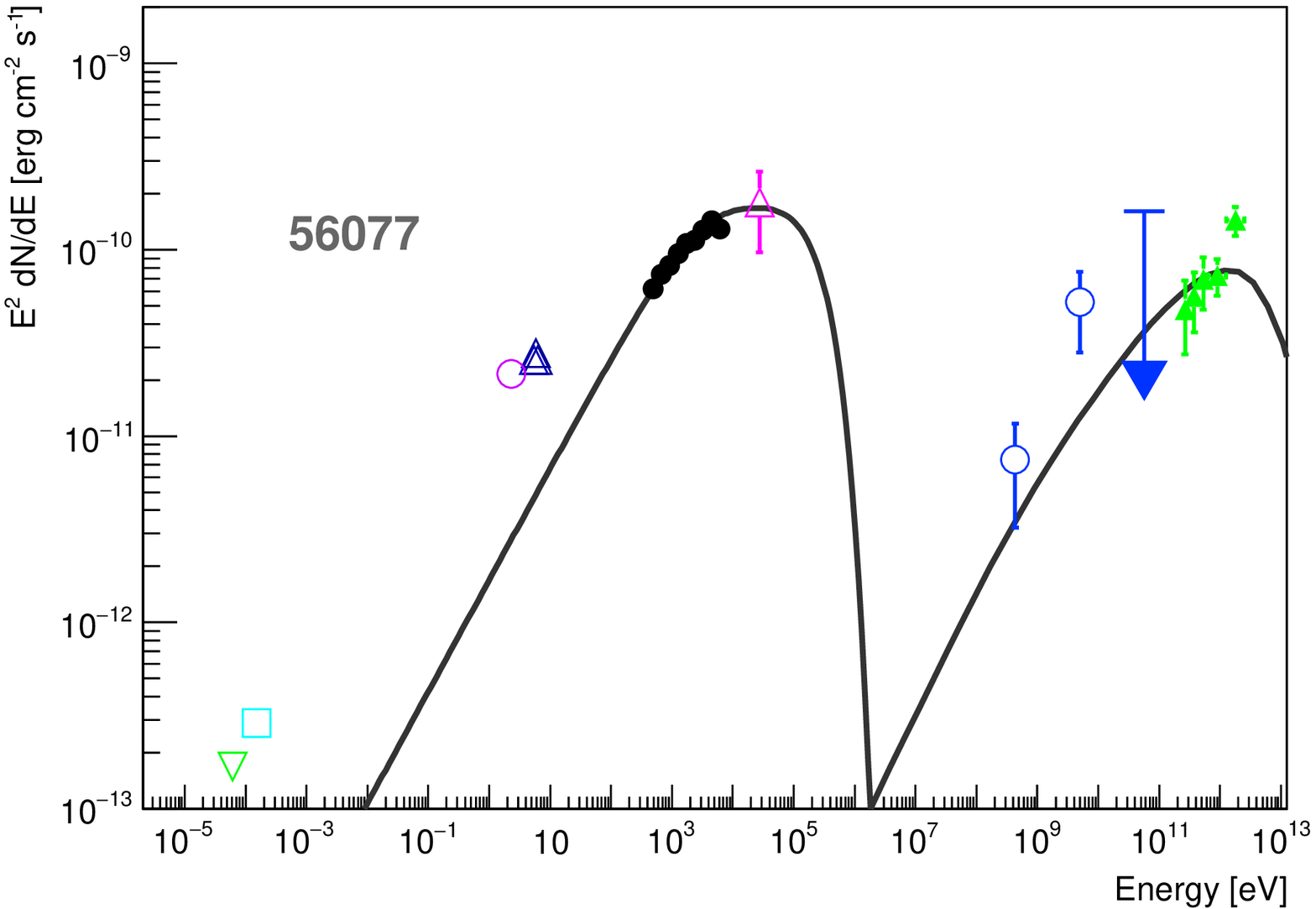}
     \includegraphics[height=11.5pc,width=21pc]{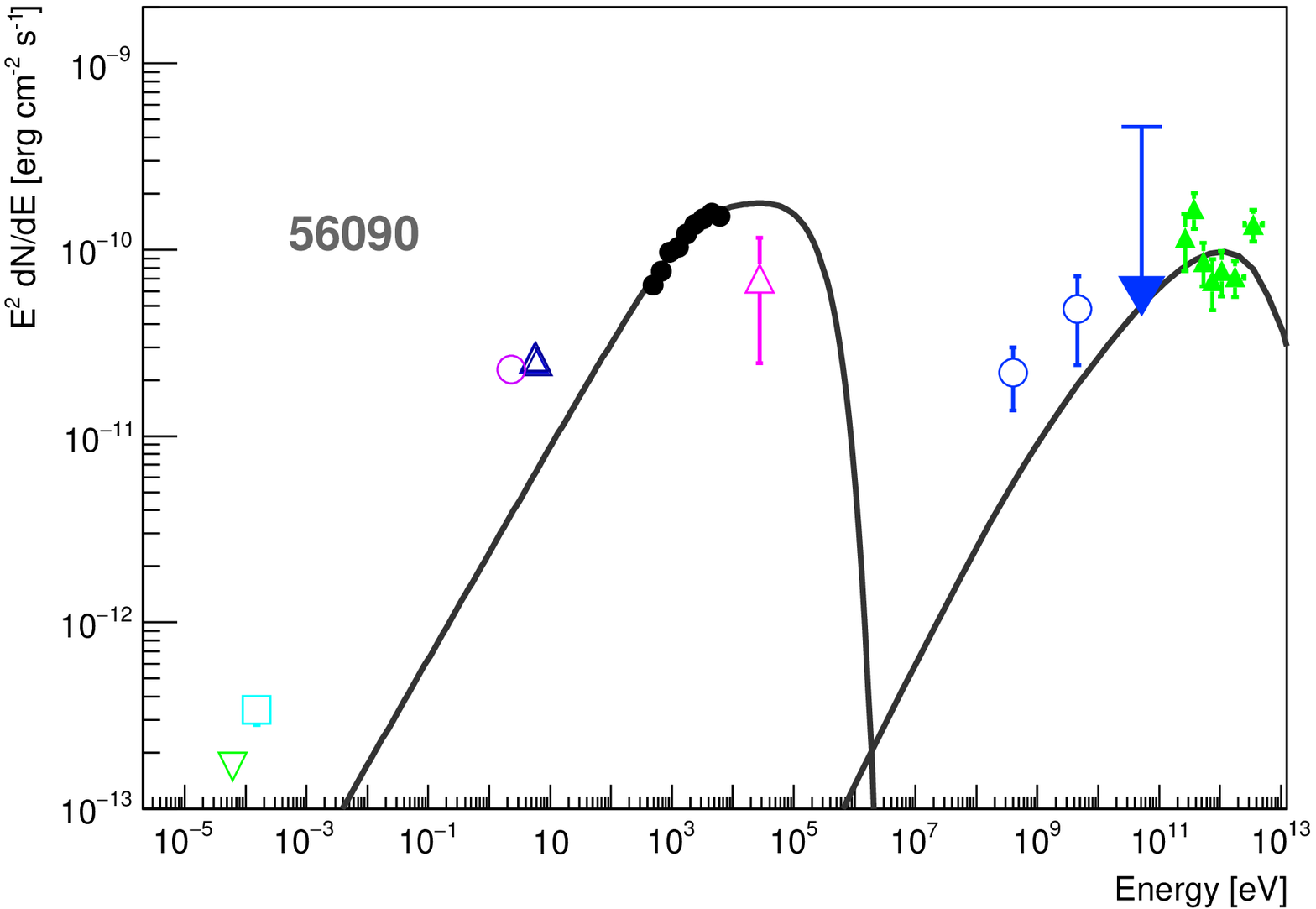}
     \includegraphics[height=11.5pc,width=21pc]{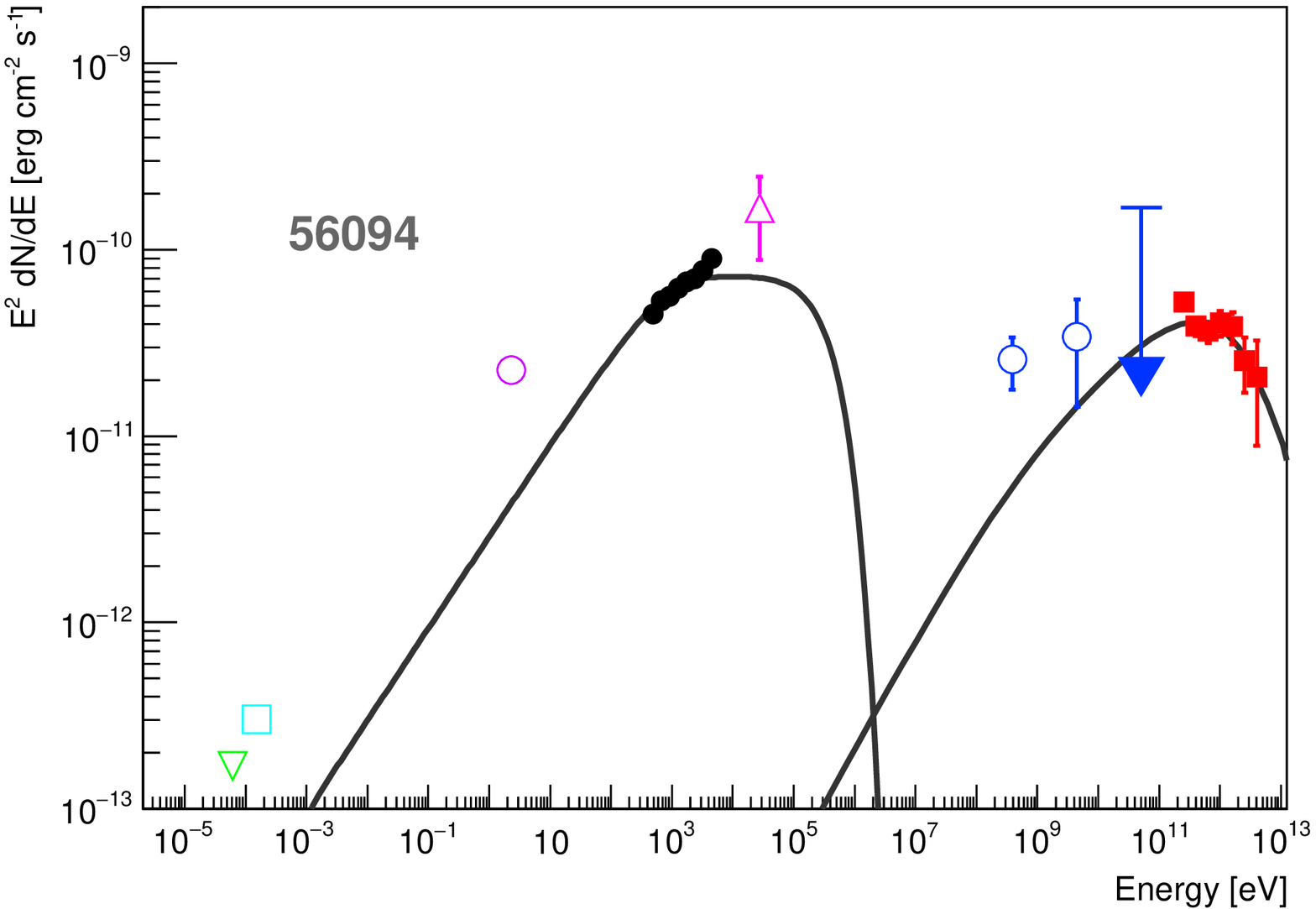}
     \includegraphics[height=11.5pc,width=21pc]{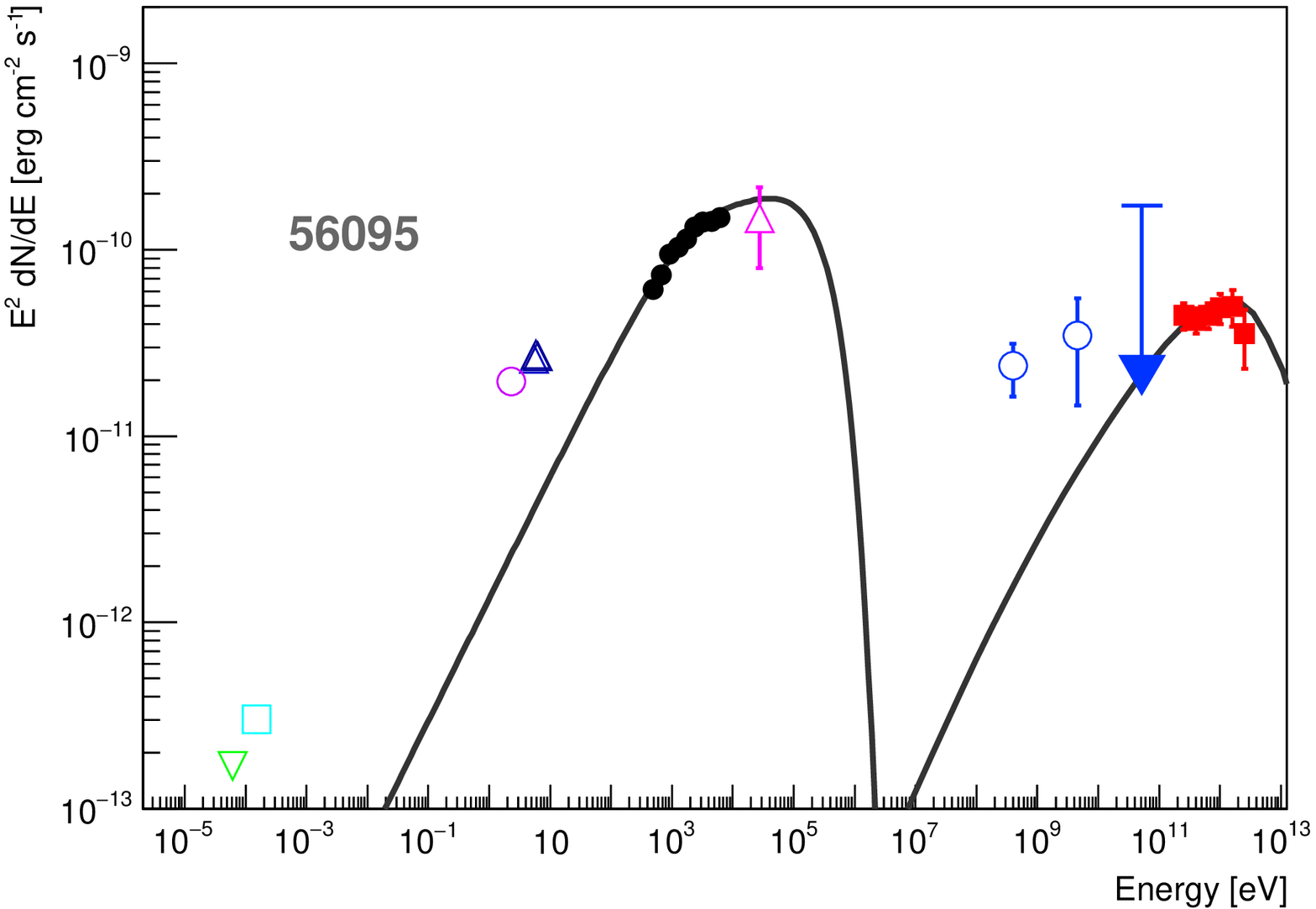}
                        
     \caption{ Spectral energy distributions (SED) for 8 observations
               between MJD~57061 and MJD 57095. See 
	       Figure~\ref{fig:sedall} for explanation of markers and
               other details. 
	       The black line represents the best fit with a one--zone SSC model,
               with the results of the fit reported in Table \ref{tab:sed1}.
             }
     \label{fig:sedall2}
   \end{figure*}

\begin{figure*}
   \centering
         \includegraphics[height=27.0pc,width=41pc]{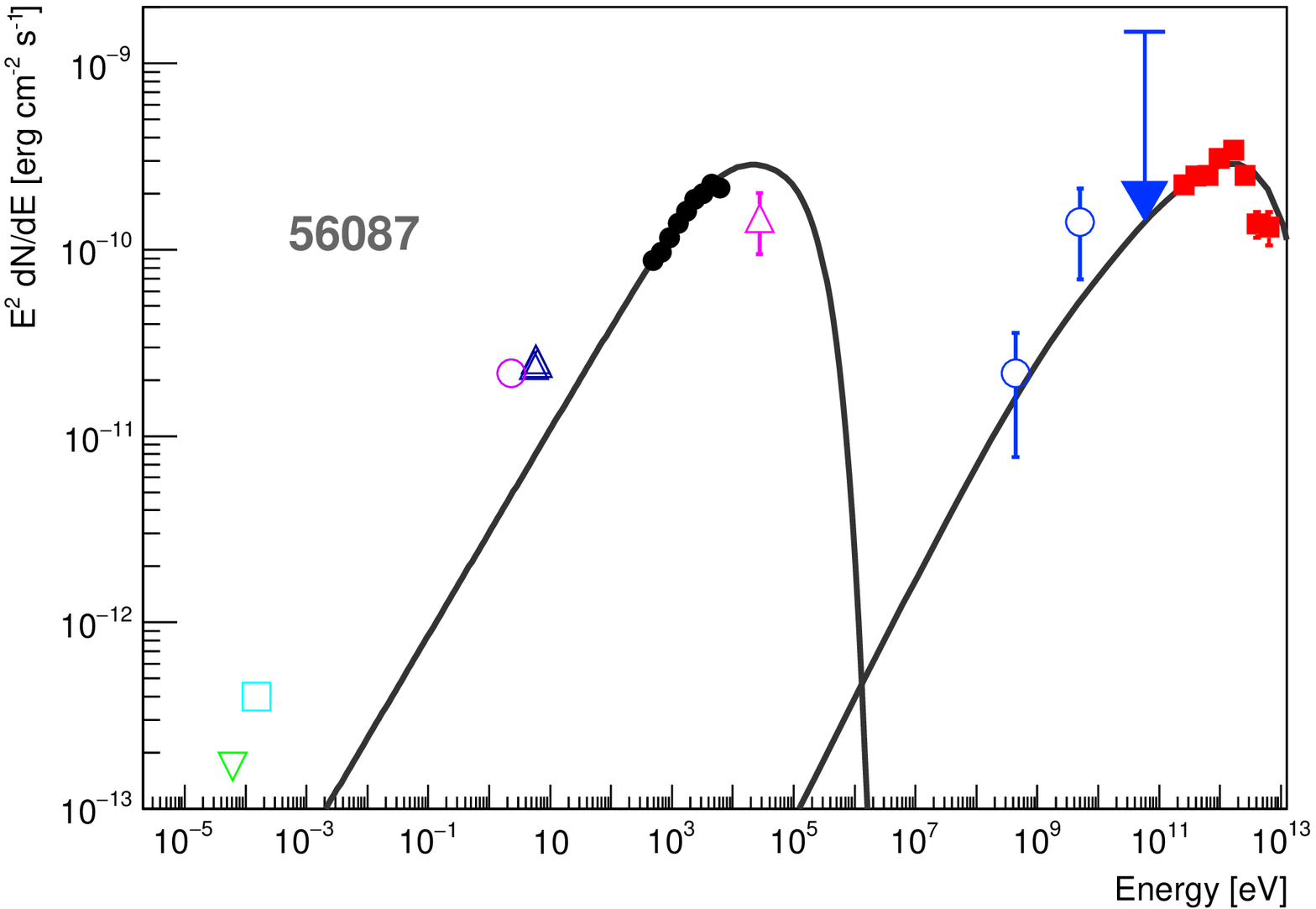}
     \includegraphics[height=27.0pc,width=41pc]{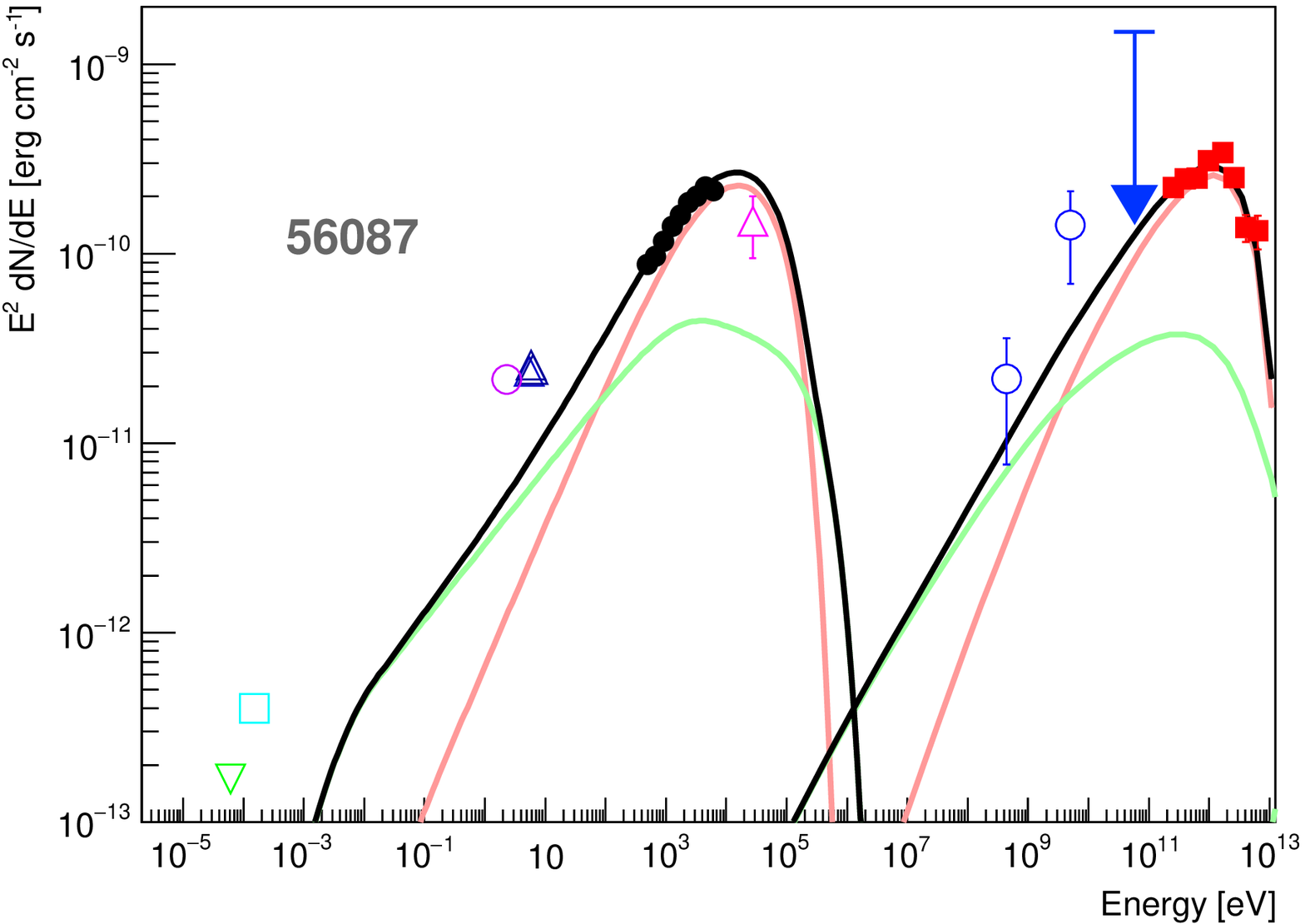}
     \caption{ 
                Broadband SED for MJD 56087 (VHE flare from 2012 June 9), fitted with a one-zone SSC model
                (top) and a two-zone SSC model (bottom). See Figure~\ref{fig:sedall} for explanation of markers and
                other details. In the bottom panel, the green line
                depicts the emission of the first (large) zone
                responsible for the baseline emission, and the red
                line the emission from the second (smaller) zone,
                that is responsible for the flaring state. The fit
                results from the one-zone SSC model fit are reported
                in Table~\ref{tab:sed1}, while those from the two-zone
                SSC model fit are
                reported in Table~\ref{tab:sed2}. 
             }
     \label{fig:sed}
   \end{figure*}

The results for the 17 broadband SEDs mentioned above can be seen in
Figures~\ref{fig:sedall}, \ref{fig:sedall2} and \ref{fig:sed}. The
corresponding SSC model parameters  are listed in Table~\ref{tab:sed1}.

\begin{table*}
\caption{One--zone SSC model results.
The following parameters were fixed: region size (\textit{R}) 2.65$\times$10$^{16}$cm, the Doppler
factor ($\delta$) 10, $\gamma_\text{{min}}$ 3.17$\times$10$^{2}$ and $\gamma_\text{{max}}$ 7.96$\times$10$^{6}$.
V refers to VERITAS and M to MAGIC observations.}
\label{tab:sed1}      
\centering                          
\begin{tabular}{l c c c c c c}        
\hline\hline                 
MJD ($\chi^{2}$/DoF) & \textit{B} & $\gamma_\text{{brk}}$ & \textit{p}$_\text{{1}}$ & \textit{p}$_\text{{2}}$ & \textit{U}$_\text{{e}}$ & $\eta$ \\
 & [10$^{-2}$ G] & [10$^6$] & & & [10$^{-3}$ erg/cm$^{3}$] & [\textit{U}$_\text{{e}}$/\textit{U}$_\text{{B}}$] \\
\hline                        
56009 V (34.0/13) & 2.26 & 0.85 & 1.90 & 2.87 & 11.96 & 	~589 \\
56015 V (29.9/11) & 2.34 & 0.81 & 1.90 & 2.87 & ~9.27 & 	~425 \\
56032 M (19.9/10) & 2.99 & 0.49 & 1.88 & 2.77 & ~5.20 & 	~146 \\
56034 V (24.3/12) & 2.22 & 0.90 & 1.86 & 2.90 & ~6.88 & 	~350 \\
56036 M (21.0/11) & 2.00 & 1.07 & 1.93 & 2.96 & 10.50 & 	~659 \\
56038 V (19.8/10) & 2.55 & 0.63 & 1.78 & 2.82 & ~4.50 & 	~173 \\
56040 M (18.8/11) & 3.00 & 0.51 & 1.91 & 2.93 & ~5.98 & 	~166 \\
56046 V (23.5/12) & 3.26 & 0.41 & 1.81 & 2.82 & ~4.30 & 	~102 \\
56061 V (24.0/10) & 2.65 & 0.65 & 1.78 & 2.82 & ~4.66 & 	~166 \\
56066 V (36.0/12) & 3.39 & 0.42 & 1.70 & 2.73 & ~5.11 & 	~112 \\
56073 V (13.3/11) & 2.00 & 1.28 & 1.93 & 2.96 & 11.70 & 	~736 \\
56076 M (19.7/10) & 2.13 & 0.81 & 1.69 & 2.70 & ~6.57 & 	~361 \\
56077 V (17.7/9)  & 1.96 & 1.07 & 1.80 & 2.82 & ~9.29 & 	~607 \\
56087 M (62.5/12) & 1.64 & 1.70 & 1.89 & 2.91 & 21.30 & 	1398 \\
56090 V (32.7/10) & 2.21 & 0.91 & 1.86 & 2.83 & 10.10 & 	~520 \\
56094 M (18.0/10) & 2.98 & 0.50 & 2.00 & 2.97 & ~7.04 & 	~199 \\
56095 M (16.8/10) & 2.25 & 0.84 & 1.68 & 2.73 & ~6.78 & 	~336 \\
\hline                                   
\end{tabular}
\end{table*}

\begin{table*}
\caption{Two--zone SED model results.
The fixed parameters are the same as in Table \ref{tab:sed1} except
for the size and the energy span of the EED for the flaring zone,
which are $R=$3.3$\times$10$^{15}$~cm, $\gamma_\text{{min}}=$ 2$\times$10$^{3}$ and $\gamma_\text{{max}}=$ 2$\times$10$^{6}$.
}
\label{tab:sed2}      
\centering                          
\begin{tabular}{l c c c c c c}        
\hline\hline                 
MJD ($\chi^{2}$/DoF) & \textit{B} & $\gamma_\text{{brk}}$ & \textit{p}$_\text{{1}}$ & \textit{p}$_\text{{2}}$ & \textit{U}$_\text{{e}}$ & $\eta$ \\
 & [10$^{-2}$ G] & [10$^6$] & & & [10$^{-3}$ erg/cm$^{3}$] & [\textit{U}$_\text{{e}}$/\textit{U}$_\text{{B}}$] \\
\hline                        
Quiescent state & 2.1 & 1.0 & 2.17 & 3.18 & 14.1 & 775 \\
56087 M (31.2/7) & 6.8 & 0.74 & 1.50 & 2.52 & 420 & 2280 \\
\hline                                   
\end{tabular}
\end{table*}

We found that the one-zone SSC model approximately describes the X-ray
and VHE gamma--ray data.
However, the model is not able to produce sufficient emission at eV energies to describe the
optical-UV emission and the soft X--ray emission with a single component. 
A similar problem in modelling the broadband SED of Mrk~501 within a
one-zone SSC framework was reported in \cite{Mrk501MW2009_Variability}.
During 2012, the variability in the optical-UV band was less
than 10\%, as reported in Section \ref{sec:variability}, and the
R-band flux was at a historical minimum (over 13 years of observations
performed by the Tuorla group), as mentioned in Section \ref{sec:lightcurves}.
It is therefore
reasonable to assume that this part of the spectrum is dominated by
the emission from a distinct region of the jet, where the emission is
slowly changing on timescales of many weeks. 
This new region, if populated by high electron density, could also contribute to the GeV emission. 
But this contribution should be characterised by low flux variability (lower than the one measured), 
as occurs in the optical emission.
For simplicity, we will not consider the description of the
optical-UV emission in the theoretical scenario presented here, which
focuses on the X-ray and VHE gamma--ray bands, i.e. the
most variable portions of the electromagnetic spectrum and where most
of the energy is emitted. 

While only the X--ray and VHE data are strictly simultaneous (within
4 hours) and therefore used in the one-zone SSC model fits, 
we note that the three-day average GeV emission (centered on the VHE
observation) measured with
\textit{Fermi}--LAT matches well most model curves on a case-by-case basis. 
The notable exceptions are on MJD 56046 and MJD 56095, where the LAT
spectral points (especially the one at the lower energy) deviate
from the theoretical curve, worsening the $\chi^{2}$/DoF of the fit from 
23.5/12 to 33.0/14
for the first day and from 
16.8/10 to 27.2/12
for the second one.
The combination of the LAT and MAGIC/VERITAS spectral points for these
two days shows a flat gamma--ray bump over four orders of magnitude (from 0.2~GeV to
2~TeV).
The $p$ values of those fits, when considering also the agreement with
the two LAT data points, are 0.3\% and 0.7\%, which is comparable to the data-model agreement from other
broadband SEDs where the LAT spectral
points match well with the model curves (e.g. MJD 56015, 56034). 
These two broadband SEDs may hint at the
existence of an additional component emitting at GeV energies, as has already
been proposed by \citet{2015ApJ...798....2S}.
However, using 
the data presented in this paper, 
the statistical significance is not large enough to make that claim, and we will not
consider additional (and variable) GeV components in our theoretical model.
On the other hand, it is also worth noticing that most of the
\textit{Fermi}--LAT data points are systematically located above (within
1--2 $\sigma$) the SSC model curves, which may be taken as another
hint for the existence of an additional contribution at GeV
energies that is constantly present at some level.

From the fit parameters, we can derive a value for $\eta$, the ratio
of the electron energy density to the magnetic field energy density,
which gives an indication of the departure from equipartition \citep{2016MNRAS.456.2374T}.
Here the values differ from unity by more than two orders of
magnitude, indicating 
that the particle population has an excess of energy compared to the magnetic field.
This is a common situation when modeling the broadband SEDs of Mrk~501
(and TeV blazars in general) with a one-zone SSC scenario 
\citep[see][]{2001ApJ...554..725T,2011ApJ...727..129A,2015A&A...573A..50A,2015ApJ...812...65F}, 
which implies more energy in the particles than in the magnetic
field, at least locally where the broadband blazar emission is
produced. It is interesting to note that \citet{2017MNRAS.464.4875B}
employ complete thermal plus
non-thermal distributions in their shock acceleration modeling
of Mrk~501 (2009 campaign) and other blazar multiwavelength spectra,
determining \textit{U}$_\text{{e}}$ consistently, and, using a \textit{B} field of
$\sim$10$^{-2}$~G,  arrive at a value of $\eta
\sim 300$ for Mrk~501, 
which is very similar (within a factor of $\sim$2) to the
values reported in Table~\ref{tab:sed1} of this manuscript.

The worst SSC model fit by far is the one for MJD 56087 (2012 June 9), 
where $\chi^{2}$/DoF=62.5/12 ($p$=$8 \times 10^{-9}$). 
This day corresponds to the large VHE gamma--ray flare reported in Section
\ref{sec:lightcurves}, for which the SED shows a peak-like
structure centered at $\sim$2 TeV.
For the sake of completeness, we attempted a fit leaving all the model
parameters free, apart from the relation between \textit{R} and Doppler factor to 
ensure a minimum variability of 1 day.  This fit yielded a $\chi^{2}$/DoF=30.3/9 (\textit{p}=4$\times$10$^{-4}$). While this fit provides a better data-model agreement, 
the obtained model is less physically meaningful because the model parameters are not related as expected in the canonical one-zone SSC 
framework (e.g. $\gamma_\text{{br}}$ and $B$, or $p_{\text{1}}$ and $p_{\text{2}}$). Moreover, 
this fit requires a $\gamma_\text{{min}}$=6$\times$10$^{4}$, which is an unusually high value for 
HBLs such as Mrk~501. Because of that, we attempted a fit with a two-zone SSC scenario with model parameters physically related as we did for 
the one-zone SSC scenario described in Section 6.1. In this framework, one relatively large zone
dominates the emission at optical and MeV energies (and
is presumed steady or slowly changing with time). 
The other, smaller zone, which is spatially separated from the first, is
characterized by a very narrow electron energy distribution and dominates
the variable emission occurring at X-rays and VHE gamma rays, and
eventually also produces narrow inverse-Compton bumps. This
scenario was successfully used to model 
a 13-day-long period of flaring activity
in Mrk~421, as reported in  \citet{2015A&A...578A..22A}.
To describe the broadband SED of Mrk~501 measured for MJD 56087, the
EED of the second region was chosen to span over three orders of
magnitude, from $\gamma_\text{{min}}=$2$\times$10$^{3}$ to $\gamma_\text{{max}}=$
2$\times$10$^{6}$,  and to have a radius $R$ of 3.3 $\times$ 10$^{15}$
cm which, for a Doppler factor of 10, corresponds to a light--crossing time of three hours,
and hence suitable to describe variability with timescales much shorter than one day. 
The broadband SED fitting using the two-zone SSC model is done in the same way as the one-zone SSC model fit described above, but now with twice as many parameters.
The resulting model fit is displayed in Figure~\ref{fig:sed}, and the model parameters reported in
Table~\ref{tab:sed2}.  The data-model agreement achieved with this
two-zone SSC scenario yielded $\chi^{2}$/DoF=31.2/7 ($p$=$6 \times
10^{-5}$) which, although this scenario still does not describe the broadband data 
satisfactorily, is still several orders of magnitude better than the p value obtained with a single-zone SSC scenario.

%% file: notes/discussion.tex
\section{Discussion}
\label{sec:discussion}

\input{notes/discussion1.tex}

\input{notes/discussion2.tex}

\input{notes/discussion3.tex}

%% file: notes/discussion1.tex
\subsection{Mrk~501 as an Extreme BL Lac object in 2012}
\label{Mrk501vsEHBL}

The BL Lac objects known to emit VHE gamma rays have the maximum of their
high-energy component typically peaking in the 1--100~GeV band,
which implies that IACTs measure soft VHE spectra (power-law
indices $\Gamma > 2$, where dN/dE $\propto$ E$^{-\Gamma}$). However, there is also a small number of VHE BL Lacs
where the maximum of the gamma-ray peak is located well within the VHE band
\citep{2011MNRAS.414.3566T}, which implies that IACTs would measure
hard VHE spectra (power-law indices $\Gamma < 2$),
once the spectra are corrected for the absorption in the EBL. These objects have the peak of their
synchrotron peak also at higher energies ($>$1--10 keV), a property which was
initially used to flag them as special sources, and categorise them as
``extreme HBLs'' (EHBLs, \citet{2001A&A...371..512C}).
Archetypal objects belonging to this class, and extensively studied
in the last few years, are 1ES~0229+200
\citep[][]{2007A&A...475L...9A,2012ApJ...747L..14V,2014ApJ...782...13A,2013arXiv1307.8091C} 
and 1ES~0347-121 \citep[][]{2007A&A...473L..25A,2014ApJ...787..155T}.
Some of the sources classified as EHBLs according to the position of
their synchrotron peak have been shown to have a very soft VHE
gamma-ray spectrum \citep[e.g. RBS~0723,][]{RBS0723}, which indicates that there is not a
uniform class of EHBLs, and hence some diversity within this
classification of sources (entirely based on observations). In this section we
focus on those EHBLs that also have a hard VHE gamma-ray
component (e.g. 1ES~0229+200), which are actually the most relevant
objects for EBL and intergalactic magnetic field (IGMF) studies \citep{2015ApJ...813L..34D,2015ApJ...814...20F}

In order to model the broadband SEDs of these EHBLs with hard VHE
gamma-ray spectral components, one
requires special physical conditions \citep[see e.g.][]{2009MNRAS.399L..59T,2011ApJ...740...64L,2014ApJ...787..155T}, such as large minimum electron
energies ($\gamma_{min} > 10^{2-3}$) and low magnetic fields
(\textit{B}$\lesssim$10--20 mG). 
Moreover, leptonic models are also challenged by the limited variability
of the VHE emission of EHBLs, which differs very much from the
typically high variability observed in the VHE emission of
HBLs. For that reason, several authors have proposed that the VHE
gamma-ray emission is the result of electromagnetic cascades occurring in the intergalactic space,
possibly triggered by a beam of high-energy hadrons produced in the
jet of the EHBLs \citep[e.g.][]{2010APh....33...81E}, or alternatively
produced within leptohadronic models \citep[e.g.][]{2015MNRAS.448..910C}.
Aside from its importance in blazar emission models, the
extremely hard gamma-ray emission allows constraints to be placed on
the IGMF  
\citep[e.g.][]{2010Sci...328...73N}, and provides a powerful tool study the absorption 
of gamma rays in the EBL \citep[e.g.][]{2013IJMPD..2230025C}.
Potential deviations from this absorption are also of interest,
as they could be related to the mixing of photons with new spin-zero bosons such as axion-like particles 
\citep[e.g.][]{2007PhRvD..76l1301D,2009PhRvD..79l3511S,2011PhRvD..84j5030D}.

Therefore, it is evident that EHBLs with hard VHE gamma-ray spectral components are fascinating objects that can be
used to study blazar jet phenomenology, high-energy cosmic rays, EBL
and IGMF. The main problem is that there are only a few sources
identified as EHBLs and detected using IACTs and that they are typically rather faint.
implying the need for very long observations, which complicates
the studies mentioned above.
For instance, 1ES~0229+200,
which is probably the most studied EHBL, has a VHE flux above 580~GeV of only
$\sim$0.02 CU, and for many years it was thought to be a
steady gamma-ray source. A 130-hour observation performed by H.E.S.S. recently
showed that the source is variable \citep{2015arXiv150904470C}, which has
strong implications for example on the lower limits derived on the IGMF.

Mrk~501 has been observed for a number of years by MAGIC and VERITAS, and it has
typically shown a soft VHE gamma-ray spectrum, with a power-law
index $\Gamma \sim$2.5
\citep[e.g.][]{2011ApJ...727..129A,2011ApJ...729....2A,2015A&A...573A..50A}.
It is known that during strong gamma-ray activity, such as the
activity in 1997 and 2005, the VHE spectra
became harder, with $\Gamma \sim$2.1--2.2
\citep{1999A&A...350...17D,1998ApJ...501L..17S,2007ApJ...669..862A} and recently 
\citet{2016A&A...594A..76A}
have also reported similar spectral hardening during
the outstanding activity in May 2009. It is worth noticing that during
the big flare in April 1997, the synchrotron peak of Mrk~501 shifted to
energies beyond 100~keV, and that Mrk~501 was identified as an EHBL by 
\citet{2001A&A...371..512C}. However, this happened only during
extreme flaring events. 
On the contrary, as displayed in Figure~\ref{fig:index_xrt_pl}, during this campaign, Mrk~501 shows very hard X-ray and VHE gamma-ray
spectra during both very high and the quiescent or low activity.
The measured VHE spectra show power-law indices harder than 2.0, which has never been measured
before, and the hardness of the VHE
spectrum is independent of the measured activity. 
A fit to the spectral indices with a constant yields $\Gamma$=2.041$\pm$0.015
($\chi^{2}$/NDF = 86/38). The left panel of Figure~\ref{fig:index_xrt_pl} 
shows an average X-ray spectral index value of 1.752$\pm$0.004 
($\chi^{2}$/NDF = 330/51). In both cases we have clear spectral variability, 
hence spectra which statistically differ from the mean value. 
In contrast to the VHE spectra, in the X-ray spectra one can
observe a dependence on the source activity, with the spectrum
getting harder with increasing flux; but Mrk~501 shows
spectra with photon index $\textless$ 2.0 even for the lowest-activity days.
We did not find any relation between the X-ray and VHE spectral indices.

In summary, during the 2012 campaign, both the X-ray and VHE spectra were persistently harder than
$\Gamma$=2 (during low and high source activity), which implies that the maximum of the synchrotron peak is above 5 keV,
and the maximum of the inverse-Compton peak is above 0.5~TeV. In other words,
Mrk~501 behaved effectively like an EHBL during 2012.
This suggests that being an EHBL may not be a permanent characteristic
of a blazar, but rather a state which may change over time.

%% file: notes/discussion2.tex
\subsection{Modeling temporal evolution of the broadband SED}
\label{Mrk501SSCModel}

The accurate description of the broadband SED of Mrk~501 and its
temporal evolution can be provided by a complex theoretical scenario involving
the superposition of several emitting regions, as reported in
Section \ref{sec:sed}. We have shown that the optical-UV emission and the soft X-ray
emission cannot be parameterized with a single synchrotron component,
something that had already been observed during the campaign
from 2009 \citep{Mrk501MW2009_Variability}. Additionally, the \mbox{3-day-integrated} GeV emission, as measured by
\textit{Fermi}-LAT, is systematically above (at 1--2 $\sigma$ for
single SEDs) the model curves and,  in two SEDs, we found indications of an additional component at MeV--GeV energies,
something which had been also reported by \citet{2015ApJ...798....2S} using observations
from 2010 and 2011. 
However, the X-ray and VHE gamma-ray bands, which
are the segments of the SED with the highest energy flux, and the most variable ones,
can be described in a satisfactory way with a simple one-zone
SSC model. This fact allows one to draw straightforward physical
conclusions with a reduced number of model parameters.

The electron spectral 
indices vary slightly across the models while the break energy 
changes by a factor of three, 
reaching the highest value during the night of the flare. The particle spectra
were found to be hard, with $p_\text{{1}}$ $\leq$ 2 for most cases, which is needed
to explain the very hard X-ray and VHE spectra. We also find a strong
positive correlation between the electron energy density, 
\textit{U}$_\text{{e}}$ (derived from the one-zone SSC model) and the VHE
gamma-ray emission measured by MAGIC and VERITAS.  The Pearson's
correlation coefficient between \textit{U}$_\text{{e}}$ and both the 0.2--1~TeV and above 1~TeV flux is
0.97$^{+0.01}_{-0.02}$, with the significance of the correlation being larger than 7$\sigma$. 
The value of \textit{U}$_\text{{e}}$ depends on the value used for $\gamma_\text{{min}}$, which is not well constrained by the data. 
But we noted that \textit{U}$_\text{{e}}$ only changes by 10--20\% when changing $\gamma_\text{{min}}$ 
by one order of magnitude. Given that the SSC modeling requires changes in \textit{U}$_\text{{e}}$ by factors of a few 
to explain the 17 broadband SEDs of Mrk~501, we consider that the dependency on the chosen value for $\gamma_\text{{min}}$ does not have any 
relevant impact in the significant correlation between the measured VHE flux and the SSC model \textit{U}$_\text{{e}}$  values.
This relation indicates that, within the one-zone SSC used here, the main cause of the broadband SED variability is the injection or
acceleration of electrons. 

The average broadband SED of Mrk~501 during the observing campaign in 2009 was successfully modelled with a 
one-zone SSC scenario, where the energisation of the electrons was attributed to diffusive first-order 
Fermi acceleration \citep{2011ApJ...727..129A}. Yet during the multi-instrument observations in 2012 we 
measured substantially harder X-ray and VHE spectra that required EEDs with harder spectra in the models. 
Such hard-spectrum EEDs may be produced through second-order Fermi acceleration 
\citep{2015MNRAS.447..530C,2011ApJ...740...64L,2009MNRAS.393.1063T,2016ApJ...832..177S}.
Additionally, the radiative cooling of a monoenergetic pileup particle energy distribution can result in a 
power-law particle distribution with index of 2
\citep{2004ApJ...616..136S}. 
These narrow distributions of particles may arise through stochastic
acceleration by energy exchanges with resonant Alfv\'en waves in a
turbulent medium as described by \cite{1985A&A...143..431S},
\cite{2008ApJ...681.1725S}, and \cite{2014ApJ...780...64A}.
In this case quasi-Maxwellian
distributions are obtained: these have been suggested by multiwavelength modeling
of Mrk~421 \citep{2015A&A...578A..22A}. 
Magnetic reconnection in blazar jets
\citep{2010MNRAS.402.1649G}, which has been invoked by \citet{2015ApJ...811..143P}
to explain the variability of Mrk~421, is another process that can
effectively produce hard EEDs \citep{2012ApJ...746..148C,2012ApJ...754L..33C}.  
Additionally, as reported in \citet{2014ApJ...789...66Z,2015ApJ...804...58Z}, through magnetic
reconnection, the dissipated magnetic energy is converted into
non-thermal particle energy, hence leading to a decrease in the magnetic field strength $B$ for increasing
gamma-ray activity and $U_e$. This trend is
also observed in the parameter values retrieved from our SSC model
parameterisation (see Table~\ref{tab:sed1}), thus supporting the hypothesis of
magnetic reconnection occurring in the jets of Mrk~501.

In principle, obtaining hard EEDs from a diffusive shock acceleration 
process is difficult, as first-order Fermi acceleration produces a power-law index with value of 2, and 
the spectrum then evolves in time due to radiative cooling and steepens further. However, Baring and 
collaborators \citep{2017MNRAS.464.4875B} have recently shown that
shock acceleration can also produce hard EEDs with indices as hard as one, primarily because of
efficient drift acceleration in low levels of MHD turbulence near
relativistic shocks: see \citet{2012ApJ...745...63S} for a complete
discussion.

As reported in Section \ref{sec:sed}, the X-ray and gamma-ray segments from the SED related to the large 
VHE flare on MJD 56087 
(2012 June 9) were modelled with a two-zone SSC model in order to better describe
the high-energy peak, with a maximum at $\sim$2~TeV. In 
this scenario, the X-ray and VHE spectra are completely dominated by the emission of a region that is 
smaller (by one order of magnitude), and with a narrower EED characterized by a very high minimum Lorentz 
factor $\gamma_\text{{min}}$. This multizone SSC scenario was successfully used to model the broadband SEDs of 
Mrk~421 that also showed peaked or multi-peaked structures during a 13-day period of flaring activity in March 2010 
\citep{2015A&A...578A..22A}. The relatively steady optical and GeV emission could be produced in a 
shock-in-jet component while the variable X-ray and VHE gamma-ray emission could arise from a component 
originating in the base of the jet and producing this relatively narrow EED. The more compact zone is probably
intimately connected to the injector site, perhaps a jet shock, thereby
more directly sampling the acceleration characteristics since there
has been less time for electrons to cool in the ambient magnetic field \citep{2017MNRAS.464.4875B}.

It is also worth noting that the large VHE flare from June 9 2012 occurred when the degree of 
polarization was at its lowest value ($\sim$2\%) during the 2012
campaign (see bottom panels of Figure~\ref{fig:lightcurvecombinedTeV}). On 
the other hand, the large VHE flare from May 1 2009 occurred when the degree of polarization was at its 
highest value ($\sim$5\%) during the 2009 campaign \citep{2016A&A...594A..76A,Mrk501MW2009_Variability}. Since enhanced
polarization is naturally anticipated in short duration flares where
smaller length scales are sampled, this observational dichotomy
complicates the picture. This observation suggests that 
there is a diversity in gamma-ray flares in Mrk~501, and at least some of them 
seem not to involve any change in the degree of polarization, which may occur naturally if the optical and
the VHE emission are produced in different regions of the jet.

%% file: notes/discussion3.tex
\subsection{Multiband variability and correlations}
\label{discussion:variability}

Section \ref{sec:variability} reports a general increase in the flux variability with
increasing energy. At radio, optical and UV bands we observe relatively
low variability ($F_{\text{var}}\leq 0.1$), but for the variability at the 37 GHz
radio fluxes from Metsahovi, which
is 0.13$\pm$0.02. This variability is not produced by a flare, or by a
slow temporal evolution (weeks or months long) of the light curve, which is
often observed at radio, but by a consistent flickering in the radio fluxes.
Such flickering is rare in blazars, but it has been
already reported in previous observing campaigns of Mrk~501
\citep[][]{2015A&A...573A..50A,2015ApJ...812...65F}. In the X-rays and GeV
gamma-ray bands we observe high variability ($F_{\text{var}}\sim
0.2-0.4$). However, we note that we do not have sensitivity to
determine the fractional variability in the
band 0.2-2~GeV (where the excess variance is negative), and hence the
fractional variability in this band could be lower than that measured at
X-rays. 
In the VHE gamma-ray band we observe very high variability
($F_{\text{var}}\sim 0.5-0.9$), i.e. about three times larger than that at X-rays.

Such a multiband (from radio to VHE) variability pattern 
was first reported with observations from 2008 in \citet{2015A&A...573A..50A} and then confirmed with more 
precise measurements from the 2009 campaign \citep{Mrk501MW2009_Variability}. The repeated occurrence of 
this variability pattern in 2012 demonstrates that this is a typical characteristic in the broadband 
emission of Mrk~501. 
On the other hand, \citet{2015ApJ...812...65F} show that, during the observing 
campaign in 2013, the multiband variability pattern was somewhat different, with the variability at X-rays 
being similar to that at VHE, thereby showing that somewhat different dynamical processes occurred in 
Mrk~501 during that year.

It is worth comparing this multi-year variability pattern of Mrk~501 with that from the other archetypical 
TeV blazar, Mrk~421. During the multi-instrument campaigns from 2009, 2010 and 2013, as reported in 
\cite{2015A&A...576A.126A}, \cite{2015A&A...578A..22A} and \cite{2016ApJ...819..156B}, Mrk~421 showed a 
double-peak structure in the plot of $F_\text{{var}}$ against energy, 
where the largest variability occurs in X-rays and VHE (instead of a
broad increase with energy, with the variability at VHE being much larger than that at X-rays).
These observations show a fundamentally different behavior when compared to that of Mrk~501.

\begin{figure}
   \centering
       \includegraphics[width=10.0cm]{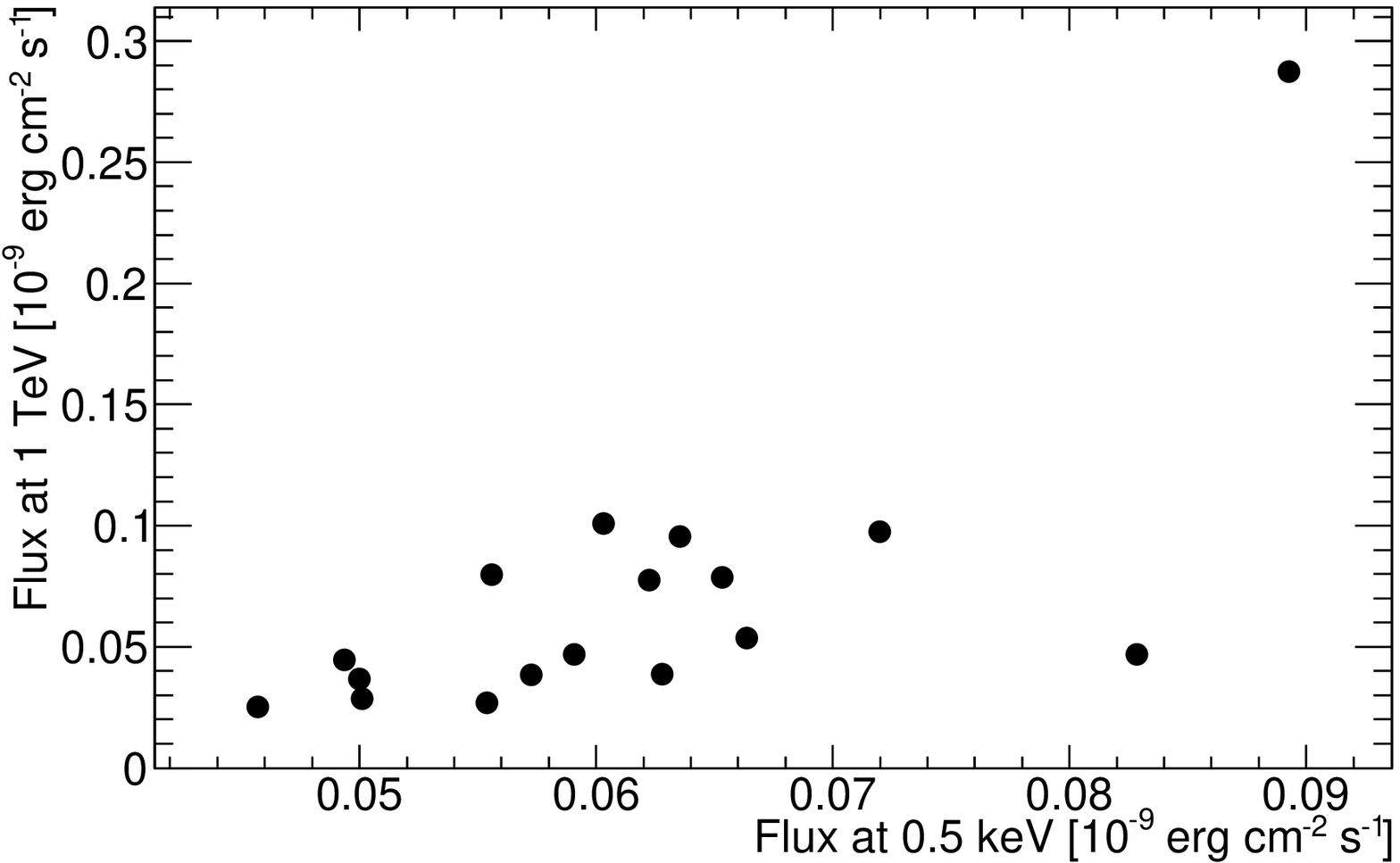}
       \includegraphics[width=10.0cm]{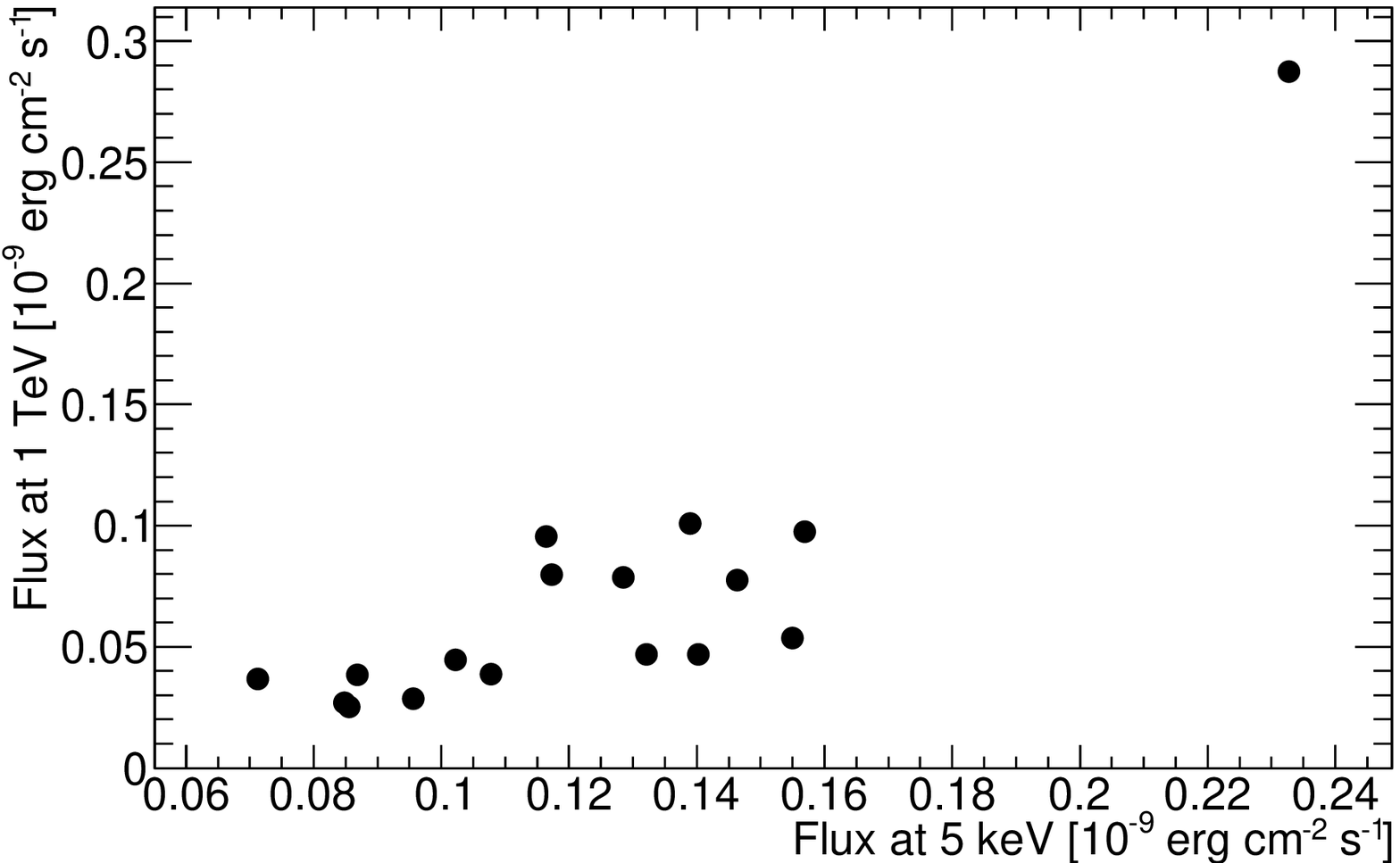}
       \includegraphics[width=10.0cm]{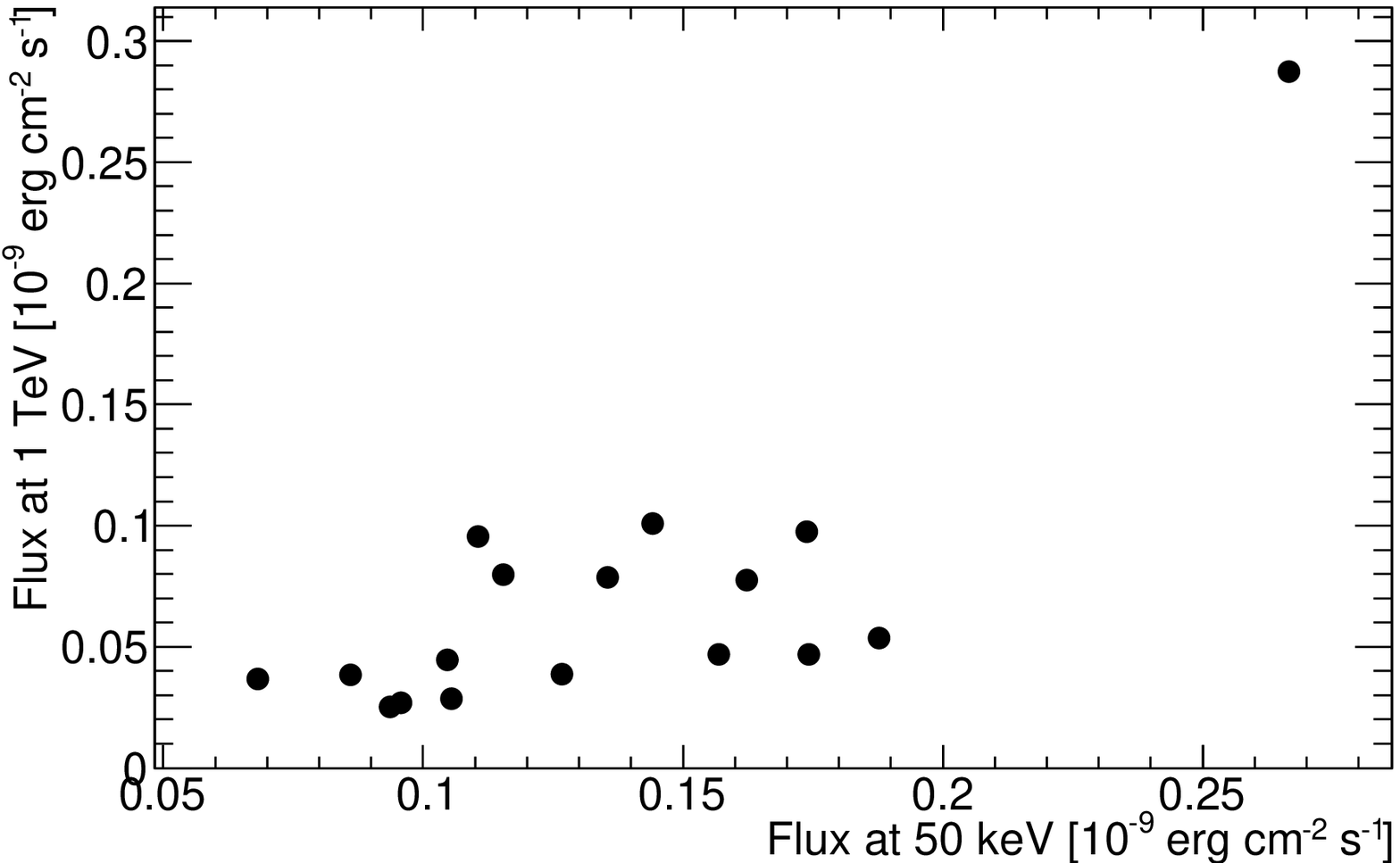}
   \caption{Flux-flux plots derived from the one-zone SSC model used
     to fit 17 broadband SEDs (see Section~\ref{sec:sed}). The SSC model flux at 1 TeV is compared to the
     SSC model flux at 0.5~keV (top), 5~keV (middle) and 50~keV
     (bottom).
     The data point with the highest X-ray and VHE activity
     corresponds to the one-zone SSC model for the June 9 flare (MJD 56087).
     }
     \label{fig:sedcorrtev}
   \end{figure}

During large VHE gamma-ray flaring activity, the X-ray and VHE gamma-ray emission of Mrk~501 have 
been found to be correlated. This occurred during the long and historical flare of 1997 
\citep[][]{1998ApJ...492L..17P,Gliozzi:2006it}, and the large few-days-long flare observed in 2013 
\citep{2015ApJ...812...65F}. During non-flaring activity, a positive X-ray/VHE correlation was reported 
at the 99\% confidence level \citep{2015A&A...573A..50A}. On the other hand, using the measurements from 
the 4.5-month-long 2009 campaign, where significant variability was observed in both X-ray and VHE bands, 
the emission from these two bands was found to be uncorrelated \citep{Mrk501MW2009_Variability}. Using the 
data collected during the 2012 campaign, with many more observations than in 2009, one also finds only marginal
correlation between the X-ray emission and the VHE gamma-ray emission. 
This is an interesting result
because, under the most simplistic and widely 
accepted theoretical scenarios, the X-ray emission and the VHE gamma-ray 
emission are produced by the same population of high-energy particles (electrons and positrons). We note 
also that the situation for Mrk~421 is radically different from that of Mrk~501. The various multi-instrument 
campaigns performed on Mrk~421 always show a clear and positive correlation between the X-ray emission and the VHE 
gamma-ray emission, during both high and low source activity 
\citep{2008ApJ...677..906F,2011ApJ...738...25A,2015A&A...578A..22A,2015A&A...576A.126A,2016A&A...593A..91A,2016ApJ...819..156B}.

\begin{table}
  \caption{Correlations derived for several combinations of X-ray
    bands and VHE flux above 1 TeV using the data (upper part) and the
    one-zone SSC theoretical model used to describe 17 broadband
    SEDs. See text for further details.}
  \label{tab:datamodelcorr}
  \begin{center}
    \begin{tabular}{ l | r } 
       \hline
         & Pearson correlation \\ 
         & coefficient ($\sigma$) \\
       \hline
         Data: 0.3--2~keV vs > 1~TeV & 0.78$^{+0.10}_{-0.15}$ (3.9) \\
         Excluding June 9 flare: & 0.39$^{+0.23}_{-0.29}$ (1.6) \\
         \vspace{-0.3cm} 
~~~~  &  ~~~~~~ \\
         Data: ~2--10~keV vs > 1~TeV & 0.88$^{+0.06}_{-0.10}$ (4.9) \\
         Excluding June 9 flare: & 0.59$^{+0.16}_{-0.24}$ (2.5) \\
          \vspace{-0.2cm} 
         ~~~    &  ~~~~ \\
	 \hline
\vspace{-0.2cm} 
~~~~  &  ~~~~~~ \\
Model: 0.5~~keV vs 1~TeV & 0.71$^{+0.11}_{-0.16}$ (3.3) \\
Excluding June 9 flare: & 0.44$^{+0.19}_{-0.25}$ (1.7) \\
\vspace{-0.3cm} 
~~~~  &  ~~~~~~ \\
Model: 5.0~~keV vs 1~TeV & 0.87$^{+0.06}_{-0.09}$ (4.8) \\
Excluding June 9 flare: & 0.63$^{+0.14}_{-0.20}$ (2.7) \\
\vspace{-0.3cm} 
~~~~  &  ~~~~~~ \\
Model: 50.0~keV vs 1~TeV & 0.77$^{+0.09}_{-0.13}$ (3.8) \\
Excluding June 9 flare: & 0.73$^{+0.11}_{-0.16}$ (3.4) \\
       \hline
      \end{tabular}
    \end{center}
\end{table}

\citet{Mrk501MW2009_Variability} put forward two scenarios to explain the measured multiband variability 
and correlations seen in the emission of Mrk~501: a) the high-energy electrons that are responsible for a large 
part of the TeV emission do not dominate the keV emission; b) there is an additional (and very 
variable) component contributing to the TeV emission, such as external inverse-Compton. 
In this manuscript, we 
use the results from our one-zone SSC modelling of the 17 broadband SEDs from 2012 to probe the first 
scenario. We compared the one-zone SSC model fluxes at 0.5~keV, 5~keV and 50~keV with 
the one-zone SSC model fluxes at 1~TeV, which are, by construction of the theoretical model, produced by 
the same population of electrons. The results are depicted in Figure~\ref{fig:sedcorrtev}, and the 
correlations obtained are reported in Table~\ref{tab:datamodelcorr}. The  X-ray
vs VHE gamma-ray correlation 
as a function of the X-ray energy is: 3.3$\sigma$ for 0.5~keV, 4.8
$\sigma$ for 5~keV, and 3.8$\sigma$ for 50~keV. As reported in section \ref{sec:sed} 
(see Table \ref{tab:sed1}), the June 9 flare (MJD 56087) is not
properly described by the one-zone SSC scenario used here. If we remove the
results derived with the SSC model for this large flare, hence providing a
more reliable description of the typical behaviour of Mrk~501 during
the campaign in 2012,  the  significance of the X-ray/VHE  correlation is 1.7$\sigma$ for 0.5~keV, 2.7$\sigma$ for 5~keV, and 3.4$\sigma$ 
for 50~keV. Therefore, the one-zone SSC model provides X-ray/VHE correlations at a level that are 
consistent with correlations obtained with the measured X-ray and VHE gamma-ray fluxes reported in 
Table~\ref{tab:Corr1} \citep{2017A&A...603A..31A}. This shows that an
additional high-energy component (e.g. external inverse-Compton) is not necessary to explain the 
variability and correlation patterns observed in Mrk~501.

This exercise also shows the importance of sampling with accuracy a large portion of the electromagnetic 
spectrum. 
In particular, sensitivity in the 50--100~keV range comparable to that currently provided by 
MAGIC and VERITAS in the 0.1--1~TeV range would greatly increase the potential for studying flux variability 
and interband correlations.
The main differences in the multiband 
variability and correlation patterns with Mrk~421 may be related to the fact that, for Mrk~421, 
the electrons dominating the emission of the $\sim$1~TeV photons also dominate the emission at 
$\sim$1~keV (the peak of the synchrotron spectrum).
This is the energy region sampled with \textit{Swift}/XRT with exquisite accuracy and extensive temporal coverage.

Recent and future publications devoted to multiwavelength campaigns on blazars will continue to benefit from a new 
generation of X-ray telescopes, including \textit{NuSTAR}\footnote{\url{http://www.nustar.caltech.edu}} and 
\textit{Astrosat}\footnote{\url{http://astrosat.iucaa.in}}, which
operate at 3--79~keV and 2--80~keV respectively. \textit{NuSTAR}
represents a significant improvement (by a factor of 100~in sensitivity)
over coded-mask instruments like \textit{Swift}/BAT, and hence
provides a much better view into the hard X-ray emission of blazars.

%% file: notes/conclusion.tex
\section{Summary and conclusion}
\label{sec:conclussion}

We have presented the results from the 2012 Mrk~501 multiwavelength campaign.
An excellent set of data was taken using more than 25 instruments over the period
covering March to June 2012. The source flux was observed
to vary between 0.5 and 4.9~CU above 1~TeV, with an average flux of $\sim$1~CU.
The highest VHE gamma-ray flux was observed on a single night, June 9.
This outburst was also observed in the X-ray band by the \textit{Swift}/XRT instrument. 

The fractional variability was seen to increase as a function of energy 
and peak in the VHE regime. 
This is similar to what has been seen in other multiwavelength campaigns targeting 
Mrk~501, but different to the behavior of Mrk~421 which peaks at X-ray energies.
This remains the case even when the flare information is removed or when we consider
only data taken simultaneously in the X-ray and VHE bands, thereby
underlining the difference already observed between these two sources.

Investigating possible correlations between X-ray and VHE bands, 
in two energy ranges each, a maximum Pearson correlation coefficient
of 4.9$\sigma$ was observed for energies above 2~keV and 1~TeV. 
The significance of this correlation drops to 2.5$\sigma$ when the flare day is excluded.
A further search for correlation
using a discrete correlation function found no evidence for a time lag between X-rays and VHE
gamma rays.  

Interestingly, the X-ray and VHE power-law index corresponded to an extremely hard spectrum during
the entire three-month period.
The VHE spectral index above 0.2 TeV, as observed by both MAGIC
and VERITAS, was around $\sim$2,
compared to the typical power-law index of 
2.5 \citep{2011ApJ...729....2A,2011ApJ...727..129A,2015A&A...573A..50A}.
The source did not show the previously observed harder-when-brighter
behavior at VHE energies.
In the X-ray domain, Mrk~501 showed a hardening of the spectral shape
with increasing X-ray flux, but  the X-ray power-law spectral
index was always less than 2 (for both low and high activity). Therefore, 
the synchrotron peak was located above 5 keV and the inverse-Compton peak above 0.5 TeV, 
making Mrk 501 an extreme HBL during the entire observing period.
This suggests that being an EHBL is a temporary state of the source, instead of an instrinsic characteristic.

We were able to form 17 SEDs, where the time difference between X-ray and VHE
data taking was less than four hours.
The X-ray and VHE data were modeled using a one-zone SSC model; however,
the model underestimated the amount of optical and UV radiation required to fit the observed light curves.
During 2012 the optical light curve was observed to be at a 10-year low and we assume that this component
comes from a different region.
The emission at GeV is systematically (within 1--2$\sigma$) above the 17 SSC model curves, two 
showing a data-model difference of $\sim$3$\sigma$, which may be interpreted as a hint for the existence of an 
additional contribution at GeV energies, and is not considered in the current theoretical scenario.
A two-zone model was also used in order to model the flare day of \mbox{June~9 (MJD~56087).} 
The two-zone SSC scenario improved the data-model agreement with respect to the one-zone SSC model and theoretical assumptions,
although it still does not describe the broadband data well.

Despite the caveats mentioned above, the one-zone SSC framework provides a reasonable description of the segments 
of the SED where most energy is emitted, and where most of the variability occurs. 
The direct relation between the electron energy density
and the gamma-ray activity shows that most of 
the variability can be explained by injection of high-energy electrons. 
The very hard EED (\textit{p}$_\text{{1}}$<2)  obtained in our fits, together with the
trend of lower magnetic field strength $B$ for higher electron energy
densities $U_e$, suggests that magnetic reconnection plays a dominant
role in the acceleration of the particles. 
However, we cannot exclude scenarios that incorporate other mechanisms
for acceleration of the radiating (high-energy) particle population,
such as efficient shock drift acceleration, or second-order Fermi
acceleration if the electrons can be effectively trapped in regions of
strong turbulence.

%% file: notes/appendix_A.tex
\section*{Appendix A}
\label{sec:appA}

\begin{figure}[h]
   \centering
   \includegraphics[width=25pc]{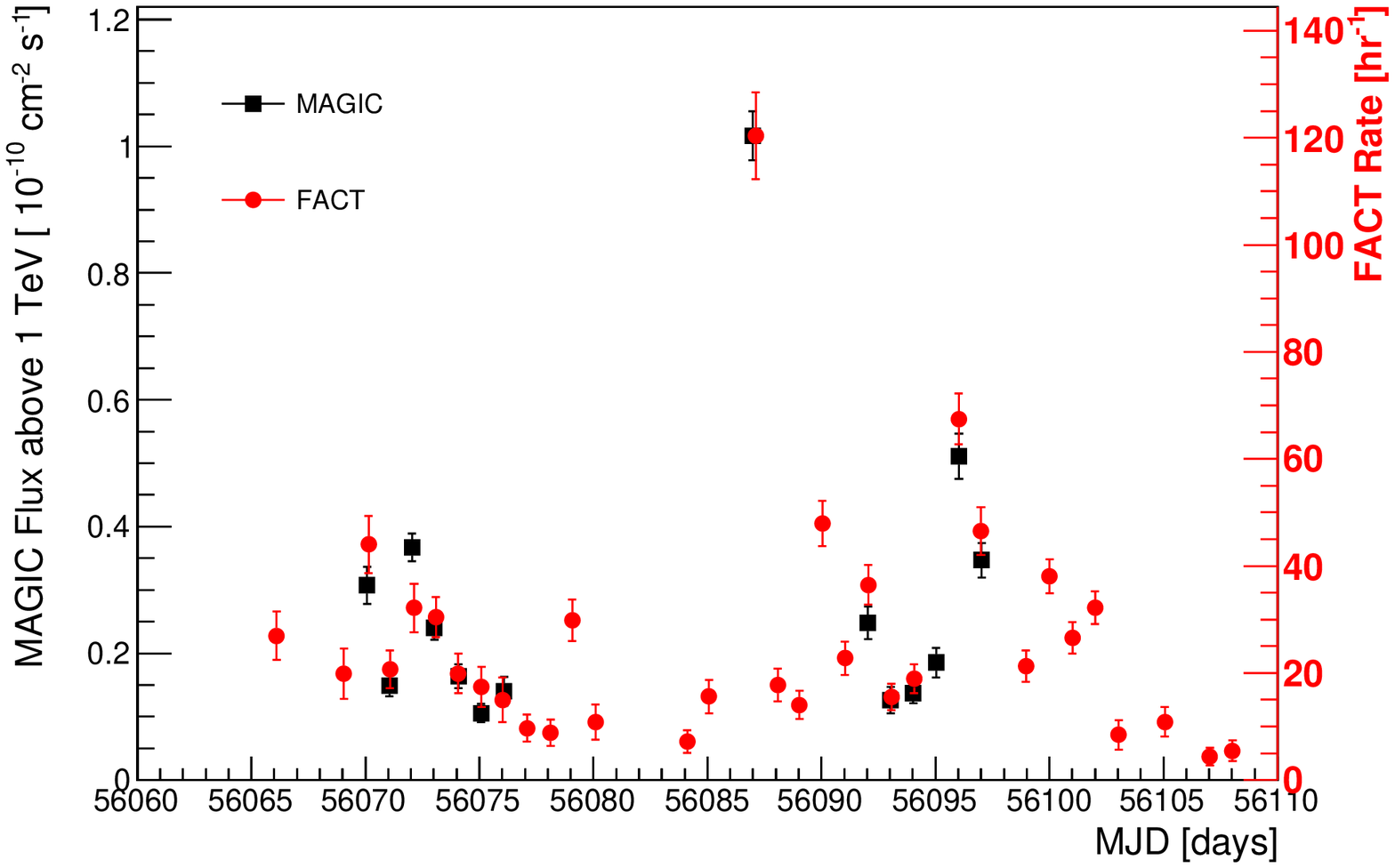}
   \includegraphics[width=25pc]{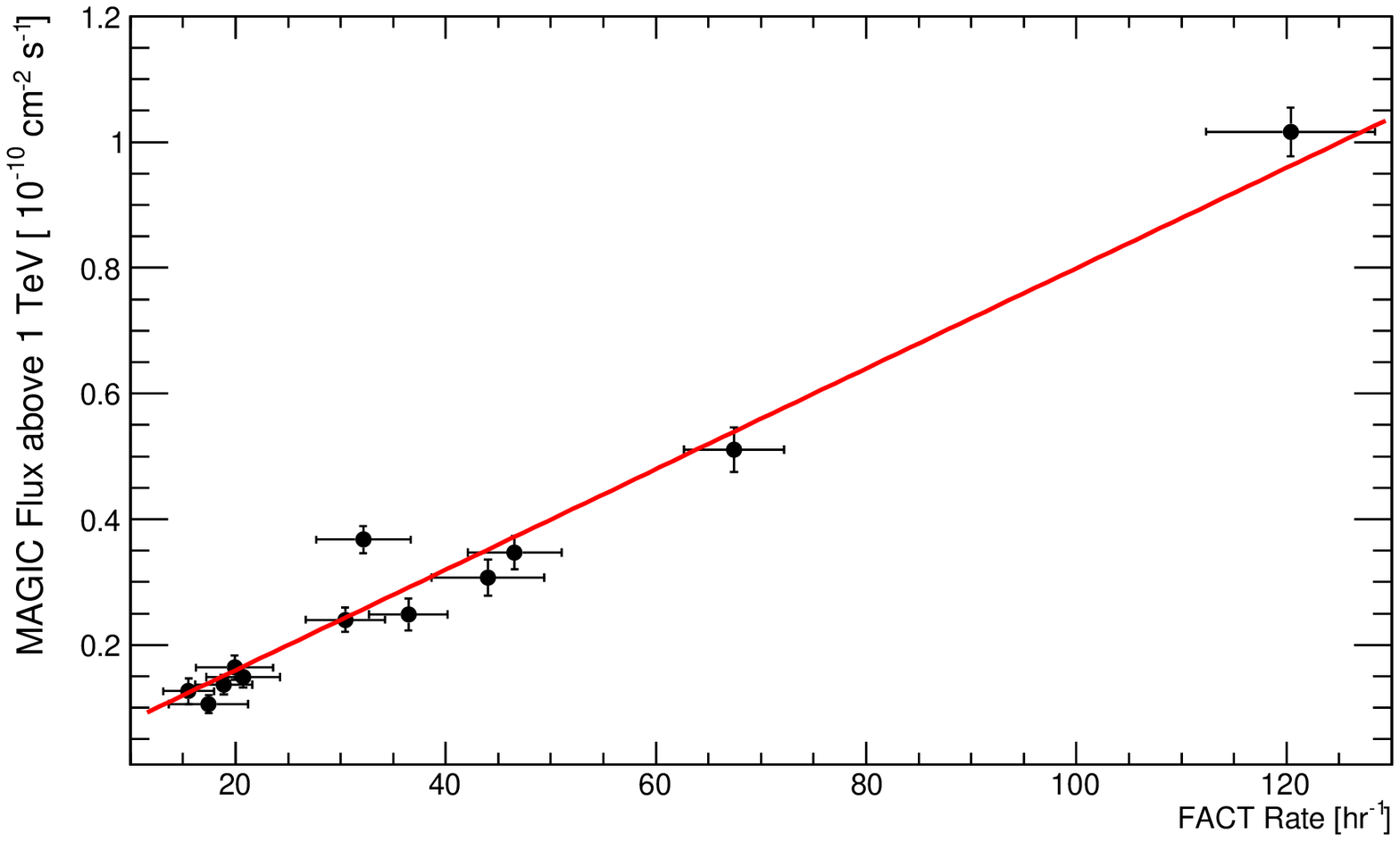}
   \caption{Top panel: 
     superposition of the FACT excess rates and MAGIC light curve above 1 TeV.
     Bottom panel: correlation between the FACT excess rate and the MAGIC flux
     above 1~TeV on the 12 days when data were taken by both instruments.
     }
          \label{fig:corFACT}
 \end{figure}

FACT and MAGIC are located within 100~m of each other and therefore 
observe under exactly the same atmospheric conditions. The
instrumentation used in these two telescopes is different (e.g. FACT
uses SiPMs as light detectors, instead of PMTs), and the observation
mode (stereo vs mono) and analysis chains are completely separate. 
It is interesting and useful therefore to compare the two experiments' light curves. 
Figure~\ref{fig:corFACT} shows the MAGIC light curve above 1 TeV superimposed with
the FACT light curve. The scale is chosen so that for the night of the highest flux the points
overlap. As FACT monitors Mrk~501 every possible night, the FACT light curve is
more densely sampled than that of MAGIC. 
The right panel of Figure~\ref{fig:corFACT} shows that there is an
excellent agreement between the MAGIC VHE fluxes above 1 TeV and the
excess rates measured with FACT. 
The flux-flux plot is fit to a first order polynomial resulting in a $\chi^{2}$/DoF of 10.4/10,
with a slope of (8.51$\pm$0.61)$\times$10$^{-13}$ cm$^{-2}$ s$^{-1}$ hr$^{-1}$ and an offset of
($-$0.0021$\pm$0.0021)$\times$10$^{-10}$ cm$^{-2}$ s$^{-1}$ hr$^{-1}$ , which is consistent with zero, as expected. 
This function was then used to normalize the FACT excess rate to the
fluxes (above 1 TeV) reported in the bottom panel of Figure~\ref{fig:lightcurvecombinedTeV}.

%% file: notes/appendix_B.tex
\section*{Appendix B}
\label{sec:appB}

This section reports the spectral parameters resulting from the fits to the X-ray and gamma-ray spectra.

\begin{table}[h]
\caption{
Parameters resulting from the fit with a power law F(E) =
N$_{0}$(E/0.5~TeV)$^{-{\Gamma}}$ to the measured MAGIC spectra shown
in Figures~\ref{fig:index_xrt_pl}, \ref{fig:sedall}, \ref{fig:sedall2}, and \ref{fig:sed}.
The fitted spectra were EBL corrected using \citet{2008A&A...487..837F}. }
\label{tab:magicIndex}      
\centering                          
\begin{tabular}{l c c c }        
\hline\hline                 
MJD & N$_\text{{o}}$ [10$^{-11}$ s$^{-1}$ cm$^{-2}$ TeV$^{-1}$] & $\Gamma$ & $\chi^{2}$/dof \\
\hline                        

56007 & ~1.90$\pm$0.25 & 1.85$\pm$0.25 &~0.70/3 \\
56032 & ~2.39$\pm$0.23 & 1.89$\pm$0.15 &~1.49/3 \\
56036 & ~4.36$\pm$0.30 & 1.88$\pm$0.08 &~0.33/4 \\
56040 & ~2.46$\pm$0.23 & 2.31$\pm$0.18 &~1.36/3 \\
56070 & ~5.15$\pm$0.25 & 1.94$\pm$0.08 &~1.99/3 \\
56071 & ~2.85$\pm$0.18 & 2.22$\pm$0.09 &~6.76/5 \\
56072 & ~5.60$\pm$0.20 & 2.02$\pm$0.04 &~8.94/6 \\
56073 & ~4.73$\pm$0.23 & 2.20$\pm$0.07 &18.37/6 \\
56074 & ~2.41$\pm$0.18 & 2.01$\pm$0.09 &~5.63/5 \\
56075 & ~1.83$\pm$0.15 & 2.11$\pm$0.11 &~5.49/5 \\
56076 & ~2.24$\pm$0.23 & 2.20$\pm$0.15 &~6.79/5 \\
56087 & 15.43$\pm$0.31 & 2.02$\pm$0.03 &96.26/6 \\
56093 & ~2.50$\pm$0.21 & 2.15$\pm$0.12 &~4.26/5 \\
56094 & ~2.60$\pm$0.17 & 2.24$\pm$0.09 &~3.09/5 \\
56095 & ~2.75$\pm$0.21 & 2.05$\pm$0.10 &~2.70/5 \\
56096 & ~7.15$\pm$0.30 & 2.09$\pm$0.05 &12.94/6 \\
56097 & ~6.17$\pm$0.29 & 2.20$\pm$0.07 &~5.82/5 \\

\hline                                   
\end{tabular}
\end{table}

\begin{table}[h]
\caption{
Parameters resulting from the fit with a power law F(E) =
N$_{0}$(E/0.5~TeV)$^{-{\Gamma}}$ to the measured VERITAS spectra shown
in Figures~\ref{fig:index_xrt_pl}, \ref{fig:sedall}, and \ref{fig:sedall2}.
The fitted spectra were EBL corrected using \citet{2008A&A...487..837F}. }
\label{tab:veritasIndex}      
\centering                          
\begin{tabular}{l c c c }        
\hline\hline                 
MJD & N$_\text{{o}}$ [10$^{-11}$ s$^{-1}$ cm$^{-2}$ TeV$^{-1}$] & $\Gamma$ & $\chi^{2}$/dof \\
\hline                        
56007 & 1.08$\pm$0.09 & 2.05$\pm$0.10 & 10.73/6 \\
56009 & 2.26$\pm$0.19 & 1.99$\pm$0.09 & 20.46/7 \\
56015 & 1.82$\pm$0.17 & 1.75$\pm$0.09 & ~6.80/6 \\
56034 & 0.89$\pm$0.11 & 1.59$\pm$0.21 & 10.82/6 \\
56038 & 0.58$\pm$0.09 & 1.91$\pm$0.24 & ~9.01/5 \\
56046 & 0.65$\pm$0.08 & 1.80$\pm$0.18 & ~4.85/6 \\
56050 & 0.70$\pm$0.08 & 2.09$\pm$0.23 & ~5.39/6 \\
56061 & 0.71$\pm$0.08 & 2.19$\pm$0.19 & ~9.15/4 \\
56066 & 1.12$\pm$0.09 & 2.10$\pm$0.09 & ~1.74/6 \\
56069 & 1.05$\pm$0.08 & 2.02$\pm$0.13 & ~4.05/4 \\
56073 & 2.27$\pm$0.12 & 2.08$\pm$0.05 & 10.22/5 \\
56075 & 2.52$\pm$0.45 & 2.12$\pm$0.22 & ~3.97/2 \\
56076 & 1.35$\pm$0.14 & 1.94$\pm$0.11 & ~1.73/3 \\
56077 & 1.62$\pm$0.28 & 1.23$\pm$0.26 & ~0.42/3 \\
56090 & 2.08$\pm$0.40 & 1.76$\pm$0.29 & 12.39/4 \\
56092 & 3.57$\pm$0.30 & 2.48$\pm$0.19 & 11.57/6 \\
56093 & 1.46$\pm$0.35 & 1.43$\pm$0.23 & ~3.39/2 \\
56094 & 2.14$\pm$0.31 & 1.81$\pm$0.29 & ~2.42/3 \\
56095 & 2.06$\pm$0.30 & 1.95$\pm$0.24 & ~2.22/4 \\
56096 & 2.55$\pm$0.31 & 1.54$\pm$0.26 & ~5.07/3 \\
56097 & 1.76$\pm$0.23 & 1.91$\pm$0.17 & ~8.97/3 \\
56099 & 0.85$\pm$0.15 & 1.66$\pm$0.20 & ~1.16/2 \\
\hline                                   
\end{tabular}
\end{table}

\begin{table*}[h]
\caption{\textit{Fermi}--LAT flux and power-law index above 0.2~GeV
  from the gamma-ray spectra shown
in Figures~\ref{fig:sedall}, \ref{fig:sedall2}, and \ref{fig:sed}.}
\label{tab:fermiIndex}      
\centering                          
\begin{tabular}{c c c c c}        
\hline\hline                 
Observation [MJD] & Flux [10$^{-8}$ cm$^{-2}$ s$^{-1}$]  & $\Gamma$ & TS \\
\hline
56007.5--56010.5 & ~6.3$\pm$2.1 & 1.7$\pm$0.2 & 43 \\
56013.5--56016.5 & ~7.3$\pm$2.4 & 1.8$\pm$0.2 & 61 \\
56030.5--56033.5 & ~4.9$\pm$2.0 & 1.9$\pm$0.3 & 34 \\
56032.5--56035.5 & ~3.6$\pm$1.9 & 1.9$\pm$0.4 & 19 \\
56034.5--56037.5 & ~5.0$\pm$2.7 & 2.4$\pm$0.6 & 16 \\
56036.5--56039.5 & ~2.2$\pm$1.2 & 1.4$\pm$0.3 & 33 \\
56038.5--56041.5 & ~2.9$\pm$1.9 & 1.6$\pm$0.3 & 28 \\
56044.5--56047.5 & ~8.4$\pm$2.7 & 2.0$\pm$0.2 & 52 \\
56059.5--56062.5 & ~1.1$\pm$1.1 & 1.8$\pm$0.7 & ~4 \\
56064.5--56067.5 & ~7.6$\pm$2.7 & 2.1$\pm$0.3 & 40 \\
56071.5--56074.5 & ~5.7$\pm$1.9 & 1.7$\pm$0.2 & 56 \\
56074.5--56077.5 & ~2.3$\pm$1.3 & 1.4$\pm$0.3 & 24 \\
56075.5--56078.5 & ~4.0$\pm$1.6 & 1.5$\pm$0.2 & 41 \\
56086.5--56087.5 & 12.0$\pm$5.0 & 1.5$\pm$0.2 & 59 \\
56088.5--56091.5 & ~8.4$\pm$2.9 & 1.7$\pm$0.2 & 71 \\
56092.5--56095.5 & ~8.1$\pm$2.5 & 1.8$\pm$0.2 & 65 \\
56093.5--56096.5 & ~7.5$\pm$2.3 & 1.7$\pm$0.2 & 72 \\
\hline                                   
\end{tabular}
\end{table*}

\begin{table*}[h]
\caption{Spectral models (log-parabola and power-law functions) fitted to the \textit{Swift}--XRT data used
  in  Figures~\ref{fig:index_xrt_pl}, \ref{fig:sedall}, \ref{fig:sedall2}, and \ref{fig:sed}.}
\label{tab:xrtIndex}      
\centering                          
\begin{tabular}{l c c c c c c c}        
\hline\hline                 
Start Time & $\alpha$ & $\beta$ & $\chi^{2}$/dof & 0.3--2~keV & 2--10~keV & $\alpha$ & $\chi^{2}$/dof \\
MJD & & & & [10$^{-10}$ erg cm$^{-2}$ s$^{-1}$] & [10$^{-10}$ erg cm$^{-2}$ s$^{-1}$] & & \\
\hline                        

55972.079 & 1.532$\pm$0.052 & 0.356$\pm$0.094 & 189/221 & 1.545$\pm$0.024 & 2.062$\pm$0.069 & 1.687$\pm$0.028 & 233/222 \\ 
55977.296 & 1.624$\pm$0.053 & 0.411$\pm$0.103 & 226/205 & 1.144$\pm$0.018 & 1.255$\pm$0.046 & 1.778$\pm$0.028 & 277/206 \\ 
55981.242 & 1.694$\pm$0.055 & 0.282$\pm$0.107 & 176/198 & 1.041$\pm$0.019 & 1.147$\pm$0.047 & 1.809$\pm$0.033 & 196/199 \\ 
55985.247 & 1.757$\pm$0.052 & 0.261$\pm$0.102 & 191/191 & 1.427$\pm$0.026 & 1.443$\pm$0.054 & 1.862$\pm$0.033 & 210/192 \\ 
55989.256 & 1.669$\pm$0.047 & 0.237$\pm$0.093 & 211/219 & 1.346$\pm$0.021 & 1.609$\pm$0.051 & 1.761$\pm$0.028 & 230/220 \\ 
55994.206 & 1.712$\pm$0.046 & 0.347$\pm$0.092 & 190/222 & 1.306$\pm$0.021 & 1.316$\pm$0.044 & 1.849$\pm$0.028 & 233/223 \\ 
56000.142 & 1.634$\pm$0.047 & 0.402$\pm$0.091 & 270/229 & 1.650$\pm$0.023 & 1.797$\pm$0.057 & 1.804$\pm$0.026 & 329/230 \\ 
56005.166 & 1.638$\pm$0.055 & 0.363$\pm$0.108 & 208/195 & 1.628$\pm$0.029 & 1.822$\pm$0.074 & 1.782$\pm$0.033 & 243/196 \\ 
56009.429 & 1.662$\pm$0.047 & 0.267$\pm$0.072 & 221/229 & 1.374$\pm$0.021 & 1.617$\pm$0.054 & 1.776$\pm$0.027 & 247/230 \\ 
56011.450 & 1.519$\pm$0.049 & 0.378$\pm$0.088 & 264/245 & 1.447$\pm$0.020 & 1.930$\pm$0.054 & 1.691$\pm$0.026 & 320/246 \\ 
56015.451 & 1.606$\pm$0.055 & 0.321$\pm$0.095 & 203/217 & 1.408$\pm$0.023 & 1.725$\pm$0.055 & 1.758$\pm$0.029 & 237/218 \\ 
56017.316 & 1.685$\pm$0.051 & 0.214$\pm$0.095 & 204/220 & 1.370$\pm$0.024 & 1.629$\pm$0.057 & 1.780$\pm$0.029 & 219/221 \\ 
56019.188 & 1.714$\pm$0.044 & 0.250$\pm$0.087 & 232/230 & 1.257$\pm$0.020 & 1.380$\pm$0.043 & 1.812$\pm$0.027 & 256/231 \\ 
56019.457 & 1.623$\pm$0.051 & 0.342$\pm$0.099 & 177/202 & 1.186$\pm$0.018 & 1.387$\pm$0.053 & 1.755$\pm$0.029 & 214/203 \\ 
56023.080 & 1.760$\pm$0.059 & 0.260$\pm$0.116 & 185/181 & 1.095$\pm$0.024 & 1.104$\pm$0.044 & 1.866$\pm$0.034 & 200/182 \\ 
56027.548 & 1.767$\pm$0.057 & 0.302$\pm$0.113 & 198/187 & 1.173$\pm$0.024 & 1.125$\pm$0.044 & 1.890$\pm$0.033 & 220/188 \\ 
56032.174 & 1.664$\pm$0.054 & 0.297$\pm$0.100 & 183/215 & 1.330$\pm$0.025 & 1.518$\pm$0.054 & 1.796$\pm$0.030 & 209/216 \\ 
56034.249 & 1.607$\pm$0.054 & 0.298$\pm$0.103 & 219/190 & 1.080$\pm$0.019 & 1.350$\pm$0.053 & 1.728$\pm$0.031 & 244/191 \\ 
56036.043 & 1.590$\pm$0.057 & 0.305$\pm$0.104 & 176/208 & 1.229$\pm$0.021 & 1.568$\pm$0.055 & 1.727$\pm$0.031 & 202/209 \\ 
56038.376 & 1.677$\pm$0.058 & 0.397$\pm$0.112 & 186/189 & 1.003$\pm$0.019 & 1.024$\pm$0.040 & 1.846$\pm$0.033 & 223/190 \\ 
56040.123 & 1.710$\pm$0.054 & 0.313$\pm$0.109 & 213/196 & 1.169$\pm$0.022 & 1.218$\pm$0.047 & 1.836$\pm$0.033 & 237/197 \\ 
56042.395 & 1.795$\pm$0.061 & 0.258$\pm$0.125 & 162/167 & 1.199$\pm$0.028 & 1.144$\pm$0.051 & 1.894$\pm$0.038 & 174/168 \\ 
56044.005 & 1.706$\pm$0.075 & 0.326$\pm$0.135 & 119/165 & 1.100$\pm$0.025 & 1.142$\pm$0.049 & 1.857$\pm$0.039 & 136/166 \\ 
56046.412 & 1.691$\pm$0.057 & 0.298$\pm$0.111 & 197/193 & 1.108$\pm$0.021 & 1.208$\pm$0.048 & 1.815$\pm$0.034 & 218/194 \\ 
56047.413 & 1.673$\pm$0.057 & 0.265$\pm$0.099 & 200/205 & 1.128$\pm$0.021 & 1.306$\pm$0.045 & 1.793$\pm$0.031 & 219/206 \\ 
56048.148 & 1.574$\pm$0.057 & 0.323$\pm$0.111 & 195/196 & 1.020$\pm$0.019 & 1.312$\pm$0.053 & 1.711$\pm$0.033 & 220/197 \\ 
56053.089 & 1.678$\pm$0.054 & 0.296$\pm$0.107 & 186/192 & 1.053$\pm$0.019 & 1.175$\pm$0.044 & 1.796$\pm$0.033 & 208/193 \\ 
56054.947 & 1.732$\pm$0.059 & 0.341$\pm$0.112 & 167/185 & 1.111$\pm$0.022 & 1.090$\pm$0.044 & 1.877$\pm$0.034 & 193/186 \\ 
56056.029 & 1.624$\pm$0.056 & 0.321$\pm$0.085 & 181/197 & 1.067$\pm$0.019 & 1.270$\pm$0.047 & 1.764$\pm$0.032 & 208/198 \\ 
56059.358 & 1.712$\pm$0.054 & 0.226$\pm$0.102 & 171/198 & 1.069$\pm$0.019 & 1.203$\pm$0.043 & 1.807$\pm$0.032 & 185/199 \\ 
56061.296 & 1.700$\pm$0.057 & 0.289$\pm$0.107 & 178/187 & 1.115$\pm$0.022 & 1.207$\pm$0.047 & 1.825$\pm$0.034 & 199/188 \\ 
56066.310 & 1.578$\pm$0.070 & 0.509$\pm$0.144 & 164/130 & 1.645$\pm$0.035 & 1.779$\pm$0.093 & 1.770$\pm$0.039 & 202/131 \\ 
56073.333 & 1.576$\pm$0.047 & 0.225$\pm$0.088 & 224/236 & 1.355$\pm$0.020 & 1.908$\pm$0.060 & 1.671$\pm$0.027 & 243/237 \\ 
56074.059 & 1.628$\pm$0.062 & 0.168$\pm$0.113 & 183/184 & 1.330$\pm$0.028 & 1.817$\pm$0.074 & 1.704$\pm$0.035 & 190/185 \\ 
56076.021 & 1.565$\pm$0.050 & 0.336$\pm$0.095 & 189/224 & 1.371$\pm$0.022 & 1.771$\pm$0.057 & 1.709$\pm$0.028 & 226/225 \\ 
56077.273 & 1.608$\pm$0.048 & 0.274$\pm$0.088 & 212/232 & 1.451$\pm$0.024 & 1.850$\pm$0.061 & 1.729$\pm$0.027 & 241/233 \\ 
56078.208 & 1.565$\pm$0.065 & 0.383$\pm$0.122 & 172/166 & 1.311$\pm$0.026 & 1.619$\pm$0.068 & 1.723$\pm$0.035 & 202/167 \\ 
56078.423 & 1.636$\pm$0.056 & 0.236$\pm$0.100 & 200/215 & 1.316$\pm$0.024 & 1.661$\pm$0.058 & 1.745$\pm$0.031 & 216/216 \\ 
56086.973 & 1.538$\pm$0.051 & 0.190$\pm$0.090 & 249/241 & 2.055$\pm$0.031 & 3.192$\pm$0.103 & 1.627$\pm$0.027 & 262/242 \\ 
56089.047 & 1.637$\pm$0.053 & 0.339$\pm$0.101 & 208/216 & 2.187$\pm$0.037 & 2.506$\pm$0.084 & 1.786$\pm$0.030 & 241/217 \\ 
56090.049 & 1.542$\pm$0.045 & 0.291$\pm$0.083 & 229/255 & 1.591$\pm$0.022 & 2.225$\pm$0.063 & 1.669$\pm$0.025 & 265/256 \\ 
56090.923 & 1.660$\pm$0.046 & 0.327$\pm$0.088 & 207/226 & 1.404$\pm$0.022 & 1.568$\pm$0.051 & 1.790$\pm$0.027 & 248/227 \\ 
56093.988 & 1.704$\pm$0.086 & 0.198$\pm$0.155 & 105/105 & 0.954$\pm$0.028 & 1.117$\pm$0.061 & 1.796$\pm$0.049 & 110/106 \\ 
56095.050 & 1.543$\pm$0.050 & 0.321$\pm$0.095 & 236/235 & 1.532$\pm$0.022 & 2.078$\pm$0.069 & 1.679$\pm$0.028 & 271/236 \\ 
56101.280 & 1.528$\pm$0.056 & 0.384$\pm$0.106 & 246/214 & 1.620$\pm$0.025 & 2.120$\pm$0.075 & 1.697$\pm$0.030 & 285/215 \\ 
56103.281 & 1.563$\pm$0.045 & 0.273$\pm$0.082 & 261/261 & 1.555$\pm$0.023 & 2.137$\pm$0.061 & 1.689$\pm$0.025 & 294/262 \\ 
56110.359 & 1.555$\pm$0.077 & 0.340$\pm$0.147 & 108/122 & 1.081$\pm$0.025 & 1.411$\pm$0.069 & 1.700$\pm$0.042 & 124/123 \\ 
56116.297 & 1.803$\pm$0.062 & 0.110$\pm$0.119 & 165/163 & 1.040$\pm$0.025 & 1.120$\pm$0.054 & 1.849$\pm$0.038 & 168/164 \\ 
56131.468 & 1.538$\pm$0.057 & 0.293$\pm$0.105 & 171/198 & 1.296$\pm$0.023 & 1.820$\pm$0.068 & 1.665$\pm$0.032 & 194/199 \\ 
56138.152 & 1.438$\pm$0.042 & 0.179$\pm$0.072 & 281/288 & 1.807$\pm$0.023 & 3.348$\pm$0.082 & 1.524$\pm$0.023 & 299/289 \\ 
55959.895 & 1.562$\pm$0.047 & 0.321$\pm$0.084 & 240/247 & 1.494$\pm$0.021 & 1.964$\pm$0.057 & 1.709$\pm$0.025 & 283/248 \\ 
55966.582 & 1.652$\pm$0.054 & 0.280$\pm$0.103 & 161/190 & 1.501$\pm$0.028 & 1.774$\pm$0.067 & 1.764$\pm$0.032 & 183/191 \\

\hline                                   
\end{tabular}
\end{table*}